\documentclass[a4paper,fleqn]{cas-dc}

\usepackage[sort&compress,numbers]{natbib} 

\usepackage{graphicx}
  \usepackage{textcomp}
  \usepackage{multirow}
  \usepackage{graphicx}
  \usepackage[noend, boxruled, linesnumbered]{algorithm2e} 
  \usepackage{algorithmic}
  \usepackage[figurename=Fig.]{caption}
  \usepackage{subcaption}
  \usepackage{amssymb}
\usepackage{pifont}
\usepackage[switch]{lineno}

\def\tsc#1{\csdef{#1}{\textsc{\lowercase{#1}}\xspace}}
\tsc{WGM}
\tsc{QE}

\begin{document}
\let\WriteBookmarks\relax
\def\floatpagepagefraction{1}
\def\textpagefraction{.001}

\shorttitle{P4TE}    

\shortauthors{D.D. Robin}  

\title [mode = title]{P4TE: PISA Switch Based Traffic Engineering in Fat-Tree  Data Center Networks (Preprint: Accepted for publication in 
Elsevier Computer Networks)}  



%

\affiliation[1]{organization={Kent State University},
            addressline={}, 
            city={Kent},
            postcode={44240}, 
            state={OH},
            country={USA}}

\author[1]{Debobroto Das Robin}[]
\ead[1]{drobin@kent.edu}
\cormark[1]
\author[1]{Javed I. Khan}[]
\ead{javed@kent.edu}













\cortext[1]{Corresponding author}



 \begin{abstract}
  This work presents P4TE, an in-band traffic monitoring, load-aware packet forwarding,
  and flow rate controlling mechanism for traffic engineering in fat-tree topology-based data center networks
  using PISA switches. It achieves 
  sub-RTT reaction time to change in network conditions, improved flow completion time, and
  balanced link utilization. Unlike the classical probe-based monitoring approach, P4TE uses an in-band
  monitoring approach to identify traffic events in the data plane. 
  Based on these events, it re-adjusts the priorities of the paths. It 
  uses a heuristic-based load-aware forwarding path selection mechanism to respond to changing network conditions and 
  control the flow rate by sending feedback to the end hosts. 
  It is implementable on emerging v1model.p4
  architecture-based programmable switches and capable of maintaining the line-rate performance. 
  Our evaluation shows that P4TE uses a small amount of resources in the PISA pipeline and achieves an improved flow completion time 
  than ECMP and HULA.
  \end{abstract}




\maketitle

\section{Introduction}

    \textbf{Why PISA switches for Traffic Engineering}:
    Traffic engineering (TE) is the prevalent mechanism to achieve optimal resource usage and performance in data center networks (DCN). 
    The traffic engineering problem is intractable in general. 
    A large set of research~\cite{awduche2002overview,mendiola2016survey,benson2011microte}
    works formulated the \textit{traffic engineering} as a stochastic network utility maximization (NUM)~\cite{palomar2006tutorial} problem over a network 
    of fixed-function switches. 
    However, these switches
    can not make a dynamic decision for packet forwarding as they
    can only forward packets over a preconfigured
    set of paths.  Here, the network control plane collects monitoring results from the data plane   and recomputes forwarding entries to maximize 
    network utility using approximation schemes or heuristic-based algorithms.  
    Then the control plane re-configures the paths in the data plane, and they remain static until the next iteration of path reconfiguration. 
    Traditionally, this task is assigned to a centralized controller, which results in a lengthy feedback cycle.
    However, the traffic in DCN can change on a fine time scale~\cite{benson2010understanding}. 
    As a result, the newly computed paths can be non-optimal.
    Due to lack of programmability, the switches can not be customized for traffic-aware decision-making in the data plane.
    The recent emergence of \textit{Protocol Independent Switch Architecture} based programmable 
    switches (commonly referred to
    as the  PISA switch~\cite{bosshart2013forwarding})  
    along with P4 programming language~\cite{p416,bosshart2014p4}, has  enabled programmability in the data plane. 
    These computationally rich PISA switches~\cite{hauser2021survey,kfoury2021exhaustive}
    can provide dynamic decision-making capability to overcome
    the limitations of traditional fixed-function switch-based traffic engineering schemes.
    
    \textbf{No Silver Bullet}: Though PISA switches provide a rich set of computational capabilities compared to the traditional fixed-function switch, 
    they are not the silver bullet for \textit{traffic engineering}.  To maintain a Tbps scale line-rate, PISA switches are built as a multi-staged pipeline with 
    few important architectural constraints (section~\ref{ProgrammableSwitchArchitectureAndLimitations}).
    Data plane programs that can be mapped over the PISA pipeline (while satisfying the architectural constraints) are capable of  
    maintaining the line-rate throughput of these switches~\cite{jose2015compiling,bosshart2013forwarding}. These switches have a strict per-packet processing and stateful memory access budget. 
    It is still not clear exactly which class of the existing traffic engineering algorithms is implementable under these constraints. 
    Moreover, \textit{traffic engineering}  is computationally expensive \footnote{Even approximation algorithms for \textit{traffic engineering} are also highly complex~\cite{chiesa2016traffic}.};
    any newly designed  system not conforming to the architectural restrictions of PISA switches can not maintain the line-rate, leading to throughput loss.
    These factors make the design and implementation of traffic engineering systems using PISA
    switch a non-trivial task.

    Though the traffic engineering problem is intractable in general, 
    the highly regular structure of DCN network topologies can help reduce 
    the computational complexity~\cite{chiesa2016traffic}.
    This suggests that, instead of focusing on traffic engineering systems for generalized topology, it is
    more practical to work on  schemes for widely used DCN topologies 
    (fat-tree~\cite{al2008scalable}, bcube~\cite{lin2012hyper}, etc.). The reduced complexity of regular topologies 
    can help to design improved traffic engineering system while maintaining PISA switch's architectural constraints. 
    In this work, we searched for such a  system for fat-tree topology-based DCN that can be deployed 
    using commodity PISA switches and provides performance improvement.

    Several PISA switch-based systems~\cite{paolucci2019p4,mostafaei2020tel}
    focusing on various aspects of traffic engineering  in regular topology-based DCN
    are proposed in recent times. But their  \textit{static path selection} strategy 
    fails to adapt to dynamic changes in link performance metrics.
    Their \textit{probe-based monitoring} mechanism often fails to identify microbursts and  
    varying network conditions at a fine-grained timescale. 
    Moreover, their reliance on
    \textit{end-host-based rate-controlling} schemes also makes
    a sub-optimal decision in flow rate handling. 
    Various point solution~\cite{joshi2018burstradar,pizzutti2019adaptive,sivaraman2015dc,liu2016one,robin2022clb,sivaraman2016programmable} exists in the 
    literature to address these issues separately. 
    However, it is still not clear how these schemes can achieve optimal resource usage and performance improvement in an 
    integrated manner while maintaining the line-rate throughput of the PISA switches.
    Any scheme having less than line-rate throughput will slow down the whole chain of switches in the network. 
    It reduces the overall throughput of the whole system, which is against the objective of TE. 
    Therefore, \textit{line-rate} performance is the fundamental test a scheme has to pass in the case of TE.

    In this work, we present \textbf{P4TE}; a \textbf{P4} supported PISA switch-based \textbf{T}raffic 
    \textbf{E}ngineering system for 
    fat-tree topology~\cite{al2008scalable} based data center networks. 
    P4TE utilizes the programmability features of PISA switches to implement 
    \textit{link performance-aware} decision-making at each switch for 
    \textit{path selection} and \textit{rate-controlling} of the flows at line-rate.    
    Studies~\cite{greenberg2009vl2,alizadeh2010data}  show that DCN workload mainly consists of a large number of 
    \textit{latency-sensitive short flows } and a small number of \textit{throughput-oriented large flows}. 
    P4TE focuses on optimizing the flow completion time for these two types of flows.
    Its fully switch-based \textit{in-band} monitoring technique identifies traffic 
    events happening in the data plane with fine-grained accuracy. It utilizes the monitored 
    information  to implement a traffic-aware forwarding 
    path selection algorithm. Moreover, P4TE’s 
    \textit{switch-assisted} TCP window size-based rate control technique achieves sub-RTT 
    reaction time to network events and  adjusts the flow rates (through sending fake acknowledgment packet to the source)
    for efficient use of link bandwidth. 
    P4TE uses a simple and efficient local CPU-based algorithm for flow path controlling. 
    It does not require a 
    centralized controller and any modification in the end-host transport 
    layer protocol stack. Altogether, these techniques enable a
    faster response to dynamic change in the network compared to the end host-based or centralized \textit{traffic engineering} systems.
    To the best of our knowledge, these techniques are investigated 
    in isolation to achieve improved performance in data center networks.
    No prior work has explored the possibility of deploying all these techniques together in data center environments using 
    PISA switches for traffic engineering purposes while maintaining the line-rate. 
    The main contribution of this work is 
    designing  a traffic engineering scheme for fat-tree topology-based DCN, which 
    is implementable using currently available PISA
    switches and  achieves the line-rate performance.
    P4TE's overhead mainly involves an increased number of feedback packets for reporting monitoring events to the control 
    plane and TCP packet retransmissions. Our evaluation shows that despite its overheads, P4TE achieves improved flow completion time (FCT)  
    compared to its PISA switch-based alternatives (equal-cost multi-path routing (ECMP)~\cite{hopps2000analysis} and 
    HULA~\cite{katta2016hula})
    for two real life data center workloads (web-search~\cite{alizadeh2010data} 
    and data-mining~\cite{greenberg2009vl2} workload). 


    The rest of the paper is organized as follows.
    In Section~\ref{RelatedWorks}, we overview the related work.
    In section~\ref{ProgrammableSwitchArchitectureAndLimitations}, we present 
    the key restrictions  behind implementing traffic engineering schemes
    using P4 supported PISA switches. 
    Then, we present the design and implementation of \textbf{P4TE} in section~\ref{P4TE}.
    In section~\ref{Deployment},
    we discuss configuring the required parameters and the
    realizability of P4TE using existing programmable switches.
    In section~\ref{PerformanceEvaluation}, we compare P4TE’s performance with ECMP~\cite{hopps2000analysis} and HULA~\cite{katta2016hula}
     using production data center traffic traces
    and network topology. Finally, we conclude the paper in
    section~\ref{Conclusion}.

    \section{Related Work} \label{RelatedWorks}

    Any technique to optimize one or more performance objectives of a network can be termed as a
    Traffic Engineering (TE) system. 
    A large number of proposals~\cite{noormohammadpour2017datacenter,mendiola2016survey,lee2002survey,zhang2020survey,wang2008overview}
    exist in literature focusing on the optimization 
    of different traffic engineering objectives. However, it remains unclear how (or whether) these proposals (or which subset of them) 
    can be realized in commodity programmable switches while maintaining the line-rate throughput. 
    Besides this,  several works have focused~\cite{wischik2011design,bensley2017datacenter,khalili2013mptcp} on improving the end host 
    transport layer protocol stack 
    for network performance improvement. 
    However, in this work we focus on important works related to traffic engineering in the context
    of DCN which can be implemented on recently emerging PISA switches and does not depend on modification in end host protocol stack.

    The performance of traffic engineering systems in DCN depends mainly on two key factors: a) monitoring performance of the paths and select 
    a suitable path for a packet b) controlling flow rates based on link bandwidth availability. Various PISA switch-based 
    systems exist in the literature  separately addressing these
    issues.

    For example, ECMP is the dominant technique for path selection in DCN using PISA switches.  
    Its stateless and traffic load-unaware hash-based path selection mechanism consumes minimal resources in the PISA pipeline 
    and provides immunity to packet re-ordering. 
    Due to these features ECMP has been used in widespread use cases~\cite{eisenbud2016maglev,miao2017silkroad}.
    However, ECMP often performs poorly in case of hash collision~\cite{benson2010understanding, he2015presto}
    and leads to uneven traffic splitting. It results in link congestion despite having a spare capacity in alternate paths~\cite{benson2010network}.
    Moreover, ECMP's incapability to consider traffic characteristics and resource usage for path selection make 
    it oblivious to various  objectives of traffic engineering. 

    Few recent works attempted to improve ECMP's performance by using a centralized controller's 
    global knowledge~\cite{turkovic2018fast,pizzutti2019adaptive}. 
    Due to their centralized nature, these systems are slow to react to fine-grained traffic variations in the data plane.  
    On the other hand, distributed schemes
    mainly rely on out-of-band probe-based results to
    choose the best path for a destination at each switch. For example, 
    HULA~\cite{katta2016hula} uses probe-based results  to find
    the best path to a destination at each switch. But, using only one best path 
    may lead to congestion in the best path rapidly. MP-HULA~\cite{benet2018mp} tries to augment this limitation by tracking top $k$ paths.
    However, it depends on the Multipath TCP (MPTCP)~\cite{wischik2011design} protocol support in the end-host protocol stack. Moreover,
    it also requires  SHA-1~\cite{standard1995federal} algorithm-based token generation for every MPTCP subflow in the switch. 
    However, such external functions are not readily available in commodity~\cite{opentofino} PISA switches.
    Contra~\cite{hsu2020contra} uses a policy-based probing technique to find performance-aware routes. 
    However, their knowledge of the best path is dependent on probe frequency. 
    But, periodic probe-based systems can not capture changing network conditions between successive probes. 
    As a result, these schemes may completely miss the microbursts in the data plane. 
    Several PISA switch-based systems~\cite{chen2019fine,van2017towards,narayana2017language,sivaraman2017heavy,harrison2018network,zaballa2021towards,kim2016band} exist 
    for in-band monitoring of various key metrics in the data plane. 
    For example, both BurstRadar~\cite{joshi2018burstradar} and snappy~\cite{chen2018catching} are capable of identifying microbursts 
    at a fine time scale.
    INT~\cite{kim2016band} can monitor both link utilization rate and queue build-up at different ports. 
    Univmon~\cite{liu2016one} can measure large flows, microbursts, and several other important traffic characteristics in the data plane. 
    However, they do not provide any straightforward mechanism to use the collected information in traffic engineering schemes.
    Moreover, they share the processing budget in a PISA pipeline with normal data packets. 
    They may violate the processing budget constraints of PISA switches when integrated with other schemes for path selection and rate-controlling.

    The majority of the existing PISA switch-based traffic engineering systems rely on the end host’s transport layer protocol stack for flow rate control.
    End host-based transport layer protocols (TCP, DCTCP~\cite{bensley2017datacenter}, MPTCP~\cite{khalili2013mptcp,ford2012tcp}, etc.) 
    need $RTT$ time to react 
    to changes in network conditions.  
    But microbursts are one of the main reasons behind the sub-optimal performance of DCN~\cite{benson2010network}, and they may last for a 
    duration shorter than $RTT$ time. Switch-assisted rate control 
    protocols~\cite{katabi2002congestion,dukkipati2006rcp} can perform better than their end-host-only counterparts by 
    calculating the fair rate for flows using aggregate knowledge of traffic collected by the switches and explicitly informing the source. 
    These protocols need to maintain explicit states~\cite{dukkipati2006rcp} about active flows in DCN. 
    But, DCN carries a large number of flows,  
    and maintaining an  up-to-date per-flow state is not scalable in a PISA pipeline. Hence, despite having interesting properties 
    (fair rate sharing, quick rate adjustment, etc.), these kinds of protocols are not scalable for large-scale data centers. 
    Another approach in literature relies on signals generated from the switch toward the
     source to control the flow rate. 
    For example, in~\cite{feldmann2019p4} authors proposed the use of explicit congestion notifications (NACK) packets toward a flow source 
    about congestion in a switch.  However, the work is unable to use spare link bandwidth in the absence of congestion. 
    Similarly, BFC~\cite{276958} monitors the queues in a switch and uses backpressure to control flow rate. However, it is not capable of selecting 
    paths for a flow in a load-aware manner.

    \begin{table}[width=\columnwidth]
      \begin{tabular}{|c|c|c|c|}
      \hline
      \begin{tabular}[c]{@{}l@{}}Feature $\rightarrow$ \\ $\downarrow$ Scheme\end{tabular}& \begin{tabular}[c]{@{}l@{}}{\hspace{3mm}In-band} \\ Monitoring\end{tabular} & \begin{tabular}[c]{@{}l@{}}Load  \\ Aware Path \\Selection\end{tabular} & \begin{tabular}[c]{@{}l@{}}\hspace{5mm}Rate \\ Adaptation\end{tabular} \\ \hline
      ECMP~\cite{hopps2000analysis}                                         & \ding{55}               & \ding{55}                    & \ding{55}                \\ \hline
      HULA~\cite{katta2016hula}                                         & \ding{55}                & \checkmark                    & \ding{55}                \\ \hline
      BurstRadar~\cite{joshi2018burstradar}                                   & \checkmark                & \ding{55}                    & \ding{55}                \\ \hline
      RCP~\cite{dukkipati2006rcp,bosshart2013forwarding}                                          & \ding{55}                & \ding{55}                    & \checkmark                \\ \hline
      Contra~\cite{hsu2020contra}                                       & \ding{55}                & \checkmark                    & \ding{55}                \\ \hline
      BFC~\cite{276958}                                       & \checkmark                 & \ding{55}                      & \checkmark                \\ \hline
      P4TE                                          & \checkmark                 & \checkmark                     & \checkmark                \\ \hline

    \end{tabular}
    \caption{{\centering{\small{Features of different PISA switch based schemes. }}}}
    \label{ComparisonTable}
    \end{table}

    The majority of the existing PISA switch-based schemes consider 
    the problem of \textit{path selection} and \textit{rate adaption} separately. 
    Moreover, the use of a \textit{probe-based} monitoring scheme
    makes them prone to miss crucial traffic events between
    successive probes. Nevertheless, for effective traffic engineering, all
    of them are equally important. Currently available PISA switch based schemes fail to 
    utilize at least one of these mechanisms, as shown in Table~\ref{ComparisonTable}.
    P4TE includes all these in its design. Its design goals include a) scalable \textit{in-band monitoring}
    of traffic events in the data plane for generating
    a better view of incoming traffic and link performances b)
    a \textit{traffic-aware path selection} scheme to avoid congestion
    buildup  c) achieving sub-RTT
    reaction time on varying network conditions and improving
    the performance of DCN through \textit{flow rate control}, and 
    d) finally, deployable using existing PISA switches to achieve line-rate throughput.

    \section{PISA \& Limitations for TE}\label{ProgrammableSwitchArchitectureAndLimitations}

    Reconfigurable match-action table (RMT)~\cite{bosshart2013forwarding} is the most 
    popular paradigm for PISA switches. 
    They are designed as 
    a pipeline of several components for packet processing.
    At first, a packet passes through a programmable packet header parser where the header 
    fields are extracted, and then the packet traverses the ingress stage of the pipeline. 
    The ingress stage consists of multiple match-action stages. 
    Each of them  can match packet header 
    fields and metadata with control plane configured values using match-action-table (MAT). Depending on the result of matching, a limited number of 
    actions (stateless or stateful) can be executed. In the ingress 
    stage of the pipeline, the egress port is determined and written in the metadata field. After the ingress stage, 
    the packet passes through the egress stage. The egress stage is similar to the ingress stage in architecture. 
    But, the egress port can not be modified 
    after a packet enters the egress stage. Moreover, egress queueing information is only available at this stage. 
    Finally, a packet goes through the deparser and scheduler 
    before being transmitted. High throughput switches deploy multiple pipelines 
    for packet processing, and the control plane runs in a separate thread.

    To maintain high  throughput (Tbps scale), 
    PISA switches are designed with several key architectural restrictions~\cite{bosshart2013forwarding,ben2018efficient}. These restrictions raise 
    several  challenges for traffic engineering algorithms. Below, we have listed a few of the most important among them. Though we have discussed these restrictions
    in the context of RMT architecture, these restrictions are highly likely to stay for future high throughput programmable switch architectures.

    \subsection{Limited Processing Power}  \label{LimitedProcessingPowerSubSubSection}
    To maintain a fixed per-packet processing delay, RMT pipelines 
    are designed with a limited number of stages and a limited number of actions per stage. 
    These devices do not support complex actions (i.e., floating-point 
    operation or division) and complex branching (nested if-else) in a single 
    stage. Existing complex algorithms 
    for traffic optimizations often require more processing budget 
    than what programmable switches can afford in the pipeline. 

    \subsection{Memory Size \& Access Limitation} \label{PerStageMemorySizeandAccessLimitationsubsubSection} 
    In PISA, a small amount of costly SRAM is dedicated to each stage in the pipeline. 
    At 1GHz speed, only one read-write can be accommodated in each stage. 
    As a result, a packet
    has to face a strict stateful memory access budget in
    the pipeline, which may not be sufficient for implementing
    existing traffic engineering algorithms.
    To avoid read-write hazard caused by concurrent memory access to the same memory location, 
    sharing access to the same memory blocks from multiple stages is prohibited. As 
    important link performance metrics
    (queue depth, port utilization rate, etc.) are available only at the egress stage, 
    and from the ingress stages, they are not accessible. 

    \subsection{Inability to Modify MAT from Data Plane} \label{InabilitytoModifyMatchActionTablesfromDataPlaneSubsubSection}
    The main mechanism of programmability in RMT architecture is \textit{MAT}, and they are 
    configurable (write/update) only from the control plane. 
    Forwarding path selection logic is expressed through MAT.  For traffic-aware path selection,
    a \textit{traffic engineering} scheme needs to adjust forwarding paths and their priorities based on the data plane's feedback.
    The CPU-based control plane is slower than the data
    plane threads and runs in separate threads. 
    It results in a delay between actual events in the data plane and the control plane's MAT modifications. 
    Moreover,  the update process is not atomic,
    which can lead to inconsistent MAT configurations.

    \subsection{Lack of High-Level Programming Constructs } \label{LackofHighLevelProgrammingConstructsSubSubSection } 
    One direct implication of the first limitation (section~\ref{LimitedProcessingPowerSubSubSection}) is,  
    loop-construct can not be supported in the data plane to maintain a small per-packet processing budget. But, various dynamic data
    structure (queue, stack, graph, etc.) or most basic CPU-based
    algorithms (sorting, searching, etc.) can not be implemented
    without a loop. Due to the lack of these constructs, the existing
    rich set of works on traffic engineering needs to be redesigned for use in
    programmable switches. Moreover, important operations for
    event-based packet processing~\cite{ibanez2019event} (i.e., timer, generic rate calculator, 
    etc. ) are yet to be
    available in commercially available programmable switches.

    \subsection{Lack of Event-Driven Programming} \label{LackofEvent-DrivenProgrammingCapabilitySubSubSection} 
    In PISA switches, any kind of activity or event identification can only be triggered by a packet
    that is undergoing processing in the pipeline. But some
    important traffic engineering algorithms need to execute operations independent of the packet in the processing pipeline. Such as periodic
    measurement of link utilization, measuring
    flow rate or queue occupancy in a specific time interval, clearing previous flow states from stateful memory once a flow
    timeout, etc. These features are not available in PISA switches~\cite{ibanez2019event}.

    \subsection{Inability to Generate New Packet}  \label{InabilitytoGenerateNewPacketSubSubSection} 
    For out-of-band communication (needed for conveying any traffic event-related information) 
    with neighbor switches or the control plane, 
    on-demand new packet generation from
    the data plane is required. In currently available PISA switches, a new packet can only be
    generated by replicating or recirculating an existing packet.
    Due to run-to-completion style architecture, newly generated
    packets need to undergo full processing in the ingress and (or)
    egress stage. If the number of packets generated by replicating or 
    recirculating existing packets are not controlled; a data packet
    and new packet generated from it, they together  can reduce the
    effective throughput of a pipeline into half or even worse.

    Altogether, these factors limit the spectrum of deployable algorithms for traffic engineering on programmable switches. 
    Either existing algorithms are needed to be carefully redesigned or new 
    algorithms have to be developed for conforming to the limitations of programmable switches. 


    \section{P4TE } \label{P4TE}

\begin{figure}
    \includegraphics[trim=0 0in 0in 0,scale=.34]{{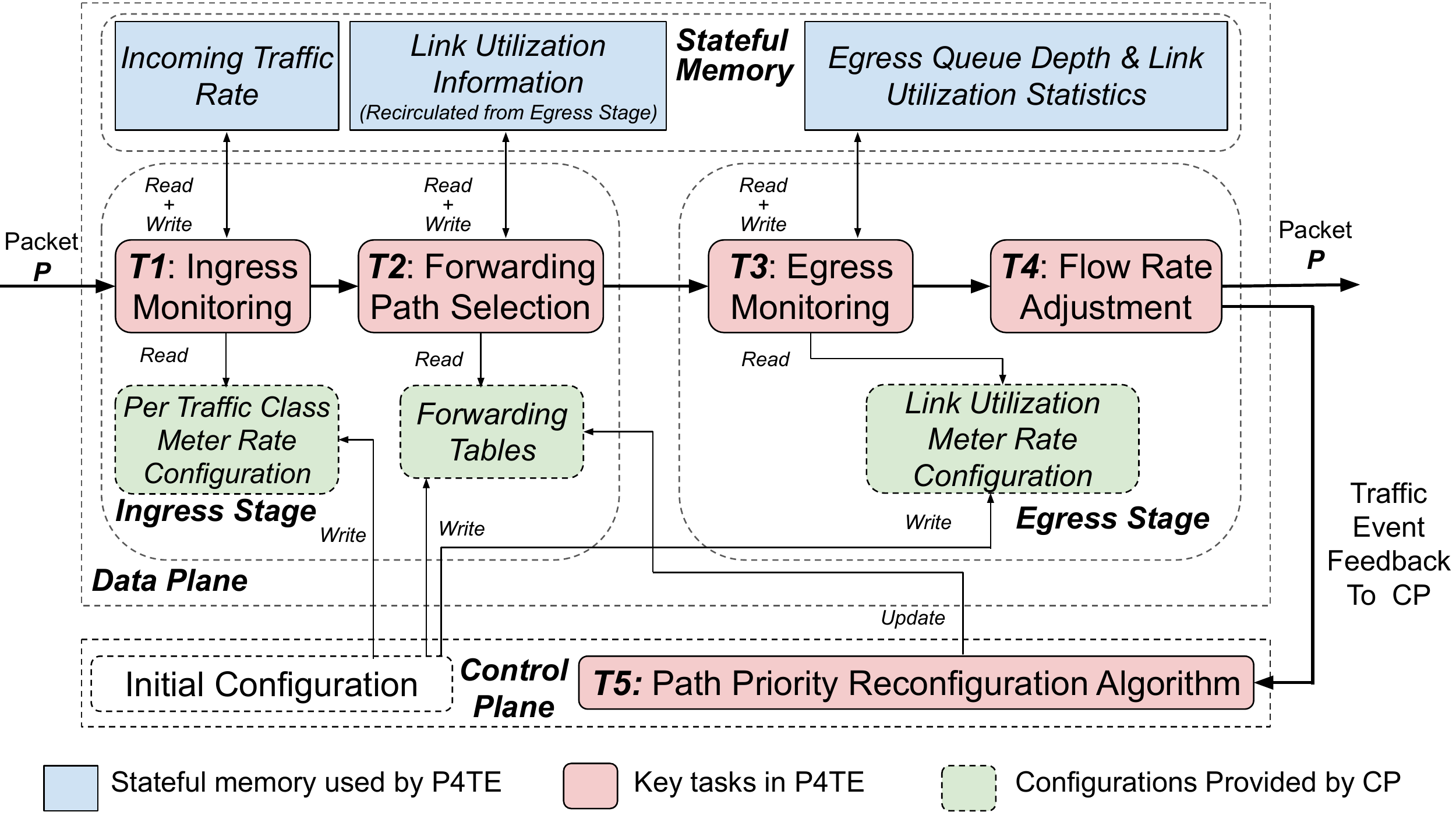}}
    \caption{\centering{\small{P4TE's Architecture}}}
    \label{fig:GenericTeModelMapping}

    \end{figure}

    \textbf{Overview}: P4TE is designed around five key tasks (fig.~\ref{fig:GenericTeModelMapping}). P4TE monitors each link's utilization rate  
    and queue depth (\textit{\textbf{T3}})  
    at each switch to identify congestion and relative qualities of the links.
    Using the monitored results, it implements a simple and fast local control plane based algorithm (\textit{\textbf{T5}})
    to maintain the priorities of the links at each switch. 
    P4TE implements 
    a congestion-aware greedy heuristic-based path selection mechanism (\textit{\textbf{T2}}) to achieve improved flow completion time.

    Now, the number of flows is large compared to the available  links to forward them. 
    They compete with each other for link bandwidth to meet their demands, and contention arises.
    Analysis of data center workloads~\cite{greenberg2009vl2,alizadeh2010data,roy2015inside} shows the majority 
    of the flows (more than 90\%) are short in size; whereas the 
    rest of the few flows are large, and they carry the majority of the bytes (more than 90\%). 
    It suggests a  flow size based mechanism can be 
    an effective choice.  P4TE
    assumes that each flow is tagged with its appropriate traffic
    class (\textit{short} or \textit{large}). 
    Various schemes~\cite{curtis2011mahout,ballani2015enabling} 
    exist in the literature for highly accurate traffic classification in the
    end host. Any of these schemes can be used
    for this purpose. However, these schemes require information on the flow size threshold for identifying large flows. 
    Operators run monitoring schemes~\cite{roy2015inside,kandula2009nature,benson2010network} to 
    collect various flow statistics (e.g., arrival time, flow count, flow size, etc.). 
    The flow size threshold can be derived from these data and used for flow classification purposes. 
    Following the previous works~\cite{alizadeh2013pfabric,zeng2019congestion,yu2021programmable} in this domain, P4TE considers any flow 
    of size greater than the 90th percentile flow size of overall traffic load as \textit{large flow}; the rest of the flows are considered as 
    \textit{short flow}.
    Once the flows are tagged, P4TE assumes every traffic class is assigned
    weights proportional to their estimated flow count (90\% and 10\% for short and large flows, respectively).
    A \textit{safe-rate} is allocated for each
    traffic class based on their weight, and the incoming traffic is monitored (\textit{\textbf{T1}}).
    In the case of resource contention (queue builds up at a port), flows from a specific traffic class are penalized (\textit{\textbf{T4}}) 
    only if their incoming traffic rate 
    is more than their \textit{safe-rate}. Conversely, a flow is awarded (\textit{\textbf{T4}}) more 
    link bandwidth if there is no resource contention and spare  link bandwidth is available.

    At a high level, P4TE works in two phases: 
    \begin{itemize}
    \item At the \textit{initialization} phase, the control plane provides preliminary 
    configuration for different components of P4TE.

    \item At the \textit{runtime} phase, 
    the data plane of each switch monitors two link performance metrics
    (link utilization rate 
    and queue depth) and reports relevant events to
    the control plane (switch local CPU). 
    The  data plane also monitors per traffic class
    (short and large flows) incoming traffic rate for each port.
    Next, the control plane reconfigures the path priorities based on these monitoring
    results. The  data plane selects an appropriate forwarding
    path for every packet using a heuristic-based distributed (switch-local) algorithm. 
    Lastly, depending on  bandwidth availability or congestion at any link, the data plane 
    increases or decreases a flow rate by sending  a fake ACK (FACK) packet to the flow source with
    a modified TCP window size. This FACK packet is generated by cloning the original data packet and modifying 
    its header fields. 

    \end{itemize}

    \textbf{Architecture}: P4TE is organized into three components: monitoring 
    (section~\ref{MonitoringSection}), path-controlling (section~\ref{PathControllingSubSection}), 
    and rate adaption (section~\ref{RateAdaptionSection}). At runtime, these three components execute the 
    five key tasks (\textit{\textbf{T1 to T5}}).   
    P4TE's high-level architecture 
    with mapping to  major components of PISA switches is shown in fig.~\ref{fig:GenericTeModelMapping}. 
    Through the rest of this section, we discuss the  
    design challenges, choices to overcome them, and the workflows of these three components.

    \subsection{Monitoring } \label{MonitoringSection}

    In P4TE, monitoring accomplishes two key tasks: 
    a) \textbf{T1}-Ingress Stage Monitoring: monitors per port incoming traffic rate of 
    two traffic classes (short and large flows) (section~\ref{TrafficClassBasedIncomingRateMonitoring} ) and 
    b) \textbf{T3}-Egress Stage Monitoring: monitors egress queue depth (section~\ref{EgressQueueDepthMonitoring}) and 
     utilization rate (section~\ref{LinkUtilizationRateMonitoring}) of each link.
    The \textit{ingress stage} monitoring aims
    to measure the link bandwidth consumption by flows from
    different traffic classes and utilize the measurement in identifying candidate flow for rate control.
    The goal of  
    \textit{egress stage } monitoring is to identify link congestion-related \textit{events} happening in the data plane 
    and provide quantitative information about  them to the control plane for comparing the relative quality of the links. 
    Achieving this in a scalable manner is challenging due to the PISA switch's architectural
    limitations  (section~\ref{InabilitytoGenerateNewPacketSubSubSection}). Next, we describe the challenges behind  
    monitoring different metrics and relevant  design choices in P4TE's design. 
    Configuration of the parameters used in the monitoring component is discussed in section~\ref{P4TEParameterSetupSection}.

    \subsubsection{Egress Queue Depth} \label{EgressQueueDepthMonitoring}
    \textbf{Challenges}: The majority of the existing solutions 
~\cite{alizadeh2010data, floyd1993random} use a fixed number of static threshold values of queue
    depth for identifying congestion over a link. 
    They consume limited resources in the PISA pipeline but unable to support P4TE's goal of identifying congestion and 
    providing quantitative 
    information about a link's performance. 

\begin{figure}
    \includegraphics[trim=0 3.25in 0 0, clip,scale=.34]{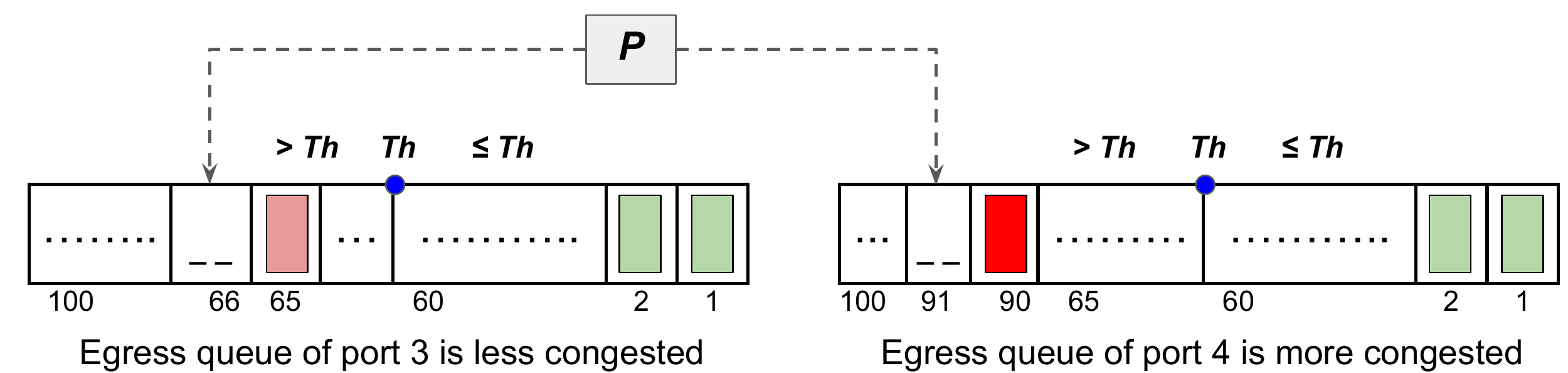}
    \caption{\centering{\small{Static threshold {\small ($Th$) } based congestion marking can not quantify congestion level.}}}
    \label{fig:EgressQueueDepthMonitoringExampleFigure}

    \end{figure}

    Consider, two  ports  (fig.~\ref{fig:EgressQueueDepthMonitoringExampleFigure}) 
    are configured with a queue capacity of 100 packets and 
    \textit{explicit congestion notification} (ECN)~\cite{floyd1994tcp} threshold ($Th$ ) of 60 packets. 
    A new packet $P$ arrives at the switch while 65 and 90 packets are waiting to be forwarded in 
    port 3 and 4's queue respectively. Forwarding $P$ through any of the two ports will  indicate congestion (queue depth $\geq Th$).
    But, port 4 is more congested, and forwarding more traffic through this port
    increases the chance of congestion. 
    A similar argument is applicable when $\leq Th$ packets are waiting in the port's queue. 
    Instead of a single threshold, using multiple static or dynamic (configured by the control plane) 
    thresholds also can not overcome these problems. 
    For example, the use of 2 thresholds (60 and 80 packets in the queue) divides a port's queue into 3 
    regions ([0,40], (40,60] and (60,100]). The problems mentioned above can occur within any  region also. 
    Moreover, comparing the current queue depth with multiple static thresholds 
    require a proportional number of ALU. It increases the amount of resource consumption in the
    PISA pipeline (section~\ref{LimitedProcessingPowerSubSubSection}) and leaves fewer resources for other functionalities in the switch.
    Altogether, static threshold-based congestion identification can not provide enough 
    information to compare multiple link’s performances at low cost.

    As a dynamic approach, the queue depth observed by each packet can be sent back to the control plane.   
    The control plane can store and compare the queue depth of multiple links to identify their relative quality. 
    However, generating a new packet to propagate the change in queue depth to the control plane 
    will reduce the pipeline's throughput  
    to half of its capacity (sec.~\ref{InabilitytoGenerateNewPacketSubSubSection}). 
    Even using a static threshold ($Th$) can not help in this case. Because when more than 
    $Th$ packets are waiting in a port's queue, inserting a new packet in the queue will indicate congestion, 
    and a new packet must be created to send relevant
    information to the control plane. This process continues until queue depth drops below $Th$.

    \textbf{P4TE's mechanism}: Considering these factors, P4TE uses alg.~\ref{EgressMonitoringAlgorithm} 
    (line~\ref{EgressQueueDepthMonitoringAlgoStart}-~\ref{EgressQueueDepthMonitoringAlgoEnd} ) 
    for monitoring the egress queue depth. 
    It stores the maximum or minimum queue depth seen by any packet at a port's 
    queue within a window of $2\Delta$. If the
    queue depth seen by a packet at a port’s queue differs from
    the previously stored value by a margin of $\Delta$, P4TE considers it as an
    \textit{egress queue depth change (increase/decrease) event} and sends relevant information to the control plane.

\begin{figure}
    \includegraphics[trim=0 0 0 0, clip,scale=.34]{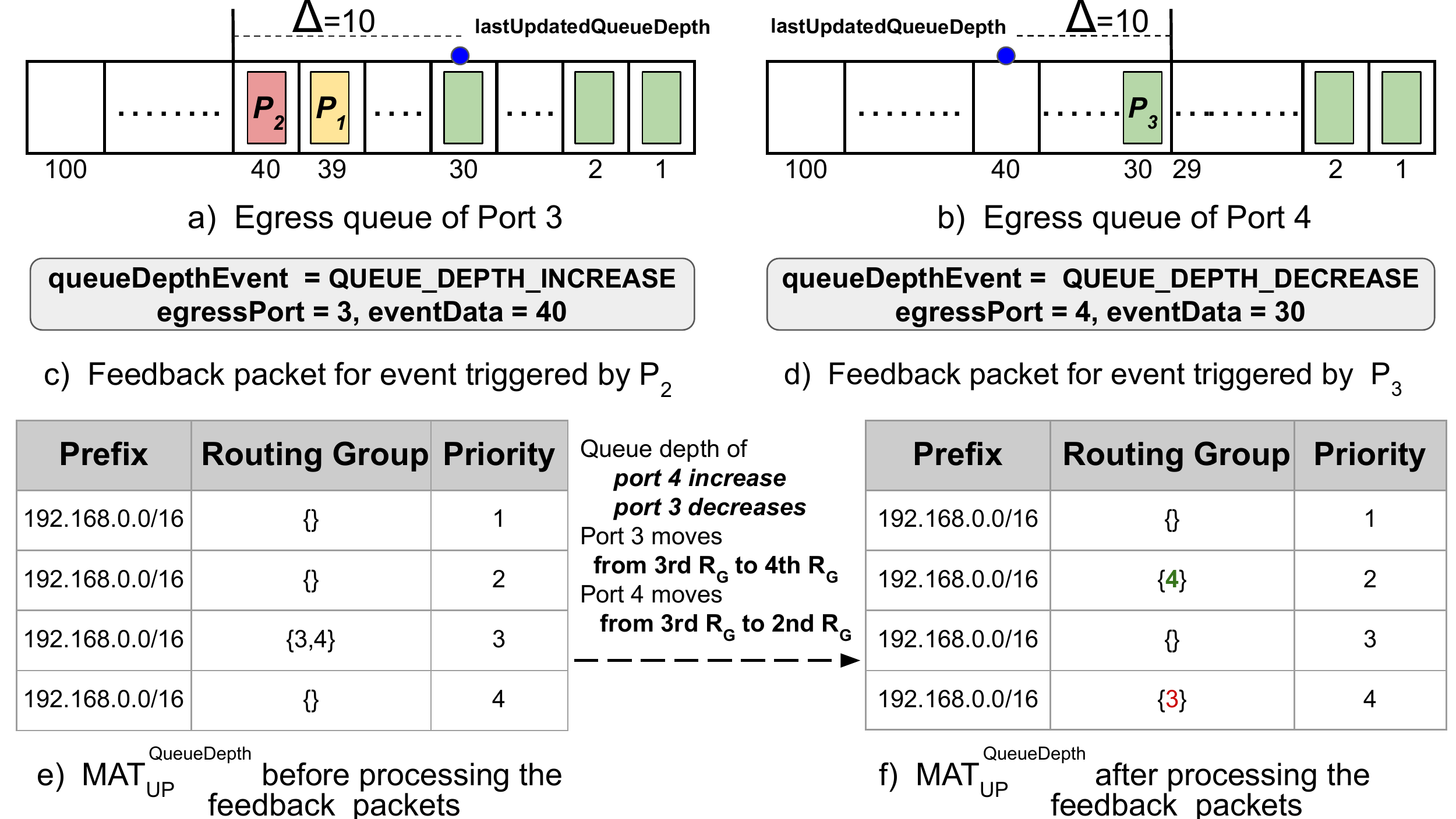}
    \caption{\centering{\small{ Egress queue depth a) increase and b) decrease event; corresponding feedback packets (c) and d)) sent to the control plane.
    ${MAT}^{QueueDepth}_{Up}$ e) before receiving the feedback packets and f) after processing the feedback packets according to 
    alg.~\ref{pathPriorityReadjustmentAlgorithm}.}}}
    \label{fig:EgressQueueDepthMonitoringExample}
    \end{figure}

    In the example of fig.~\ref{fig:EgressQueueDepthMonitoringExample}a and~\ref{fig:EgressQueueDepthMonitoringExample}b, 
    three packets $P_1$, $P_2$ and $P_3$ enter into the switch;
    $P_1$ and $P_2$ are set to be forwarded through port 3 and $P_3$ through port 4. 
    At port 3, $P_1$ observes egress queue depth 39, which doesn't differ 
    with $lastUpdatedQueueDepth$ by  $\Delta=10$. Hence no event is identified by the switch. 
    Whereas $P_2$ observes queue depth 40 at port 3 and  $P_3$  observes queue depth 30 at port 4, respectively. 
    These two are $\Delta$ increase and decrease from the $lastUpdatedQueueDepth$; 
    P4TE considers them as egress queue depth \textit{increase} and  \textit{decrease} event, respectively. 
    Information for these events is sent to the control plane through a feedback 
    packet (fig.~\ref{fig:EgressQueueDepthMonitoringExample}c and ~\ref{fig:EgressQueueDepthMonitoringExample}d).

    Firstly,  this algorithm generates one feedback packet 
    for $\Delta$ data packets, which helps  avoid an excessive reduction in packet processing throughput of the pipeline. 
    Secondly, relevant queue
    depth and port information are sent to the control plane once a traffic event is identified.  The control plane can use these values to 
    compare the quality of the links. 
    Altogether, this algorithm enables scalable 
    monitoring of egress queue depth events and collecting relative quality of 
    the links in the data plane.

  \begin{algorithm}
      
  \begin{algorithmic}[1]
    \footnotesize{
\item [\textbf{Input}: Packet P]
\item [\textbf{Output}: Packet P (with modified header fields)]
\item [\textbf{Procedure}: EgressStageMonitoringAlgorithm (P)]

\STATE queueDepthEevent = None  \label{EgressQueueDepthMonitoringAlgoStart} \linebreak
\tcp{ Every element of lastUpdatedQueueDepth \linebreak are  initialized with 0}
\STATE  lastUpdatedQueueDepth = oldQueueDepths[P.egressPort] 

\IF{P.metadata.queueDepth $\geq$ \linebreak lastUpdatedQueueDepth + $\Delta$}
  \STATE queueDepthEvent = QUEUE\_DEPTH\_INCREASE 
  \STATE eventData = P.queueDepth  
  \STATE oldQueueDepths[P.egressPort] = P.queueDepth
\ELSIF{P.metadata.queueDepth  $\leq$ lastUpdatedQueueDepth \\ - $\Delta$ }
  \STATE queueDepthEvent = QUEUE\_DEPTH\_DECREASE 
  \STATE eventData = P.queueDepth  
  \STATE oldQueueDepths [P.egressPort] = P.queueDepth
\label{EgressQueueDepthMonitoringAlgoEnd}
\ENDIF \linebreak
\STATE linkUtilizationRateEvent = None  \label{LinkUtilizationMonitoringAlgoStart} 
\STATE lastUpdatedPacketColor = oldPacketColors[P.egressPort]  \linebreak
\tcp{Every element of lastUpdatedPacketColor  \linebreak are  initialized with GREEN}
\STATE newPacketColor = match P.egressPort with ${MAT}^{LinkUtil}_{Rate}$ , \linebreak execute matching direct meter and get the color 
\STATE oldPacketColors[P.egressPort] = newPacketColor
\IF{ lastUpdatedPacketColor < newPacketColor}
  \STATE linkUtilizationRateEvent = \linebreak UTILIZATION\_RATE\_INCREASE 
\ELSIF{ lastUpdatePacketColor > newPacketColor}
\STATE linkUtilizationRateEvent = \linebreak UTILIZATION\_RATE\_DECREASE 
\label{LinkUtilizationMonitoringAlgoEnd} 
\ENDIF
\IF {{queueDepthEvent != None} \textbf{or} \linebreak {linkUtilizationRateEvent != None}}
      \STATE $P^\prime$ = make a copy of P
      \STATE copy queueDepthEvent, eventData, \linebreak linkUtilizationRateEvent, and P.metadata.egressPort to  $P^\prime$
      \STATE forward $P^\prime$ to Control Plane 
  \ENDIF
\STATE return P
}
\end{algorithmic}
\caption{\small{Egress Stage Monitoring Algorithm  }}
\label{EgressMonitoringAlgorithm}
    
    \end{algorithm}
    

    \subsubsection{Link Utilization} \label{LinkUtilizationRateMonitoring} 
    \textbf{Challenges}:
    Issues arising in queue depth monitoring (section~\ref{EgressQueueDepthMonitoring}) also arise here. Moreover, instead
    of monitoring the exact rate, current programmable switches only
    support fixed-rate (1 or 2 rates) based meters (section~\ref{LackofEvent-DrivenProgrammingCapabilitySubSubSection},~\cite{heinanen1999two}). 
    Rate configurations for these meters can be updated only from the control plane (section~\ref{InabilitytoModifyMatchActionTablesfromDataPlaneSubsubSection}). 
    But, the update process is not atomic. Hence, the meters can 
    provide inconsistent information during the update.
    As an example, assume 
    port 3 and 4 of a switch are configured with 40 Gbps \textit{Committed Information Rate} (CIR) and
    80 Gbps \textit{Peak Information Rate} (PIR). At some point, these rates are updated to 50 and 90 Gbps for both ports.
    Consider the case where the new rates are updated successfully for port 3,
    and the update for port 4 is still ongoing. At this moment, 
    if both the link utilization rate is 45 Gbps, the packet forwarded through port 3 will be marked in 
    GREEN color while the packet forwarded through port 4 will be marked in YELLOW color. 
    This leads to a wrong perception that port 3 is underutilized compared to port 4. 
    Moreover,  P4-meters do
    not provide any quantitative information about a flow; they only color mark the
    packets.
    As a result, monitoring and comparing
    link qualities based on utilization rate in a manner similar to monitoring queu-depth (section~\ref{EgressQueueDepthMonitoring}) 
    is not possible.

    \textbf{P4TE’s mechanism}: P4TE uses \textit{2 rate 3 color}-based meter~\cite{heinanen1999two} 
    for measuring  the utilization rate of each port. To
    avoid the problem of the inconsistent update, rate (CIR and PIR) configurations are kept \textit{fixed} over the whole 
    life cycle of P4TE. 
    For identifying \textit{link utilization rate} related events, P4TE follows 
    alg.~\ref{EgressMonitoringAlgorithm} (line~\ref{LinkUtilizationMonitoringAlgoStart}-~\ref{LinkUtilizationMonitoringAlgoEnd}). 
    Due to fixed-rate configurations, this algorithm can not adapt to varying traffic loads like queue depth monitoring (section~\ref{EgressQueueDepthMonitoring}). 
    But, P4 meters support burst (\textit{Committed Burst Size} (CBS) and 
    \textit{Peak Burst Size} (PBS)) handling capability near CIR and PIR.  
    Burst rates in meter configuration serve the same purpose as $\Delta$ in queue depth monitoring (section~\ref{EgressQueueDepthMonitoring}). 
    It triggers only one traffic event for a traffic burst of size  CBS (or PBS) near CIR (or PIR). This prevents 
    a reduction in throughput caused by an excessive number of feedback packets near the CIR or PIR.  
    Fig.~\ref{fig:fat-tree-topology-intiial-configurations}c shows an example of meter rate configurations for the ports of a switch. 

    P4TE's path selection algorithm (section~\ref{pathSelectionSubsection}) uses the information about a link's 
    utilization in the ingress stage.  
    However, this information is only available in the egress stage. 
    However, PISA switches do not allow cross-stage stateful memory access 
    (section~\ref{PerStageMemorySizeandAccessLimitationsubsubSection}).
    Hence, information about the  utilization  of the links  is not available in the ingress stage.   
    To make them available at the ingress stage, on identifying link utilization rate related events, besides sending feedback to the control plane, 
    P4TE recirculates a copy of the feedback packet to the ingress 
    stage. 
    Once the recirculated feedback packet reaches the ingress stage, P4TE stores the link utilization information of the port in an array of 
    stateful memory ($portUtilizations$) for use in the path selection algorithm (see alg.~\ref{PathSelectionAlgorithm}).

    \begin{figure}
      \includegraphics[trim=0 0in 0in 1.5in, clip,scale=.34]{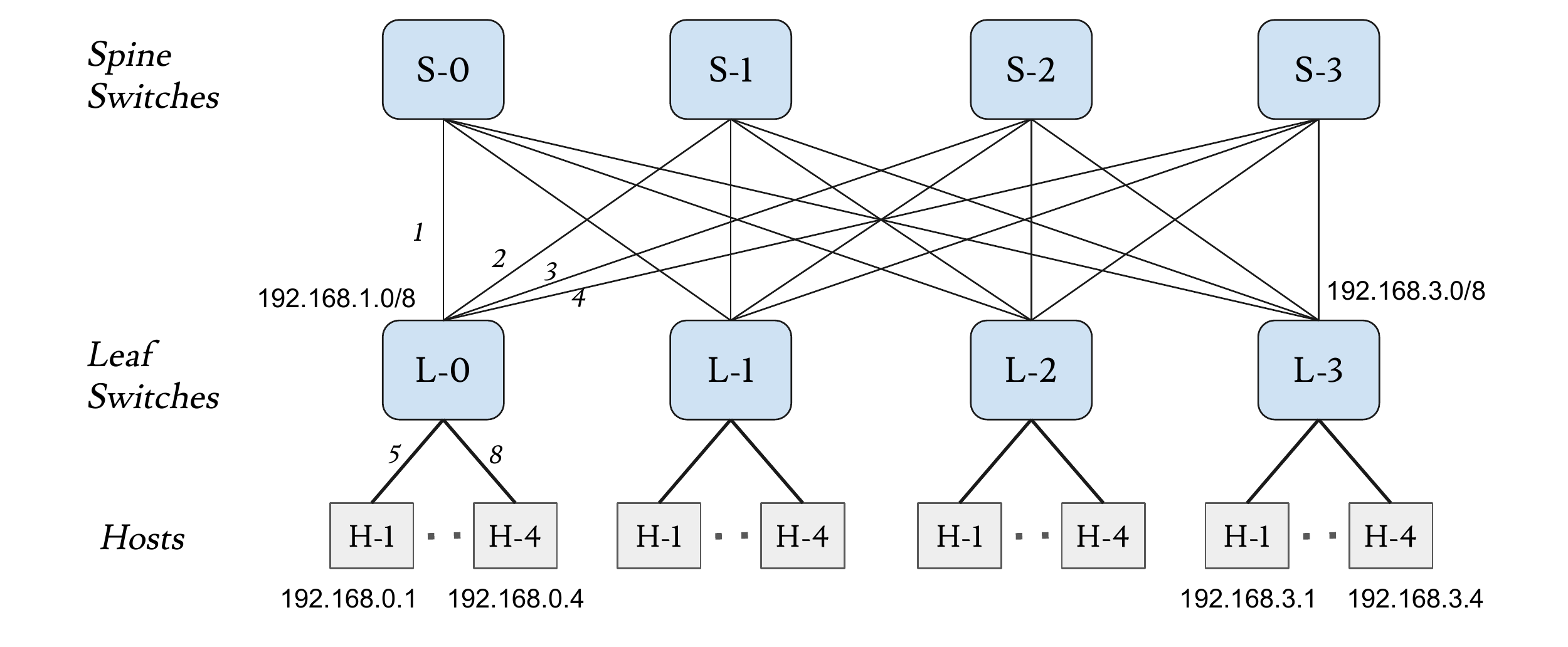}
      \caption{\centering{\small{ A fat-tree topology with 8 port switches  and hierarchiacal IP address assignment.}}}
      \label{fig:fat-tree-topology}
      \end{figure}
    
      \begin{figure}
      \includegraphics[trim=0 0in 0in 1.75in, clip,scale=.34]{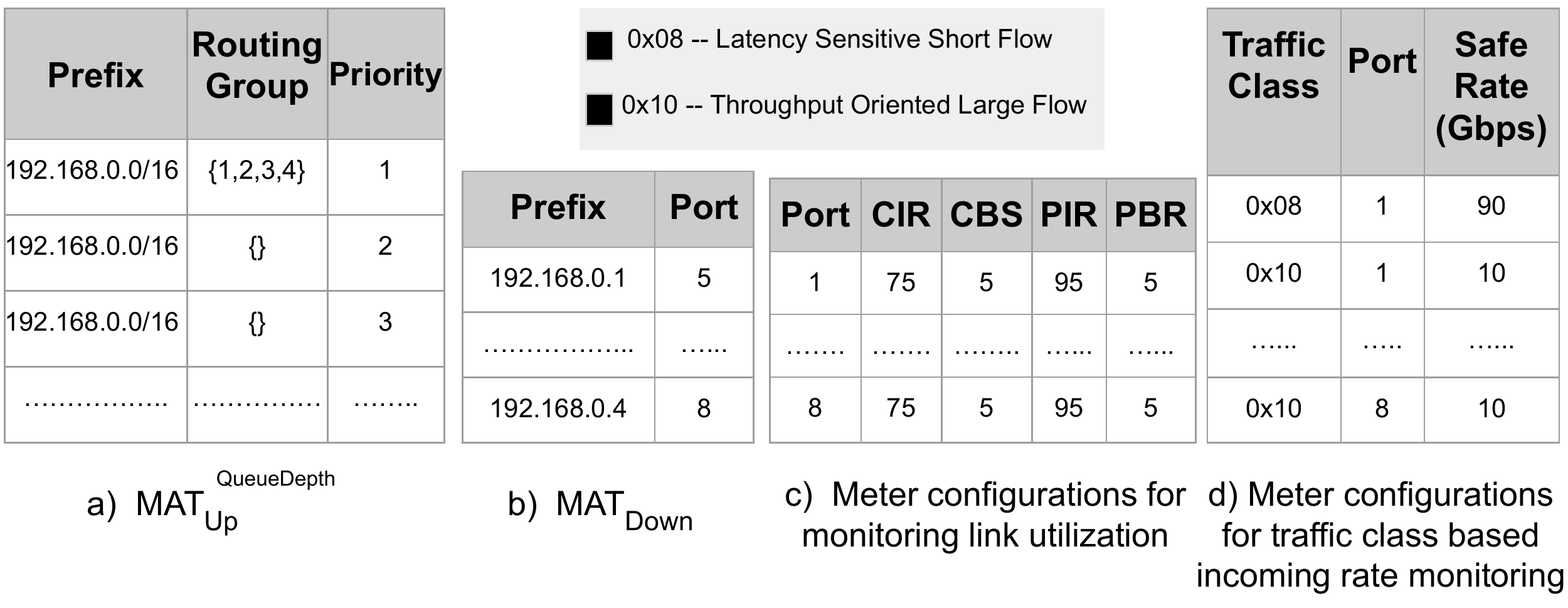}
      \caption{\centering{\small{ Example initial configurations for switch \textit{L-0} of topology (Fig.~\ref{fig:fat-tree-topology}) with 100 Gbps link speed.}}}
      \label{fig:fat-tree-topology-intiial-configurations}
      \end{figure}

    \subsubsection{Traffic Class Based Incoming Rate} \label{TrafficClassBasedIncomingRateMonitoring}
    P4TE assumes each flow is tagged with its appropriate traffic class (either short or large flow). 
    If the end-hosts do not tag the packets with a traffic class, 
    any existing PISA switch based system~\cite{sivaraman2017heavy,harrison2018network} to identify 
    large flows can be used for this purpose. 
    If flows from a traffic class consume too much
    resource, it can starve flows
    of other traffic classes in case of resource contention. 
      To avoid this,  P4TE follows a traffic class-based \textit{max-min} fairness approach.  
    Here each traffic class is 
    assigned a weight   proportional to their estimated flow count (section~\ref{IngressMonitoringRateParameterSetup}).
    An incoming link's capacity is divided among different traffic classes according to their weight and allocated as 
    their \textit{safe-rate}. P4TE monitors  whether  flows from a specific traffic class are crossing the \textit{safe-rate} or not.
    Issues arising in \textit{link utilization} monitoring (section~\ref{LinkUtilizationRateMonitoring}) also arise here, 
    and P4TE uses a meter based monitoring here also.

    P4TE uses 1 rate 2 color-based meters to monitor the incoming traffic 
    rate of flows from different traffic classes. 
    At the \textit{initialization phase}, P4TE configures  per traffic class rate information 
    (fig.~\ref{fig:fat-tree-topology-intiial-configurations}d) 
    for the meters of each port.
    Fig.~\ref{fig:fat-tree-topology-intiial-configurations}d shows example  
    rate configurations for a switch with 100 Gbps link/port bandwidth. Here, \textit{short} and \textit{large} flows are assigned 90 Gbps
    and 10 Gbps  \textit{safe-rate}, respectively. 
    At the \textit{runtime phase}, if incoming flows through a port cross the safe-rate for its traffic class, 
    P4TE marks  the packet 
    in YELLOW (GREEN otherwise) color to indicate the flow as \textit{unsafe (safe otherwise)}. 
    The color mark is stored in packet metadata ($incomingPacketColor$) for use in rate control (section~\ref{RateAdaptionSection}).

    \subsection{Path-Controlling} \label{PathControllingSubSection}
    Path-controlling component includes two tasks of P4TE: a) \textbf{T2}: path selection for a packet (section~\ref{pathSelectionSubsection})
    b) \textbf{T5}: priority reconfiguration (section~\ref{PathPriorityReconfigurationSubSection}) of available paths.  
    P4TE uses distributed (using switch local CPU) algorithm for both of them. 

    \subsubsection{Path Selection} \label{pathSelectionSubsection}
    P4TE uses a modified version of the existing two-level route lookup algorithm~\cite{al2008scalable}. 
    Next, we describe the algorithm in two parts: downward  and upward. 

    \paragraph{\textbf{Downward}}:  
    Fat-tree topology contains no path
    diversity in the downward direction, and P4TE uses IP 
    prefix-based matching here.
    The single administrative authority of a DCN enables hierarchical IP address 
    assignment~\cite{al2008scalable} with a single common IP prefix for the DCN. 
    As a result, the devices connected to the downward ports of a switch  remain in a single subnet. 
    Similar to existing works~\cite{al2008scalable},
    at the \textit{initialization phase}, P4TE configures the 
    match-action-table (${MAT}_{Down}$) for  downward route lookup with IP prefix and  port 
    of the connected end hosts or subnets at each switch.   
    Fig.~\ref{fig:fat-tree-topology-intiial-configurations}b 
    \footnote{P4TE implementation supports IPv6; IPv4 configuration is used in the examples for readability.}
    shows an example ${MAT}_{Down}$ 
    for switch \textit{L-0} of the topology in fig.~\ref{fig:fat-tree-topology}.  
    For path selection, a packet is at first matched against the ${MAT}_{Down}$; 
    If a matching entry is found, the packet is forwarded through the corresponding 
    port. Otherwise, the packet goes through the upward routing algorithm (described next) for path selection.

    \paragraph{\textbf{Upward}}: \label{UpwardRoutingParagraph}
    \linebreak
    \textbf{Challenges}: Fat-tree topology provides path diversity in the upward direction at each switch. 
    A traffic engineering scheme's success depends on the traffic-aware split of traffic over multiple paths 
    in the existence of path diversity. In programmable
    switches, this is a non-trivial task due to several reasons. Assume the simplest 
    scenario at any switch; there are $l$ available
    links toward a packet's destination with only one link performance metric to monitor. 
    Based on the monitoring result, maintaining the links in sorted order and selecting the
    best one entirely in the data plane requires loop support, 
    which is unavailable (section~\ref{LackofHighLevelProgrammingConstructsSubSubSection }) 
    in programmable switches. The only viable
    option is sending monitored information to the control plane.
    Then the control plane can
    arrange the $n$ links in the match-action-tables (MAT) in sorted
    order, and the data plane selects the best one from them. 
    However, the process of updating the relative order of $n$ links in the data plane from the control plane
    is not atomic (section~\ref{InabilitytoModifyMatchActionTablesfromDataPlaneSubsubSection}) and may lead to inconsistent behavior.

    Now, assume for a flow ($f$), the best path selected using a MAT is  $f_{path}$. Due to the reconfiguration 
    delay (section~\ref{InabilitytoModifyMatchActionTablesfromDataPlaneSubsubSection}), $f_{path}$ may not be the best path 
    for achieving traffic engineering objectives.
    To \textit{verify}, whether using $f_{path}$ will achieve the objectives or not requires the following steps:
    a) select $f_{path}$ for $f$ from stateful memory b) read $f_{path}$'s condition according to performance metrics for verifying whether it 
    is conforming to the traffic engineering objectives or not  
    c) if $f_{path}$'s use works against achieving the traffic engineering objectives, 
    then select a new path ($\overline{f_{path}}$) for the flow and, d) finally, 
    update $f_{path}$ or $\overline{f_{path}}$'s state (relative order among all available paths) in stateful memory 
    (used in step \textit{a}) so that the next packets can use the best path. 
    Clearly, steps a), b), and d) create \textit{true dependency} and 
    can not be rewritten in any other order. 
    But a PISA switch with 2 memory ports in each stage
    can afford only one read-write~\cite{bosshart2013forwarding} in a single stage at 1 GHz clock 
    speed (section~\ref{PerStageMemorySizeandAccessLimitationsubsubSection}). 
    As a result, this \textit{true dependency}
    is impossible to overcome in the data plane. 
    Hence, the fundamental
    problems for the traffic-aware path selection and verification
    will always remain the same.

    \textbf{P4TE's path selection mechanism}: 
    P4TE maintains the upward ports in  two link performance metrics (egress queue depth and link utilization rate) based \textbf{Routing-Groups ($R_G$)}. 
    The key idea is dividing the range of a link performance metrics into $r$ predefined sub-ranges 
    and allocating one \textit{routing-group} for each sub-range.
    Each \textit{routing-group} is a set of ports/links through which a packet can be forwarded. Every port in the same \textit{routing-group} has 
    an equal priority of being selected (ECMP like hash-based selection). 
    However,
    each of the $r$ \textit{routing-group} has a different priority, which determines their order of being selected in MAT.
    Membership of the links in a \textit{routing-group} is determined based on performance metrics  collected by the monitoring 
    component (section~\ref{MonitoringSection}).
    Performance metrics to \textit{routing-group} mappings are defined independently of the data plane logic. Hence, each port can be  
    mapped to any \textit{routing-group} independently of other ports. This removes the problem of inconsistent state caused by \textit{non-atomicity} 
    of updating MAT.

    P4TE uses two separate forwarding tables (${MAT}_{Up}^{QueueDepth}$ and ${MAT}_{Up}^{LinkUtil}$) 
    to arrange upward links in routing-groups according to the two monitored link performance metrics: egress queue depth and link utilization rate.
    For ${MAT}_{Up}^{QueueDepth}$, the buffer length is divided into $r$ (choice of $r$ is discussed 
    in section~\ref{DeltaForEgressQueueDepthMonitoring}) sub ranges. One routing-group for each sub range is allocated.
    Whereas, for the ${MAT}_{Up}^{LinkUtil}$ three routing-groups are used because, every link can be marked in one of the 
    three color (GREEN, YELLOW, RED) by the link utilization monitoring algorithm (see alg.~\ref{EgressMonitoringAlgorithm}). One routing-group for each color is sufficient. 
    These two MAT provides the opportunity to select the best  path for a flow based  on egress queue depth ($lowQueueDepthPort$) and 
    link utilization rate ($lowUtilPort$). 
    Finally, instead of using static hash-based or only one best path~\cite{katta2016hula} for all flows, P4TE
    uses a heuristic-based strategy to select suitable from these two paths for the short and large flows.

    \begin{figure}
    \includegraphics[trim=0 0in 0 0, clip,scale=.34]{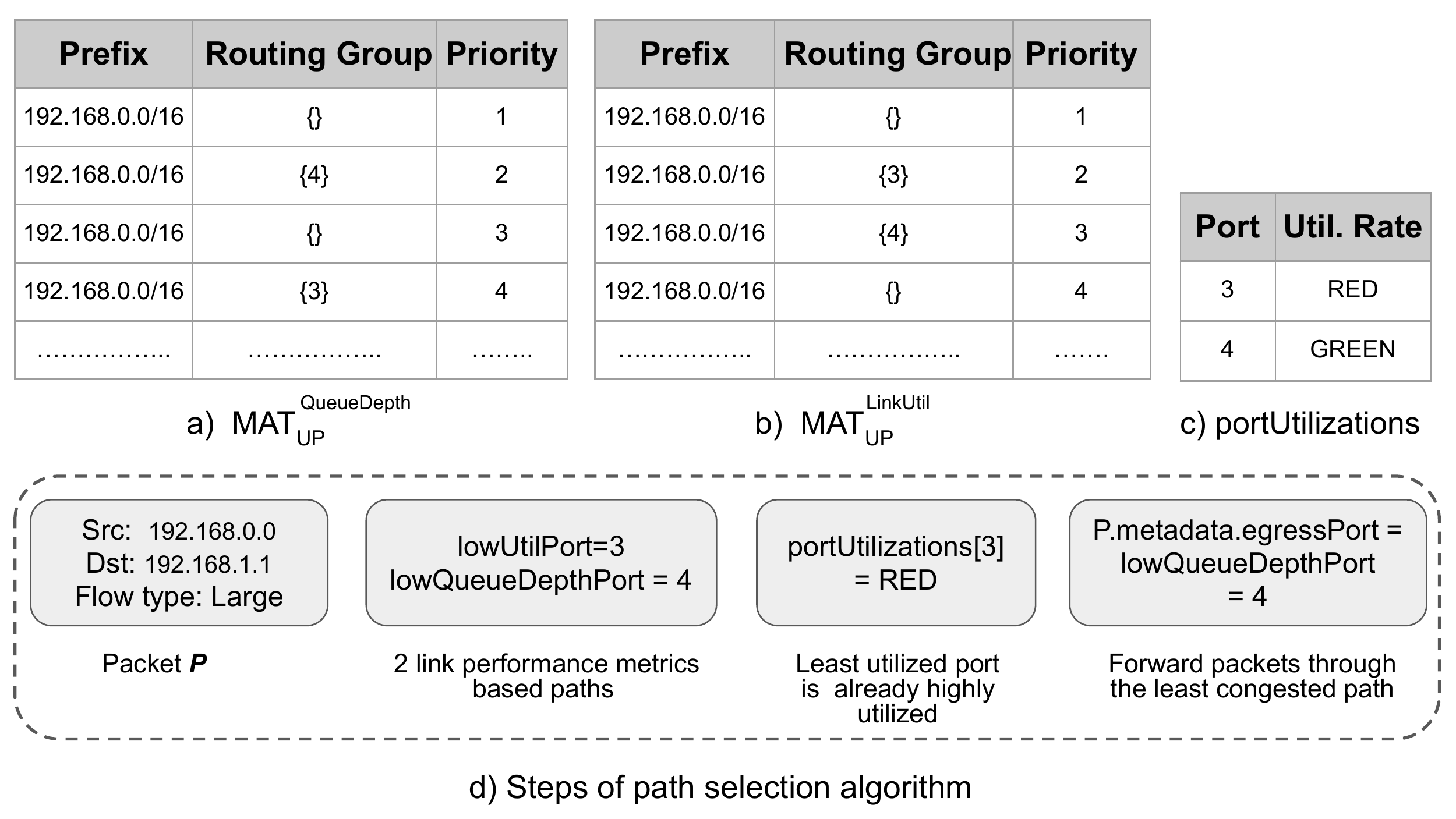}
    \caption{\centering{\small{Upward path selection algorithm example. }}}
    \label{fig:PathSelectionAlgo}

    \end{figure}

    At the \textit{\textbf{Initialization Phase}}, P4TE initializes the two match-action-tables (MAT) for upward routing: 
    ${MAT}_{Up}^{QueueDepth}$ and ${MAT}_{Up}^{LinkUtil}$. 
    For each of the MAT, 
    P4TE creates $r$ empty \textbf{Routing-Groups ($R_G$)}.  
    For each of these routing-groups, a ternary match entry containing the common IP prefix of the whole DCN, a pointer to the routing-group, and 
    priority of the routing-group  is inserted in the MAT. 
    Initially, there is no traffic flow through the upward ports. Hence, P4TE considers all upward 
    ports of the same performance at startup. These ports are inserted in the highest priority routing-group, and 
    all other routing-groups remain empty. 
    Fig.~\ref{fig:fat-tree-topology-intiial-configurations}a shows the structure of the ${MAT}_{Up}^{QueueDepth}$ after the initialization phase.

    At the \textit{\textbf{Runtime phase}}, P4TE executes alg.~\ref{PathSelectionAlgorithm} to select a forwarding path for a packet.
    Per packet-based forwarding schemes can quickly adapt to changes in  path utilization. However, 
    they may lead to packet reordering due to the use of multiple paths for packets of a flow. 
    It adversly~\cite{vanini2017let} impacts the performance of end-host transport layer protocols 
    (TCP, DCTCP~\cite{bensley2017datacenter}, etc.). 
    Considering this issue, numerous works~\cite{katta2016hula,alizadeh2014conga,perry2017flowtune,zhang2018load} 
    have explored the effectiveness of path selection at \textit{flowlet}~\cite{sinha2004harnessing}   granularity. 
    In this mechanism, when the consecutive packets from a flow are separated by a large enough time 
    gap (called \textit{flowlet interval}), a new flowlet is started, and a fresh path is selected for the flow. 
    P4TE's path-selection  algorithm also works at flowlet
    granularity to reduce the impact of packet reordering.
    It   always tries to use a path for a flowlet that has a low chance of overutilization.
    In selecting a path for \textit{latency-sensitive short flows}, 
    P4TE tries to use the link with the least queue build-up. The assumption behind this is if a link is 
    more than 95\% utilized, a short flow of size 50-200KB can be sent through the link without creating queue build-up. 
    However, if that port has already seen high utilization (YELLOW or RED status), the flowlet is 
    assigned the least utilized port. Both the selection reduces the chances for overutilization of a link.
    On the other hand, for
    \textit{throughput-oriented large flows}, P4TE gives the highest priority
    on selecting the least utilized port with the hope that this will
    leave enough link capacity for pushing more packets of the
    flow without over-utilizing the link. However, if the least utilized port's color mark is not GREEN, that implies all the ports are 
    facing high utilization. In that case, P4TE assigns the port with the least queue build-up to the flowlet. 
    To verify whether a link is overutilized or not, it relies on the monitoring component's feedback stored 
    in $portUtilizations$ (section~\ref{LinkUtilizationRateMonitoring}). This feedback packet is generated 
    from the egress stage and recirculated to the ingress stage without any additional delay. 
    Hence, it gives the most up-to-date information about a link's utilization rate. 
    Fig.~\ref{fig:PathSelectionAlgo} shows an example
    explaining how the path for a flowlet of \textit{throughput-oriented large flow}
    is selected.

    \begin{algorithm}[h]
      \begin{algorithmic}[1]
        \footnotesize{
        \item [\textbf{Input}: Packet P]
        \item [\textbf{Output}: Packet P (with modified egressPort)]
        \item [\textbf{Procedure}: PathSelectionAlgorithm (P)]

    \STATE lowQueueDepthPort = select matching port \linebreak from ${MAT}^{QueueDepth}_{Up}$ 
    \STATE lowUtilPort = select matching port from ${MAT}^{LinkUtil}_{Up}$ 
    \STATE flowletId = calculate 5 tuple (src IP, src Port, dest IP, \linebreak dest Port, flow label) based hash code  
    \STATE lastSeenTime = lastSeenTimeRegisterArray[flowletId] 
    \STATE lastSeenTimeRegisterArray[flowletId] = P.metadata.ingressTimestamp 
    \STATE flowletInterpacketGap = P.metadata.ingressTimestamp - lastSeenTime

    \IF{flowletInterpacketGap  $\geq$ FLOWLET\_INTER\_PACKET\_GAP\_THRESHOLD}
      \IF{P.trafficClass == SHORT\_FLOW}
          \STATE status = portUtilizations[lowQueueDepthPort]  
          \IF{status == GREEN}
              \STATE  P.metadata.egressPort = lowQueueDepthPort 
          \ELSE
              \STATE P.metadata.egressPort = lowUtilPort  
          \ENDIF
          
          \STATE lastUsedPorts[flowletId] = P.egressPort 
      \ELSIF{P.trafficClass == LARGE\_FLOW}
          \STATE status = portUtilizations[lowUtilPort]  
          \IF{status == GREEN}
              \STATE P.metadata.egressPort = lowUtilPort 
          \ELSE
              \STATE P.metadata.egressPort = lowQueueDepthPort 
          \ENDIF
          \STATE lastUsedPorts[flowletId] = P.metadata.egressPort 
      \ENDIF

    \ELSE
      \STATE  P.metadata.egressPort = lastUsedPorts[flowletId] 
    \ENDIF
    \item return P
      }
      \end{algorithmic}
    \caption{\centering{\small{Upward Path Selection Algorithm (Packet P) }}}

    \label{PathSelectionAlgorithm}

    \end{algorithm}

    \subsubsection{Path Priority Reconfiguration} \label{PathPriorityReconfigurationSubSection}
    
    Initially, P4TE assumes all upward ports have the same quality, and they are configured in the highest priority routing-group. With
    traffic flow, the relative performance of the links change,
    and the monitoring component reports these events (increase/decrease in queue depth or link utilization) to the control plane
    through feedback packets. 
    The control plane executes alg.~\ref{pathPriorityReadjustmentAlgorithm} to evaluate a link's priority, based on the 
    event (\small{M\_INCREASE or M\_DECREASE}) for link performance metrics 
    $M$ ($M \in$ \{\small{UTILIZATION\_RATE}, 
     \small{QUEUE\_DEPTH}\}).
    This algorithm moves a port to a lower (or higher) priority routing-group 
    if the link performance metrics decrease (or increase). 
    The mapping from performance metrics to routing-group is defined by how the whole 
    range of a metrics is divided into sub-ranges (section~\ref{DeltaForEgressQueueDepthMonitoring} and~\ref{LinkUtilizationParameterSetup}). 
    For example, fig.~\ref{fig:EgressQueueDepthMonitoringExample}c-\ref{fig:EgressQueueDepthMonitoringExample}f shows on receiving 
    queue depth increase and decrease event information for port 3 (fig.~\ref{fig:EgressQueueDepthMonitoringExample}c) and 
    port 4 (fig.~\ref{fig:EgressQueueDepthMonitoringExample}d), the control plane move port 3 to lower priority group and 
    port 4 to a higher priority group.

    When all the links are facing high load, they are assigned to the lowest priority routing-group. Similarly, all links are assigned to 
    the highest priority routing-group (or any other routing-group) in low load. In both cases, all the links exist in the same routing-group, 
    and the forwarding path for a packet is selected using hash-based mapping. Hence, in such cases, P4TE behaves like ECMP. It  ensures
    P4TE does not perform worse than ECMP when extreme congestion occurs on any switch in the DCN.

    \textbf{Algorithm Complexity}: 
    P4TE divides the whole range of link utilization rate and a port's queue depth  
     into a fixed number of predefined ranges (section~\ref{LinkUtilizationParameterSetup} and ~\ref{DeltaForEgressQueueDepthMonitoring}).
     The  control plane can store these range information in a 2-D array or tree-based data structure.
    P4TE's data plane monitoring algorithm (see alg.~\ref{EgressMonitoringAlgorithm})  
    provides quantitative information (utilization rate and queue length) about a link's current quality.  
    The information in every feedback packet directly maps a link to one of the ranges. 
    For each range of link utilization rate and queue depth segment, a separate routing group is configured into the 
    ${MAT}_{Up}^{LinkUtil}$ and ${MAT}_{Up}^{QueueDepth}$, respectively. 
    On receiving the feedback packets from the data plane, the control plane algorithm (see alg.~\ref{pathPriorityReadjustmentAlgorithm})  needs to find (line~\ref{routingGropfindingLine})
    the corresponding routing group. It can be executed very efficiently ($O(log n)$) using any tree search schemes. 
    Removing a port from a routing group and inserting it into another group does not involve rearranging any TCAM entries. Hence,
    it also takes a small amount of time. 
    Besides this, P4TE's monitoring algorithm (sec.~\ref{EgressMonitoringAlgorithm})  generates only one feedback packet when a
     microburst is identified and another when the microburst is cleared (sec.~\ref{OverHeadAnalysis}).
     Therefore, the load on the control plane for a microburst is always constant. 
    The local CPUs executing the control plane threads are connected to the 
    data plane pipelines using dedicated circuitry~\cite{opentofino}. 
    Therefore overall delay between the time of a  \textit{traffic event} in the data plane and the completion of the corresponding measure 
    taken by alg.~\ref{pathPriorityReadjustmentAlgorithm} is also small. 

  \begin{algorithm}
  
  \begin{algorithmic}[1]
    \footnotesize{
    \item [\textbf{Input}: Event E (port,eventData)]
    \item [\textbf{Output}: New routing-group for E.port]
    \item [\textbf{Procedure}: PathPriorityReconfiguration (E)]
  
  \SetNoFillComment
  \STATE  oldRoutingGroupId = ${{portToRoutingGroupMap}^{{M}}}$[E.port] 
  \STATE  newRoutingGroupId = find matching routing-group based on E.eventData; \label{routingGropfindingLine}
  \IF{newRoutingGroupId != None}
  \STATE Use P4Runtime API to remove E.port from the \linebreak  routing-group with oldRoutingGroupId
  \STATE  Enter E.port into routing-group with id \linebreak newRoutingGroupId;
  \STATE  ${{portToRoutingGroupMap}^{{M}}}$[E.port] = \linebreak  newRoutingGroupId;

  \ENDIF
  }

  \end{algorithmic}

  \caption{\centering{\small{Path Priority Reconfiguration Algorithm (Event $E$)}}}
  \label{pathPriorityReadjustmentAlgorithm}
  \end{algorithm}

    \subsection{Rate Adaptation} \label{RateAdaptionSection}
    P4TE’s path selection algorithm tries to avoid overutilization of the links. But congestion starts to build up
    when the total incoming traffic volume is greater than the outgoing capacity of the links and/or 
    an overutilized link is selected for further traffic. 
    While forwarding a packet, if any P4TE enabled switch senses congestion or 
    spare capacity available to forward more packets, it sends  \textit{fake acknowledgment packets (FACK)} to
    the sender of the flow with decreased/increased window size for flow rate adjustment \textbf{T4}. 
    As the feedback is generated from a switch in the \textit{source-destination-source} path, 
    it helps the sender to react to changing network
    conditions faster than $RTT$ times. Before sending a FACK, a switch needs to answer two crucial 
    questions: a) how to select a flow for rate control, and b) what should be the appropriate rate for a flow?

    \subsubsection{Flow Selection}

    Assume, for a packet $P$ from flow $f$ of traffic class $TC$  the egress port selected by the path selection 
    algorithm (section~\ref{pathSelectionSubsection}) is $i$. 
    To determine whether $f$'s sending rate should be controlled or not, 
    P4TE uses two pieces of information collected by the monitoring component:
    a) whether all the incoming flow of a traffic class $TC$ is crossing the safe-rate for that class (section~\ref{TrafficClassBasedIncomingRateMonitoring}) 
    and 
    b) utilization rate of link $i$  (section~\ref{LinkUtilizationRateMonitoring}).

    If $f$ finds the \textit{safe-rate}  for its traffic class and the incoming port has crossed ($incomingPacketColor$ is YELLOW)
    and the link $i$ is overutilized ($newPacketColors[i] == RED$), P4TE considers this as a sign of congestion. 
    P4TE enabled switch immediately tries to reduce
    the flow’s  rate by sending a FACK with reduced
    window size. 
    Now crossing the safe-rate for traffic class $TC$
    implies that the traffic has already used the allocated
    portion of that class. However, there may be no other competing
    flow for the selected egress port. In this case,
    if port $i$ is not overutilized ($newPacketColors[i] == GREEN$ ), 
    P4TE tries to utilize
    the selected link's spare capacity by increasing the flow’s
    rate through sending a FACK with increased window size.
    On the other hand, if flows from traffic class $TC$ have not crossed the safe-rate ($incomingPacketColor$ is GREEN), but the port is 
    $i$ overutilized, P4TE does not take any rate control action.
    The justification behind this is that flows of traffic class $TC$ have not used their allocated rate; here, flows from other competing traffic 
    classes may cross their safe-rate and also find port $i$ is overutilized. 
    In such a case, P4TE will reduce the rate of flows from competing traffic classes,
    or an uncongested path will be selected for the next flowlet.

    To not increase the rate of a flow too much or penalize a
    flow unfairly compared to other flows of the same class, 
    P4TE uses an approach similar to DCTCP~\cite{bensley2017datacenter} and standard TCP~\cite{ramakrishnan2001addition}. 
    But instead of end-host, all leaf
    switches (directly connected to hosts) maintain a record  of the last TCP
    sequence number indicates when rate control was applied on a flow. 
    The leaf switch marks a flag in the header to not apply rate control on the flow for the next $B$ bytes.
    If the flag is set, all other switches in the path do not apply rate
    control on this flow.
    Moreover, if a switch decides to reduce (or increase) the rate of a flow, it marks the header with a special 
    flag to prevent other switches in the path from applying rate control 
    on the flow. On the other hand, when a leaf switch receives a FACK packet sent toward a host (connected to the downward ports of the leaf switch),
    the leaf switch updates the TCP sequence number.
    These ensure a flow is not penalized or gets an unfair advantage multiple times within a \textit{rate-control window} of $B$ bytes.

    \subsubsection{Rate Control}
    Flow rate control needs participation from the protocol stack of both end-host and switch. 
    Sending FACK from a switch with a modified window size reaches 
    the sender of a flow earlier than the usual $RTT$ time. 
    Besides this, the sequence number
    in the FACK packet is not a real reflection of bytes acknowledged
    (also bytes in flights) by the receiver of the flow.
    Both  factors play a crucial role in determining the source congestion
    window by end-host transport layer protocol stacks.  
    Hence, sending FACK does not provide exact information 
    to the sender, impacting the  
    congestion window calculation. Various end-host-based transport layer protocol stacks
    adjust flow rates using different formulas~\cite{zhang2020survey}. 
    Controlling the broad dynamics of the end-host-based
    protocol stack's congestion control scheme from the switch
    is a complex topic. We leave it as future research
    scope. 
    P4TE aims to  work in conjunction with existing TCP protocol stacks and does not make any assumptions about 
    which one is running in the end hosts. To keep P4TE's rate adaptation scheme in sync with these TCP stacks 
    P4TE follows the AIMD (additive-increase/multiplicative-decrease)~\cite{yang2000general} scheme,  
    where the  window size is reduced more
    aggressively and increased more cautiously. 
    The window resize ratios are discussed in section~\ref{RateAdaptationParameterSetup}.


\section{Robustness and Overhead} 

\subsection{Robustness}
    Link failure and high traffic load are two of the main reasons behind the 
    \textit{traffic-engineering} system's underperformance in DCN. \textit{Routing-Group} based 
    mechanism of P4TE handles each link independently and does not need a readjustment of other links due to
    failure in one link. 
    P4TE requires a small amount of resources in the PISA pipeline; it can be used in conjunction with 
    existing PISA switch based schemes~\cite{tan2019netbouncer,li2022p4resilience} for link failure detection. 
    When a link fails, the operator can 
    remove the link from its current routing-group; P4TE will
    automatically readjust the load on other links. On the other hand, all the links can fall into the lowest priority
    routing-group of corresponding metrics at a high load. In this case, P4TE behaves similarly to ECMP, and performance
    is equivalent to ECMP.

    \subsection{Overhead Analysis} \label{OverHeadAnalysis}

The main overhead of P4TE is the recirculated packets created by  
a) the \textit{egress stage monitoring algorithm} (see alg.~\ref{EgressMonitoringAlgorithm}) and 
b) the FACK packets generated by the \textit{rate adaptation} component (section~\ref{RateAdaptionSection}). 
Here, we discuss 
how P4TE can practically achieve high throughput despite its recirculation overhead.

\textbf{Case a}: When the utilization rate of a link crosses the configured rates, the egress stage monitoring algorithm
(see alg.~\ref{EgressMonitoringAlgorithm}) generates two extra feedback packets. One is recirculated to the ingress stage 
to be used in \textit{path-selection} algorithm (see alg.~\ref{PathSelectionAlgorithm}) and another one to send a feedback packet to 
the control plane. This algorithm also generates a single feedback packet to the control plane when  
the queue builds up (queue depth is reduced) in a port's buffer. 
Consider the case of incoming microbursts and a packet from flow $f_1$ first observes that CIR of a port 
is crossed and one feedback packet is generated. The next packet $P'$
(either from $f_1$ or another flow)  also observes that the CIR of the same port is crossed.  
 The immediate previous packet has experienced link utilization rate change and reported it to the control plane.  
Hence, P4TE's monitoring algorithm (see alg.~\ref{EgressMonitoringAlgorithm}) does not consider this as an event.  P4TE generates only one feedback packet for a
microburst on a single port, and another feedback packet is generated when the microburst is cleared. A similar argument is also applicable for 
the case of queue build up on a port's buffer. Hence, P4TE creates a negligible overhead (bounded to only one feedback packet per microburst on 
one port) on a  pipeline running at 1 GHz speed.

\textbf{Case b}: 
P4TE's \textit{path-selection} algorithm for upward direction always attempts to select the least utilized port for a packet . 
Despite this, an uncongested port may not be found during heavy congestion. Similarly, an uncongested  port may not be 
found for forwarding a packet in the downward direction. Moreover, the feedback packet generated by the egress stage 
will reach the ingress stage after a small delay (approximately $\approx$75ns, which is around 11.5\% of 
the port-to-port latency of a normal packet~\cite{wu2019accelerated}). Due to this delay, a port may be falsely 
marked as uncongested to the \textit{path-selection} algorithm. 
In these cases, P4TE generates a FACK  toward the flow source 
by replicating the current packet and does not generate more FACK for the next $B$ bytes. 
These replicated packets consume resources in the PISA pipeline. 
However, the average link
utilization rate in the data center network remains below 95\%~\cite{roy2015inside}. 
It  leaves enough resources in the pipeline to handle the FACK packets for a large number of flows without hampering the normal 
data packets. For example, consider  the case of a leaf switch in a DCN with 64$\times$10Gbps PISA switch~\cite{bosshart2013forwarding} running 
a 1 GHz pipeline (960M packets per second processing rate). If all the links are handling packets at 95\% of their capability, the 
pipeline needs to handle $960M \times 0.95 = 816M$ packets per second. Such a pipeline can process $960-816=144$M FACK packets per second. 
In the worst-case scenario, if every flow faces P4TE's \textit{rate-adaption} action, a 1 GHz pipeline can handle 144M flows. 
This number is small compared to the total number of flows~\cite{roy2015inside} handled by a switch in data centers. 


    \section{Deployment} \label{Deployment}

    In this section, we discuss how to configure the required parameters and 
    P4TE's deployability using on PISA switches.

  \subsection{P4TE Parameters Setup}  \label{P4TEParameterSetupSection}

  The \textit{monitoring} component needs three configurable parameters. They control P4TE's granularity of monitoring and 
  the number of routing-groups required in the path selection component. 
  Besides this, the \textit{path-controlling} and \textit{rate-adaptation} components also need a few parameters. 
  We considered several important empirical measurements on data center traffic characteristics as the key guiding 
  tool in configuring these parameters. 
  Next,  we discuss the configuration of these parameters.

    \subsubsection{$\Delta$ for Egress Queue Depth Monitoring} \label{DeltaForEgressQueueDepthMonitoring}
    In ECN~\cite{floyd1994tcp} based congestion control schemes, the egress queue depth threshold is configured as proportional to the
    round-trip time ($RTT$) $\times$ bottleneck link with capacity ($C$). Measurements~\cite{wu2012tuning} from production datacenters 
    show nearly 200\% difference in 25th and 90th percentile RTT. In~\cite{wu2012tuning},  the authors proposed and experimentally validated 
    the lower and upper  bound of  the queue depth thresholds for ECN based schemes. P4TE  utilizes these bounds and 
    divides  the whole range of queue depth into four ranges ($[0,\Delta], \cdots ,[3\Delta +1, rest]$).
    The length of each portion is used as $\Delta$ in alg.~\ref{EgressMonitoringAlgorithm}. 
    For each of the four ranges, one routing-group is used in the queue depth  based forwarding table (${MAT}_{Up}^{QueueDepth}$) by 
    the path-controlling component (section~\ref{PathControllingSubSection}). The routing-group with the smallest queue depth (least congested links) value is assigned the highest priority.
    On receiving control messages from the data plane, the path priority reconfiguration alg.~\ref{pathPriorityReadjustmentAlgorithm} 
    maps the links to the routing-groups according to the observed queue depth value.
    Any link observing queue depth more than  the upper bound ($4\Delta$) is mapped to the least priority routing-group.

    \subsubsection{Meter Rates for Link Utilization Monitoring} \label{LinkUtilizationParameterSetup}
    The goal of monitoring link utilization rates (section~\ref{LinkUtilizationRateMonitoring}) at the egress stage
    is to decide whether a link is safe for increasing a flow's rate or not. 
    Empirical measurement~\cite{kandula2009nature} shows congestion occurs in data center links 
    when they face around 75\% utilization. Whereas a link becomes overutilized when 100\% of its bandwidth is used.  
    Following these observations,  we configured the CIR and PIR of the meters for monitoring link utilization as 75\% and 95\% 
    (with 5\% burst handling capacity as CBS and PBS) of a link's bandwidth. P4TE considers a link as stable between these two rates and does not take any
    rate control actions in the flows. 
    For these three ranges, P4TE maintains three routing-groups in 
    link utilization rate based forwarding table (${MAT}_{Up}^{LinkUtil}$) (section~\ref{PathControllingSubSection}).
    The path priority reconfiguration algorithm (see alg.~\ref{pathPriorityReadjustmentAlgorithm}) maps the links to these routing-groups similar 
    to section~\ref{DeltaForEgressQueueDepthMonitoring}.

    \subsubsection{Meter Rates for Traffic Class-Based Incoming Rate Monitoring} \label{IngressMonitoringRateParameterSetup}

    Monitoring incoming traffic (section~\ref{TrafficClassBasedIncomingRateMonitoring}) aims to identify candidate flows for rate control based on their link bandwidth consumption. 
    Analysis of data center workload~\cite{alizadeh2010data,greenberg2009vl2} 
    shows the majority of the flows (more than 90\%) are short in size, whereas the rest of 
    the few flows are large, and they carry the majority of the bytes (more than 90\%). 
    Following these observations, we configured 90\% of a link capacity as the safe-rate for the short flows and the rest 
    for the large flows. This gives higher priorities to the short flows in case of resource contention and helps them to 
    achieve improved flow completion time.


    \subsubsection{Flowlet-Interval for Path Selection} \label{FlowletIntervalParameterSetup}
    P4TE's \textit{path-selection} algorithm selects a new path at  the granularity of \textit{flowlet}. 
    The \textit{flowlet-interval} can be configured to $RTT$, which is a large value ($T_f^{max}$), and there will be only one 
    flowlet for a flow. It diminishes the advantages of the flowlet mechanism. On the other hand, it can be configured to 
    a small value of $T_f^{min}=$1/maximum traffic burst rate (in packets per second), which creates a new flowlet for 
    every packet during a traffic burst. 
    It increases the number of packet reordering. To avoid these issues, 
    similar to existing works~\cite{alizadeh2014conga,vanini2017let,katta2016hula}, 
    we have used experimental results to configure the 
    \textit{flowlet-interval} time from the range $[T_f^{min}, T_f^{max}]$. In our experiments, we have used a flowlet interval of 40 ms. 
    Details of the experiments are discussed in the appendix~\ref{FlowletConfigAppendix}.

    \subsubsection{Rate-Adaptation Related Parameters} \label{RateAdaptationParameterSetup}
    P4TE's \textit{rate-adaptation} component requires two parameters: a) \textit{rate-control window}  $B$
    and b) the window size increase and decrease ratio for the AIMD scheme used in flow \textit{rate control}. 

    Packets of a flow can face P4TE's \textit{rate control} action at any switch in the path; the earliest at the leaf 
    switch ($l_s$) directly 
    connected to the source host ($h_s$)  and the latest at the leaf switch ($l_d$) directly connected to the destination host ($h_d$). 
    The result of the \textit{rate control} action requires $\delta = 2\times \: delay \: between \: l_s \: and h_d$ time 
    to reach  the source host. The maximum number
    of inflight data within this time period is
    $maximum \: flow \: rate \times \delta$ bytes. This gives the value of the \textit{rate-control window} $B$. 
    Similar to the \textit{observation window} of DCTCP~\cite{bensley2017datacenter}, 
    P4TE does  not react to change in link utilizations more than once for every
    window of $B$ byte data. However, computing the value of $\delta$ requires the implementation of
    extra logic in the switches. It consumes costly resources in the PISA
    pipeline. To avoid this, we experimentally calculated the average $RTT$
    and used it instead of $\delta$. 

    Different variations of TCP use different schemes for resizing the congestion window size. 
    However, not all schemes are implementable in PISA switches at low cost due to the lack of various necessary 
    action support on packet fields (i.e., floating-point operations and arbitrary division operations are not 
    supported in all PISA switch environments). Moreover, it also requires maintaining
    different per-flow statistics (previous window size, packets in flight, etc.).
    Developing PISA switch-based traffic-aware window sizing schemes are out of the scope of this work. 
    Instead, we followed a simple scheme in P4TE. 
    On identifying
    a candidate flow for rate control, P4TE reduces its window 
    size by $\frac{1}{2}$   or increases its window size by
    $\frac{1}{4}$ of its current size. Both the values are less than the thresholds used by standard TCP protocols and 
    cheap to implement (requires only shift operation) in PISA hardware. 
    We leave the goal of dynamically controlling the TCP window size from PISA switches as a future research goal.


    \subsection{Mapping to Hardware} \label{MappingtoHardwareSubsection}
    P4 is the dominant programming language for programming runtime behavior of RMT paradigm-based PISA switches. 
    We implemented P4TE's data plane program using P4 (v16~\cite{p416}) programming language. 
    The P4 compiler plays a key role in deciding the deployability of a P4 program over a PISA switch. 
    The compiler front-end analyzes the P4 program and provides a \textit{Table Dependency Graph} (TDG). 
    Every node in this TDG represents a logical match-action table. The deployability of a P4 program depends 
    on the successful mapping of these logical tables to physical match-action tables of a PISA switch. 
    The device architecture-specific compiler backend executes this task. 
    Finding an efficient logical to physical match-action table mapping is a computationally intractable problem~\cite{jose2015compiling}. 
    Typically compiler uses different heuristic-based algorithms for this purpose. 
    As a result, a compiler can spuriously reject a P4 program as they can not find a mapping to the switch, even though there exists a 
    possible mapping.

    V1model~\cite{v1model} is the most widely available hardware architecture for RMT paradigm-based PISA switches.
    In this work, we considered P4TE's deployability over this architecture and considered the hardware configuration described in~\cite{bosshart2013forwarding} as the benchmark. Unfortunately, there is no openly available compiler toolset to find 
    the logical to physical match-action table mapping of a P4 program for a given RMT paradigm-based PISA switch. 
    We developed a Python program that analyzes P4TE's intermediate TDG representation provided by the open-source P4 compiler~\cite{p4c}
    for BMV2~\cite{bmv2} based implementation of v1model architecture. 
    It uses the branch-removal technique used by various compilers to convert the conditional statements (if-else)
    into a logical match-action table. 
    It also converts  BMV2's multi stage register read-write operation into one single stage as it is supported by the 
    real PISA hardware (such as Toffino~\cite{opentofino}).
    Then finally, it computes a header field mapping and logical to physical match-action table mapping.
    As the goal is only to confirm P4TE's deployability using PISA switches, we do not consider
    the optimality of the mappings. 
    
    The leaf and non-leaf switches of P4TE execute the same algorithms, except the leaf switches need to 
    maintain the per-flow TCP sequence number used in flow rate adaptation. Hence, the leaf switches need 
    to accommodate more complex P4 programs in the same hardware compared to the non-leaf switches. Due to space concerns, 
    we only discuss the hardware mapping of the leaf switch P4 program over the benchmark hardware.

    \subsubsection{Mapping Header Fields to Packet Header Vector}\label{HeaderFieldMappingSubSubSection}
    The ingress portion of the P4 program requires 
    30, 8, 8, 7 and 3 words of 8, 16, 32, 48 and 
    128b width; whereas the egress portion requires 13, 13, 1, 14, 4 and 4 words of 8, 16, 24, 32, 48 and 128b width. 
    The benchmark hardware can accommodate a total of 4096b in the packet header vector using 64, 96, and 64 words of 8, 16, and
    32b width header fields. Two or more fields can be merged together to form a larger field. 
    The header field requirement by the ingress and egress portion of the P4 program can be 
    fulfilled using 46, 32, and 61 words of 8, 16, and 32b wide header fields. 
    Altogether, they consume 69.15\% space of the packet header vector available in the benchmark hardware.

    \subsubsection{Mapping Logical Match-Action Tables to Physical Tables}\label{hardwareMappingofMAT}
    The TDG's critical path length for both the ingress and egress stage of the P4 program is 26, whereas the benchmark 
    hardware contains 32 match-action stages. Therefore, P4TE's  P4 program can be safely embedded over this benchmark hardware using 26 stages. 
    The remaining six hardware stages can be reserved for the traffic classification schemes. Existing traffic classification schemes~\cite{turkovic2020sequential}
    can provide highly accurate (less than 5\% false negative) traffic classification using only six stages and a  small amount of stateful memory per stage.

    The header fields used by P4TE's match-action tables can be accommodated within the available header vector space (section~\ref{HeaderFieldMappingSubSubSection}) 
    and each stage can execute one instruction per header field in each stage of the hardware.
    But the benchmark hardware can match 1280 bits wide header fields
    (640b for each  TCAM and SRAM based hash table) as table key and execute instructions on 1280 bits (including both operands and result) 
    header fields at each stage at once. 
    The ingress and egress stage of the P4 program shares the same pipeline.  
    Therefore the logical to physical match-action table mapping needs to 
    maintain the per stage match key and action bit width limitations while mapping both the stages of the P4 program. 
    We embedded the nodes of the TDG to the 
    physical match-action stages according to their topological order. 

\begin{figure}
      \begin{subfigure}[b]{.48\textwidth}
        \centering
        \includegraphics[trim=0.0in 0in 0in 0in, clip,scale=.7]{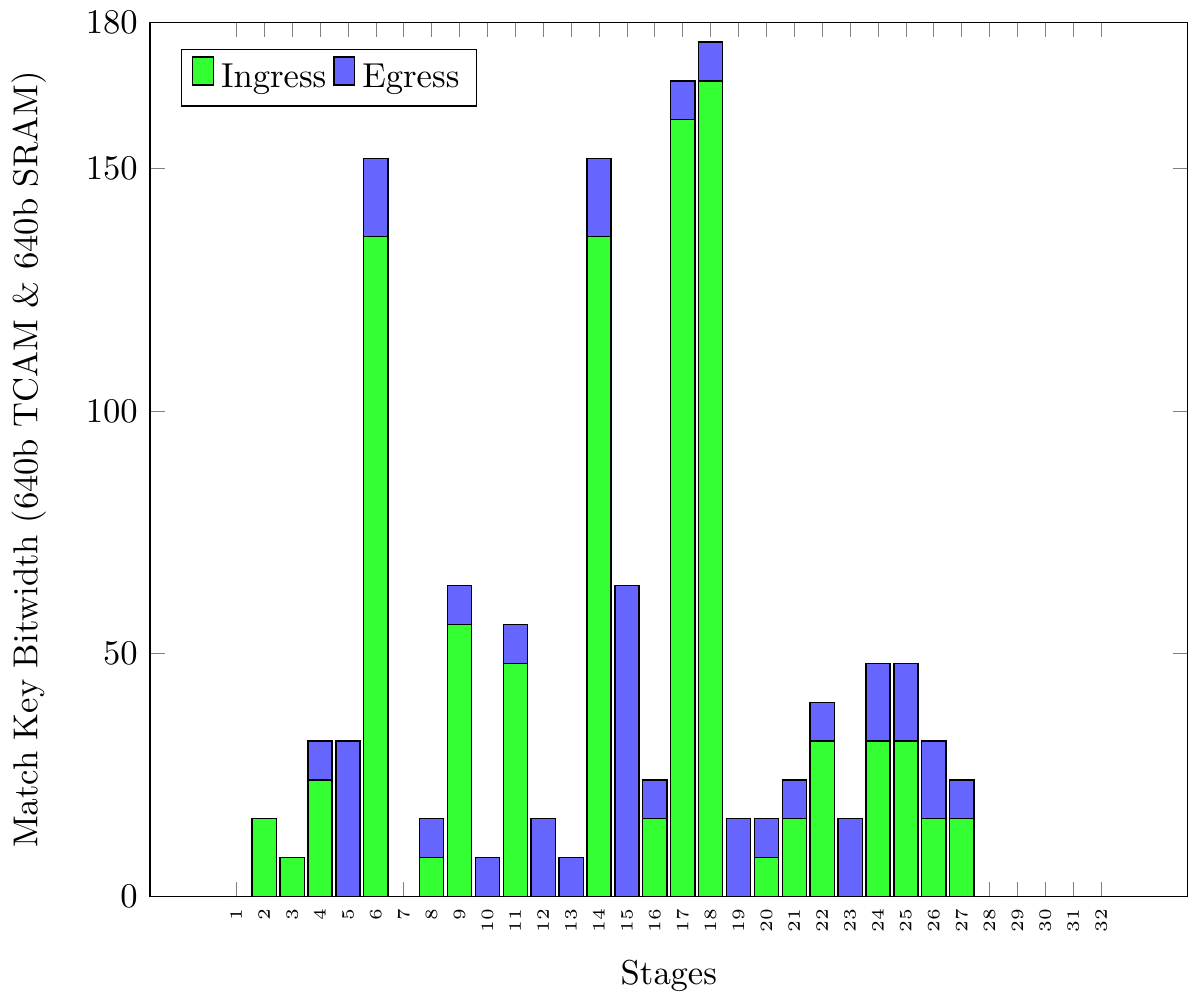}
        \caption{\centering {\small{Bitwidth of match-action table keys required to map the leaf switch's P4 program }}}
        \label{fig:P4teMatKeyUsage}
      \end{subfigure}
      \begin{subfigure}[b]{.48\textwidth}
        \centering
        \includegraphics[trim=0.0in 0in 0in 0in, clip,scale=.7]{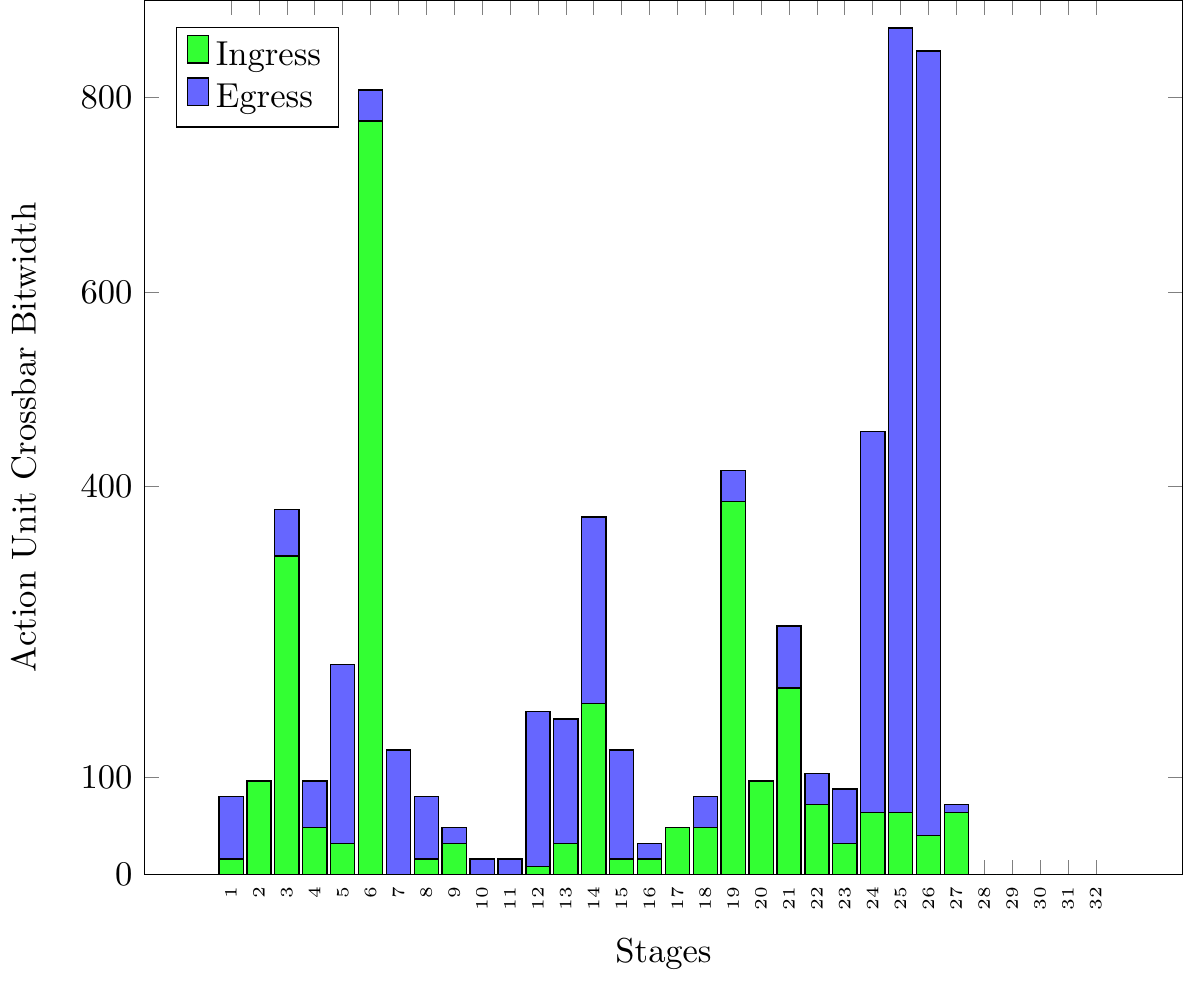}
        \caption{\centering {\small{Bitwidth of action fields required to map the leaf switch's P4 program}}}
        \label{fig:P4teActionUsage}
      \end{subfigure}
      \caption{\centering {\small{Match-action-table key and action bitwidths required for P4TE's leaf switch P4 program (unused part not shown). }}}
    \end{figure}

    Fig.~\ref{fig:P4teMatKeyUsage} shows the total bitwidth of the match fields required for concurrent execution of the 
    ingress and egress portion of the P4 program in each stage. 
    P4TE uses only TCAM based match-action tables for match purpose, and every stage 
    requires less then 640b wide match key. Similarly, the maximum bitwidth of the header fields required for concurrent 
    execution of the  actions for both the ingress and egress portion of the P4 program is shown in fig.~\ref{fig:P4teActionUsage}. 
    None of the stages require more than  
    the available bitwidth (1240b) of the header fields to be used as the operand of the instructions. 
    Therefore, the TDG of P4TE's P4 program can be safely embedded 
    over the physical match-action stages of the benchmark hardware.

    \subsubsection{Memory Consumption}
    Consider $k=$ total number of ports in a switch, $C=$ total number of traffic classes, and 
    $r=$ the total number of \textit{routing-group} for a link performance metrics.

    \textbullet \textbf{Monitoring Component (section~\ref{MonitoringSection})}: 
    Alg.~\ref{EgressMonitoringAlgorithm} requires one register in $oldQueueDepths$, one TCAM entry (with one attached direct meter) 
    to store  meter configurations for 
    monitoring link utilization (fig.~\ref{fig:fat-tree-topology-intiial-configurations}c) and one register in $oldPacketColors$ for each port. 
    Similarly, for traffic class based 
    incoming traffic rate monitoring (section~\ref{TrafficClassBasedIncomingRateMonitoring}), 
    two entries (one each for short and large flows) are required for each port in the MAT. 
    Each of them are assigned to different physical stages in the hardware. 
    For a $k=1024$ port switch individually they consume less than 1\% of  SRAM and 13\% of  TCAM memory avaialble in each stage.

    \textbullet \textbf{Path-Controlling Component (section~\ref{PathControllingSubSection})}: The match-action table for the downward route 
    lookup table  ($MAT_{Down}$) requires one TCAM entry for each downward port (at most $k$ entries). 
    The maximum number of upward ports in a fat-tree topology is $k/2$.
    ${MAT}^{LinkUtil}_{Up}$ requires three \textit{routing-groups} with the capacity to store $K/2$ ports in each group. 
    Therefore it  requires $4k/2$ entries.
    Only one stage of PISA hardware can accomodate the entries required  for ${MAT}^{QueueDepth}_{Up}$ table. For example, 
    consider the case of a switch~\cite{agrawal2020intel} with 1 MB buffer capacity for a port. 
    With 64B packet size, 1 MB buffer can accommodate at most 
    16 K packets for each port.
    The worst case arises when $\Delta$ is small. 
    For example, assume $\Delta = 20$ in alg.~\ref{EgressMonitoringAlgorithm}, the maximum number 
    of routing-groups required is around 800. For a $k=128$ port switch, a total of $800 \times 64 =$ 51K entries are required to store 
    the port information for the routing-groups of ${MAT}^{QueueDepth}_{Up}$. 
    With 128b IPv6 prefixes, traffic class, and other necessary information as the match fields, 
    it requires a 160b match key  for ${MAT}^{QueueDepth}_{Up}$. 
    Storing the IP prefix based match entries for 80 routing groups require less than 10\% of the available TCAM capacity 
    in each stage~\cite{bosshart2013forwarding}. 
    The port information of every routing-groups are stored as an \textit{action group} in PISA switches~\cite{PSA}. 
    These action groups are stored in a separate hash based table in every stage and PISA switches maintain a 
    pointer to these \textit{action-group}s from the prefix match tables. 
    Storing the 51K entries of port information (9 bits are required to store a port information) in an SRAM based hash table 
    requires around 57KB of SRAM. Hence the total SRAM consumption for such a table is less than 
     20\% of the available SRAM capacity in each stage~\cite{bosshart2013forwarding}.

    \textbullet \textbf{Rate-Adaptation Component (section~\ref{RateAdaptionSection})}:
    Finally, register arrays are used to store the time of last packet arrival (48b timestamp), last used port (9b port information), and 
    the TCP sequence number (32b) used in rate adaption for each flow. They are mapped to three different hardware stages. 
    For storing information of 1M flows, these arrays individually consume less than 70\% of available SRAM in each stage. 

    Besides these, some other common tasks for packet replication, neighbor discovery protocol~\cite{narten1998neighbor} packets processing, carrying information on conditional logic to next stages, etc., are required for each packet~\cite{P4TEGitRepo}. 
    They require a small portion of resources in the pipeline. 
    Therefore, the benchmark hardware can accommodate the P4 program used in the leaf switches. 
    The P4 program for the non-leaf switch is a simplified version of the leaf switch P4 program. The only difference between them is, 
    that the non-leaf switches do not need to store any per-flow state for 
    rate-adaptation. Therefore, their P4 program is less complex than the leaf switch P4 program and requires even less stateful memory.  It requires even fewer resources in the PISA pipeline.
        Thus, P4TE is  deployable on existing PISA switches and maintains the line-rate.  
    (Due to space concern, full analysis of mapping the leaf and non-leaf switch P4 program are  omitted from this paper, and 
    it is accessible at~\cite{P4TEHwMapping}.)

    \section{Performance Evaluation } \label{PerformanceEvaluation}
    The goal of P4TE is to improve the data center network's performance through traffic engineering 
    using only commodity PISA switches. P4TE does not require any modification in the end-host transport layer protocol stack. 
    Several works exist in the literature that relies on explicit 
    support from end-host protocol stack~\cite{benet2018mp} or requires customized ASICs~\cite{alizadeh2014conga} for DCN performance 
    improvement. Hence, we have not evaluated them in this work. 
    In this section, we evaluate the performance of P4TE and compare it with widely used schemes implementable using 
    PISA switches:  ECMP~\cite{hopps2000analysis}     and 
    HULA~\cite{katta2016hula}. 
    We simulated an 8-port switch-based layer-2 fat-tree DCN (leaf-spine topology) using Mininet~\cite{mininet} 
    and P4 BMV2~\cite{bmv2} software switch (v1model.p4 architecture~\cite{v1model}). 
    The simulated topology contains four leaf switches, four spine switches and four hosts connected to every  leaf switch. 
    The switch's data plane
    and control plane program  was developed 
    using P4\textsubscript{16}~\cite{p416} and Python language. For the control plane to the data 
    plane communication P4Runtime framework~\cite{p4runtime}
    has been used\footnote{The source code for P4TE is available at~\cite{P4TEGitRepo}.}. 
    To create a 2:1 link oversubscription ratio, the link bandwidth capacity of end-host to 
    leaf switches and leaf to spine switches were set to 40 pps and 20 pps (packets per second), respectively. 
    The buffer capacities of the links were
    configured as 0.2$\times$link bandwidth capacity with a tail drop policy. 
    All the parameters required by P4TE were configured as described in section~\ref{P4TEParameterSetupSection}.
    We measured the queue depths at different switches under ECMP (with ECN~\cite{floyd1994tcp}) scheme and the 90-th percentile value is 6. 
    Following this, we configured $\Delta=2$ for alg.~\ref{EgressMonitoringAlgorithm}. It creates three 
    sub-ranges [0-2], [3-4], [5-6] and [7-rest], and we configured one routing group for each of them in ${MAT}_{Up}^{QueueDepth}$. 
    On the other hand, for ${MAT}_{Up}^{LinkUtil}$ we configured three routing groups, 
    and the meter rate configurations
    for every link were configured  as CIR=90\% and PIR = 100\% of the link bandwidth. 
    The \textit{flowlet-inteval} for all the experiments was configured to 40ms (appendix~\ref{FlowletConfigAppendix}).
    All the experiments have been conducted on a virtual machine with six cores of 
    Intel(R) Core(TM) i7-9750H CPU (all six cores were configured at clock speed 2.4 GHz), 10 GB RAM, running Ubuntu 20.04. 
    All the experiments were repeated five times, and  the average results are presented in different graphs.

    DCN traffic pattern is dependent on various factors. 
    An exact simulation of  various traffic distribution patterns using
    a large number of flows to recreate the exact DCN traffic
    scenario is impossible. 
    Instead, similar to existing works~\cite{alizadeh2014conga,katta2016hula,bensley2017datacenter}, 
    we used two empirical workloads found in production data centers: a) a web-search workload~\cite{alizadeh2010data} 
    and b) a data-mining workload~\cite{greenberg2009vl2}. 
    Both workloads are heavy-tailed; the majority of the flows are small, and a small number of large flows carry 
    the majority of the data. 
    Besides this,  the partition-aggregate communication
    pattern is common in modern data centers. 
    This traffic pattern can lead to TCP incast problem~\cite{lee2015flow,chen2009understanding}. 
    To analyze P4TE's performance in TCP incast scenario, we generated an artificial traffic pattern 
    (similar to previous works~\cite{alizadeh2014conga}).

    Similar to prior works~\cite{dukkipati2006flow,katta2016hula}, we consider FCT as the main performance metric. 
    Our goal was to focus on whether P4TE can improve FCT of 
    the short flows ($\leq$ 90-th percentile of flow size distribution) 
     without degrading FCT for large flows ($>$ 90-th percentile of flow size distribution)  too much.
    As P4TE uses different heuristics for selecting a path for short and large flows, it is important to 
    compare P4TE's performance improvement for short and large flows separately. 
    Due to space concern, we have compared the performance in terms of the average FCT. However, 
    we have also included the cumulative distribution function (CDF) of the FCT in the appendix~\ref{CDFAppendix}. 
    As a side effect of the rate-control scheme applied by P4TE,
    FACKs are sent to the sender of a flow. It can increase the number
    of retransmissions. To analyze this impact, we also compared the total number of retransmissions 
    in each case. 
    A DCN contains multiple paths (every upward link corresponds to a path) 
    in the upward direction, congestion 
    starts to build up when a link is utilized more than other links. 
    A traffic engineering scheme should distribute the incoming load over the 
    upward links in a balanced manner. To analyze this aspect,  we compared P4TE’s ability to distribute the load over
    multiple upward paths and compared it to ECMP
    and HULA.

    \subsection{Empirical Workload} \label{EmpiricalWorkload}
    Fat-tree topology based DCNs contains multipath capability at every layer. Appropriate use of the 
    multipath capability~\cite{katta2016hula,robin2022clb} scheme can bring performance improvement here. However, 
    P4TE aims to utilize both multipath aware forwarding and in-network \textit{rate adaptation} mechanism 
    to achieve performance gain. Here we at first experimentally evaluate (section~\ref{ImpactOfRateAdaptation}) whether P4TE's 
    \textit{rate adaptation} scheme brings performance improvement over its multipath aware forwarding scheme. 
    Then we evaluate (section~\ref{PerformanceImprovement}) how does P4TE perform  compared
    to  ECMP and HULA.

    \begin{figure*}

      \begin{subfigure}[]{.48\columnwidth}
        \centering
        \includegraphics[trim=0.25in 1.25in 1in .25in, clip,scale=.7]{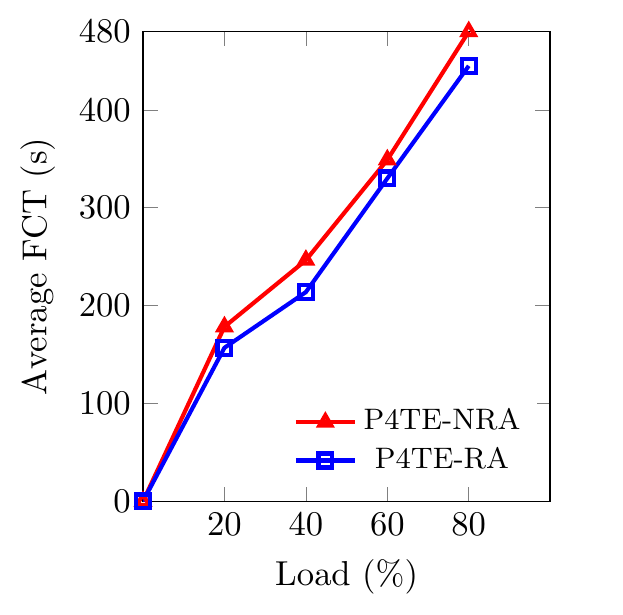}
        \caption{\centering {\small{Avg. FCT for short flows for web search workload}}}
        \label{fig:ShortFlowRateControlImpactFctComparisonWebsearch}
      \end{subfigure}
      \begin{subfigure}[]{.48\columnwidth}
        \centering
        \includegraphics[trim=0.25in 1.25in 1in .25in, clip,scale=.7]{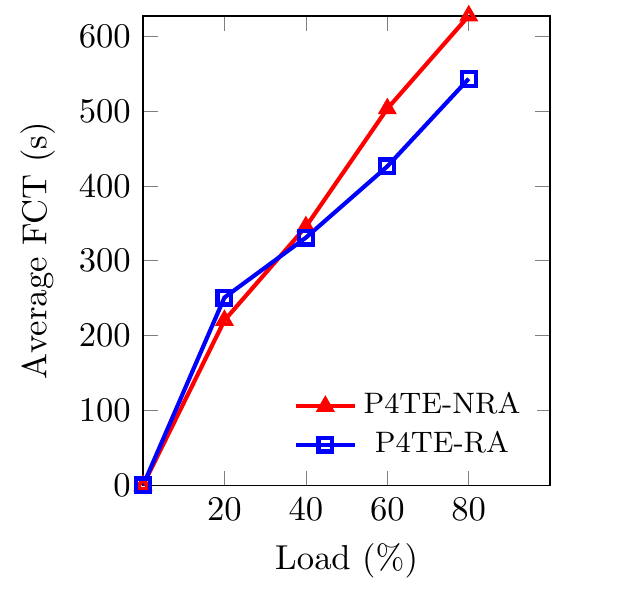}
        \caption{\centering {\small{Avg. FCT for large flows for web search workload}}}
        \label{fig:LargeFlowRateControlImpactFctComparisonWebsearch}
      \end{subfigure}
        \begin{subfigure}[]{.48\columnwidth}
          \centering
          \includegraphics[trim=0.25in 1.25in 1in .25in, clip,scale=.7]{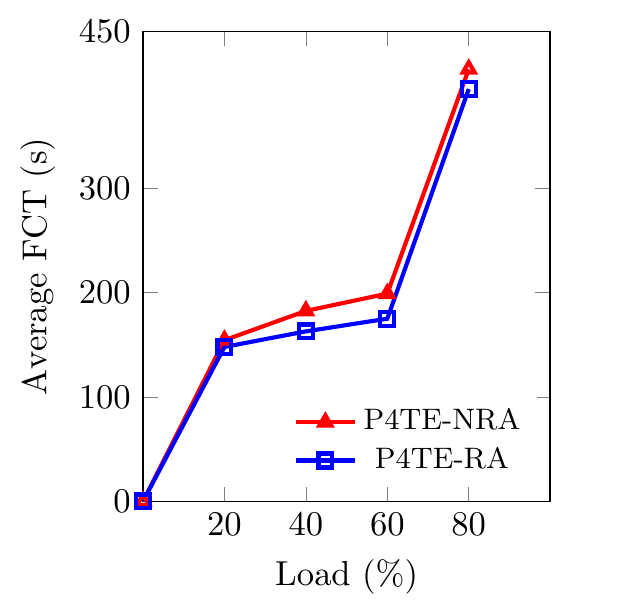}
          \caption{\centering {\small{Avg. FCT for short flows for data mining workload}}}
          \label{fig:ShortFlowRateControlImpactFctComparisonDatamining}
        \end{subfigure}
        \begin{subfigure}[]{.48\columnwidth}
          \centering
          \includegraphics[trim=0.25in 1.25in 1in .25in, clip,scale=.7]{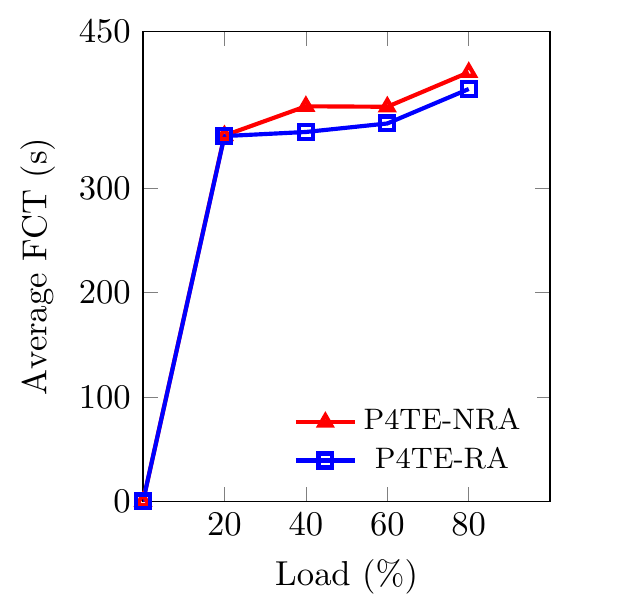}
          \caption{\centering {\small{Avg. FCT for large flows for data mining workload}}}
          \label{fig:LargeFlowRateControlImpactFctComparisonDatamining}
        \end{subfigure}
       
        \caption{\centering {\small{P4TE achieves improved FCT with \textit{rate adaptation} (P4TE-RA) compared to  
        without \textit{rate adaptation} (P4TE-NRA) for both web search and data mining workload.}}}
        \label{fig:ImpactOfRateAdaption}
      \end{figure*}

    \textbf{Traffic Design}:
    Both of the benchmark empirical workloads
    (web-search~\cite{alizadeh2010data} and data-mining~\cite{greenberg2009vl2} workload) found in
    real-life data center networks are heavy-tail characteristics and majority of the 
    data is carried by a small number of large flows. 
    In our experiments,
    The source and destinations of the flows were
    selected according to the stride~\cite{al2010hedera} pattern.
    The $i$ th host connected to $j$ th leaf switch sends data to the $(i+1)\: \% \: n$ th host connected to 
    $(j+1) \: \% \: n$ th leaf switch; $n=$ port count in the switches/2, stride index = 5. 
    The flow arrival rates
    were selected from a Poisson distribution to generate different level of load (20-80\% capacity of the aggregation layer) over the 
    simulated leaf-spine topology based data center network. 
    The flow sizes were drawn from the  distribution of the mentioned empirical workloads~\cite{greenberg2009vl2,alizadeh2010data}. 
    For both the workloads, the experiments lasted 500 seconds for every load level.
    The flows were started from Mininet simulated hosts using Iperf~\cite{mortimer2018iperf3} 
    tool, and the flow types (short or large) were tagged in the IPv6 traffic class field. 
    All the measurement (FCT and retransmissions) presented in this work is collected from an average of 5 experiments. 


    \subsubsection{Impact of Rate Adaptation}\label{ImpactOfRateAdaptation}
    To evaluate whether P4TE's in-network rate adaptation scheme provides an extra benefit over its path selection scheme,
    we experimented with two versions of P4TE: a) P4TE without rate adaption and 
    b) P4TE with rate adaptation. Fig.~\ref{fig:ImpactOfRateAdaption} shows that with the rate 
    adaptation scheme turned on (\textit{P4TE-RA}), the web search and the data mining workload require  
    less time to complete the flows compared to P4TE without rate adaption (\textit{P4TE-NRA}).
    At low load (0-20\%), the improvement is less significant because already enough link bandwidth is available to 
    accommodate the flows in these scenarios. However, at a high load, the performance improvement rate is 5-12\% 
    for both the short and large flows. 
    This behavior confirms the effectiveness of P4TE's rate adaptation mechanism.

    \begin{figure*}
        \begin{subfigure}[]{.48\columnwidth}
          \centering
          \includegraphics[trim=0.25in 1.25in 1in .25in, clip,scale=.7]{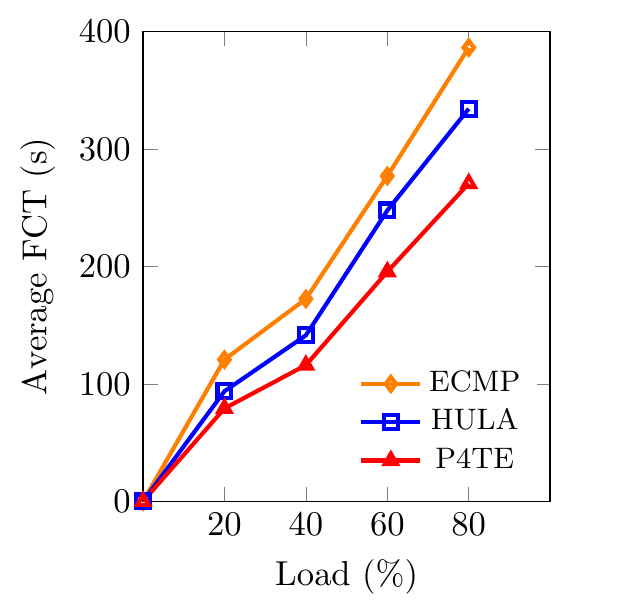}
          \caption{\centering {\small{Average FCT for \hspace{\textwidth} short flows}}}
          \label{fig:WebSearchFCTforshortFlow}
        \end{subfigure}
        \begin{subfigure}[]{.48\columnwidth}
          \centering
          \includegraphics[trim=0.25in 1.25in 1in .25in, clip,scale=.7]{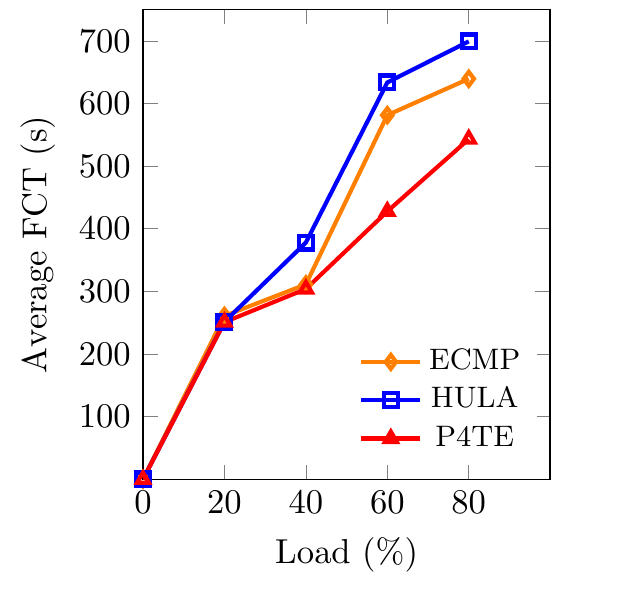}
          \caption{\centering {\small{Average FCT for \hspace{\textwidth} large flows}}}
          \label{fig:WebSearchFCTforlargeFlow}
        \end{subfigure}
        \begin{subfigure}[]{.48\columnwidth}
          \centering
          \includegraphics[trim=0.25in 1.25in 1in .25in, clip,scale=.7]{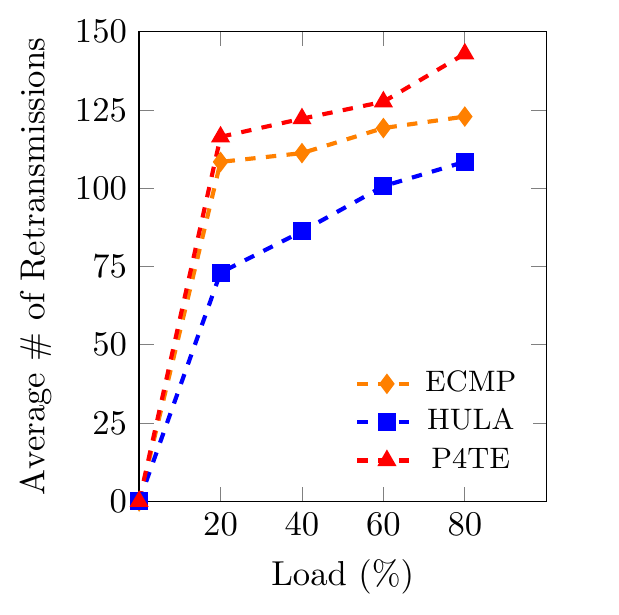}
          \caption{\centering {\small{Average retranmissions for short flows}}}
          \label{fig:WebSearchRetransforShortFlow}
        \end{subfigure}
        \begin{subfigure}[]{.48\columnwidth}
          \centering
          \includegraphics[trim=0.25in 1.25in 1in .25in, clip,scale=.7]{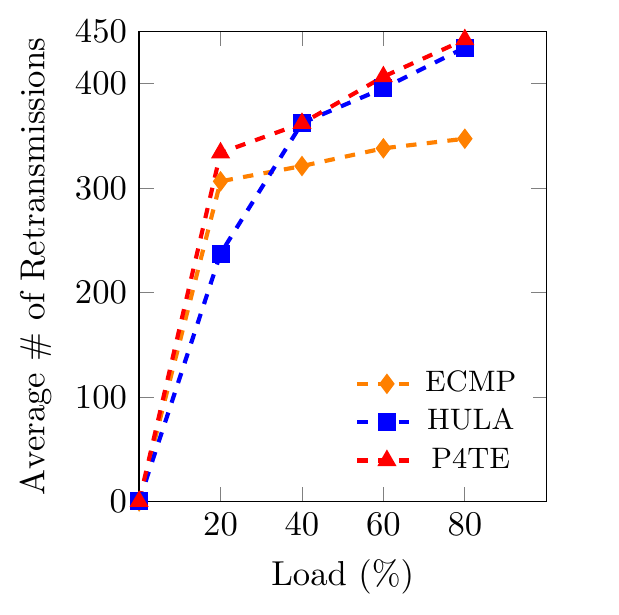}
          \caption{\centering {\small{Average retranmissions for large flows}}}
          \label{fig:WebSearchRetransforLargeFlow}
        \end{subfigure}
        \caption{\centering {\small{For web-search workload, P4TE achieves better average FCT for 
        \textit{short} and \textit{large} flows but needs more retranmissions.}}}
    \label{fig:WebsearchPerformanceComparison}
      \end{figure*}

    \begin{figure*}
        \begin{subfigure}[b]{.48\columnwidth}
          \centering
          \includegraphics[trim=.25in 1.25in 1in .25in, clip,scale=.7]{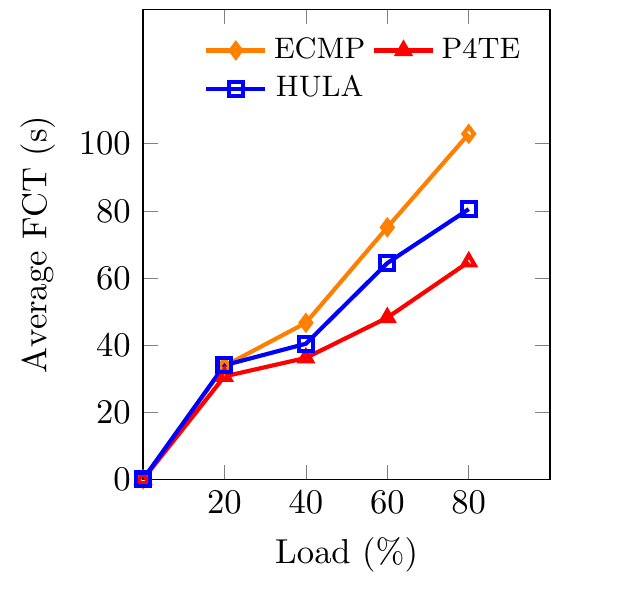}
          \caption{\centering {\small{Average FCT for \hspace{\textwidth} short flows}}}
          \label{fig:DataMiningFCTforshortFlow}
        \end{subfigure}
        \begin{subfigure}[b]{.48\columnwidth}
          \centering
          \includegraphics[trim=0.25in 1.25in 1in .25in, clip,scale=.7]{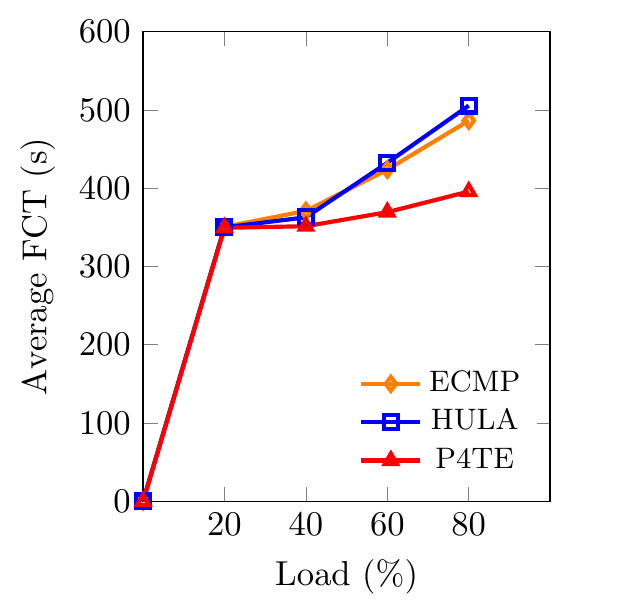}
          \caption{\centering {\small{Average FCT for \hspace{\textwidth} large flows}}}
          \label{fig:DataMiningFCTforlargeFlow}
        \end{subfigure}
        \begin{subfigure}[b]{.48\columnwidth}
          \centering
          \includegraphics[trim=0.25in 1.25in 1in .25in, clip,scale=.7]{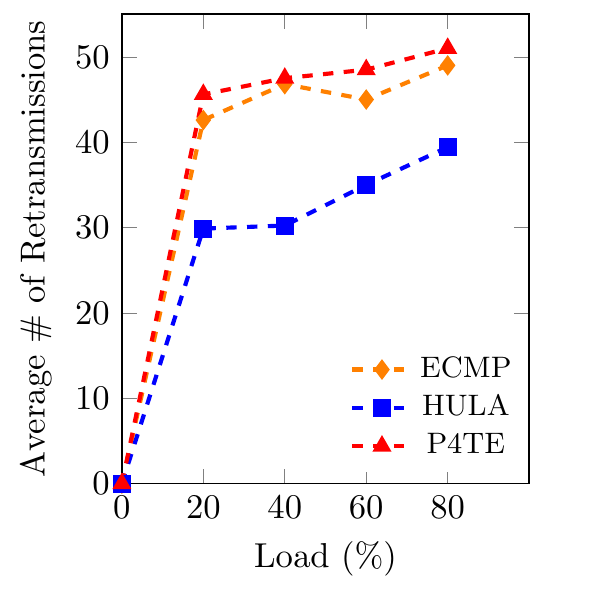}
          \caption{\centering {\small{Average retranmissions for short flows}}}
          \label{fig:DataMiningRetransforShortFlow}
        \end{subfigure}
        \begin{subfigure}[b]{.48\columnwidth}
          \centering
          \includegraphics[trim=0.25in 1.25in 1in .25in, clip,scale=.7]{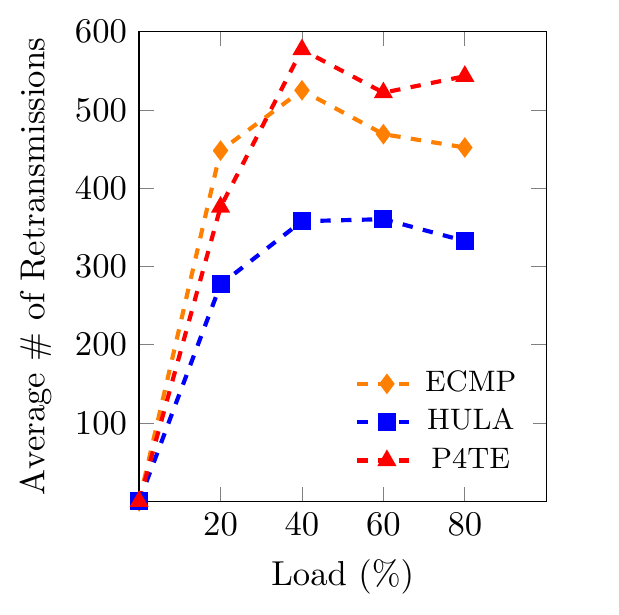}
          \caption{\centering {\small{Average retranmissions for large flows}}}
          \label{fig:DataMiningRetransforlargeFlow}
        \end{subfigure}
        \caption{\centering {\small{For data-mining workload, P4TE achieves better average FCT for 
        \textit{short} and \textit{large} flows but needs more retranmissions.}}}
    \label{fig:DataminingPerformanceComparison}
      \end{figure*}

    \subsubsection{P4TE's Performance Improvement}\label{PerformanceImprovement}

    Now we evaluate P4TE's performance improvement compared to ECMP and HULA~\cite{katta2016hula}. 
    ECMP selects path based on the hash
    value of a flow's five-tuples (src and dst IP, src and dst port
    number, and protocol number). 
    On the other hand, HULA probes every \textit{leaf-to-leaf} switch  path in every regular 
    interval to find the least utilized path for a destination. 
    All three schemes select path at the granularity of flowlets~\cite{sinha2004harnessing} to reduce 
    packet reordering in TCP.

    \paragraph{\textbf{Flow Completion Time and Retransmission}}:
    Fig.~\ref{fig:WebSearchFCTforshortFlow} and~\ref{fig:WebSearchFCTforlargeFlow} separately show 
    the average flow completion time (FCT) for both the short and large flows for the web-search workload at different load levels. 
    Compared to ECMP's link load unaware \textit{path-selection} algorithm, P4TE uses a link load and flow type aware scheme. 
    Hence, ECMP always experiences a higher FCT compared to P4TE. 
    The behavior is less significant at lower loads (0-40\%) as enough link bandwidth is available for all flows.  
    With an increase in network load, ECMP performs poorly.  
    Overall, compared to P4TE, ECMP requires approx. 40-55\% and  4-35\% more time to finish the short and large flows, respectively. 
    HULA always selects the least utilized path for a destination, enabling it to perform better than ECMP. 
    Always selecting the least utilized path gives the short flows a chance to finish quickly. 
    However, HULA does not differentiate among 
    short and large flows hence both types of flows compete with each other over the same link, and the 
    large flows are penalized more.
    On the other hand, P4TE forwards 
    short flows through the same path until queue build up starts and forwards large flows through a 
    least utilized path to push more packets without 
    hampering the short flows. Hence, it is less prone to the tendency of penalizing one type of flow for the other. 
    Fig.~\ref{fig:WebSearchFCTforshortFlow} and~\ref{fig:WebSearchFCTforlargeFlow} shows, at 60-80\% load, the large flows 
    require 29-47\% more time, and the short flows require 23-40\% more time to finish than P4TE. 
    This validates P4TE's superior performance and HULA's tendency to penalize the large flows more. 
    Overall, HULA requires 2-60\%  and 2-40\% more time than P4TE to finish the short and large flows, respectively.     
    Similar to the web-search workload, P4TE also achieves 
    improved performance (lower average FCT)  for both small and large flows in the case of data-mining workload 
    (Fig.~\ref{fig:DataMiningFCTforshortFlow} and~\ref{fig:DataMiningFCTforlargeFlow}).
    Similar to web search workload, the performance improvement in high load (40-80\%) factor is more visible. 
    At high load, under HULA, the large flows face more performance degradation compared to the short flows. 
    At high load,  ECMP requires 6-24\%, and  HULA requires 3-22\%   more time than P4TE to finish the large flows;
    Similarly, for the short flows,  ECMP requires 27-56\%, and HULA requires 11-33\% more time than P4TE. 
    Besides this, the CDFs of the FCT are presented in appendix~\ref{DataMiningCDF}  and~\ref{WebSearchCDF}. 
    Appendix~\ref{DataMiningCDF} shows that P4TE achieves shorter FCT than the alternatives for large flows in the data mining workload (except at a very low 
    load of 20\%). 
    Also, for the short flows and overall flows, it shows  comparable or shorter FCT than the alternatives (except for a small percentage of flows at the tail). 
    Appendix~\ref{WebSearchCDF} shows,   in web search workload,
    P4TE achieves shorter FCT than the alternatives for short flows at all load levels and large flows at 60\% or above load levels. 
    Similar to the data mining workload,  it shows  comparable or shorter FCT than the alternatives (except for a small percentage of 
    flows at the tail at a low load level of 20\%). 
    Considering the performance improvements, 
    P4TE can be a potential candidate traffic engineering system for improving FCTs in real data center
    networks.
    However, this benefit comes at the cost of a higher number of retransmissions compared to both ECMP and HULA; 
    and  more uneven load distribution over upward links (when available) compared to HULA. 
    We discuss these two overheads in the next two paragraphs.

    
    To achieve better performance, P4TE tries to select the path
    with a low chance of congestion, and in case of congestion, it
    controls the flow rates through FACK packets.
    P4TE uses the last sequence number in FACK packets for which the sender has received acknowledgment from the receiver.
    This sequence number is not a correct reflection of real data acknowledged by the receiver. 
    End-host protocol stacks can initiate retransmission when receiving  FACK packets with already acknowledged sequence numbers.  
    As a result, total number of retransmissions under P4TE can increase. 
    Fig.~\ref{fig:WebSearchRetransforShortFlow},~\ref{fig:WebSearchRetransforLargeFlow},~\ref{fig:DataMiningRetransforShortFlow},
    and~\ref{fig:DataMiningRetransforlargeFlow} show a higher average number of retransmissions under P4TE for both short and
    large flows in both web-search and data-mining. 
    In the case of web-search workload, for the short flows, P4TE requires 6-16\% and 27-54\% more 
    retransmission compared to ECMP and HULA, respectively.
    For the large flows, it requires 8-26\% more retransmission than ECMP. 
    At 20\% load, HULA requires 37\% less retransmission compared to HULA. However, 
    at high load (40-80\%) the gap narrows down to 3-7\%. The reason behind that is, HULA uses 
    the same path for both short and large flows. It leads to congestion on the same path and 
    results in an  increased number of retransmission. 
    But, P4TE uses the least utilized path for large flows. Hence, the large 
    flows can push more packets without facing P4TE's \textit{rate-adaption} scheme. 
    It results in less retransmission (almost close to HULA). 
    In the case of data mining workload, a large portion of flows are small in size (80th percentile 
    of flow size around 50\% of web search workload), and they 
    perform better under HULA. On the other hand, P4TE imposes \textit{rate-adaption} on the flows more frequently, 
    and they face more retransmission. 
    Under P4TE, at different load levels, the average number of 
    retransmissions in data-mining workload (fig.~\ref{fig:DataMiningRetransforShortFlow}) is around half
    of the retransmissions in web-search workload (fig.~\ref{fig:WebSearchRetransforShortFlow}) at the same load level.
    As the majority of the flow size in data-mining workload is half the
    size of web-search workload; this behavior shows P4TE’s ability to
    follow the traffic pattern. As the short flows are smaller in the data mining workload, they finish early 
    and leave more link bandwidth for a small number of large flows. Here, HULA forwards them through the least utilized path 
    and requires less retransmission. P4TE also uses the least utilized path for these flows. 
    However, its rate controlling mechanism increases the number of retransmissions (fig.~\ref{fig:DataMiningRetransforlargeFlow}).

    \paragraph{\textbf{Link Utilization}}: \label{LinkUtilizationPara}
    In fat-tree topology, end hosts in all the subnets (except the own subnet) defined by a switch 
    is reachable through any of its upward links. Hence,  at every switch, an ideal 
    traffic engineering scheme should evenly distribute the 
    traffic toward other subnets over multiple upward links (when available) to avoid congestion. However, 
    in the case of ECMP, HULA, and P4TE,  two of the main factors influencing this distribution are the following. 
    Firstly, their path selection algorithm working at the granularity of flowlet level pins a flow to a specific 
    link for the period of flowlet-interval time even though the link is 
    facing more utilization compared to the alternative links. It contributes to uneven traffic distribution over the links. 
    Secondly,   a link utilization-aware path selection algorithm can help to reduce further imbalance by the appropriate selection of a link. 
    Separately measuring the load imbalance by these two factors is not possible. Moreover, as ECMP, HULA, 
    and P4TE follow a flowlet-based approach; the load imbalance over multiple links caused by the first factor 
    can be considered the same. Therefore, the overall load imbalance over the links can be attributed mainly to the path selection algorithm.

\begin{figure}
      \centering
      \includegraphics[trim=0in 1.25in 1in .25in, clip,scale=.7]{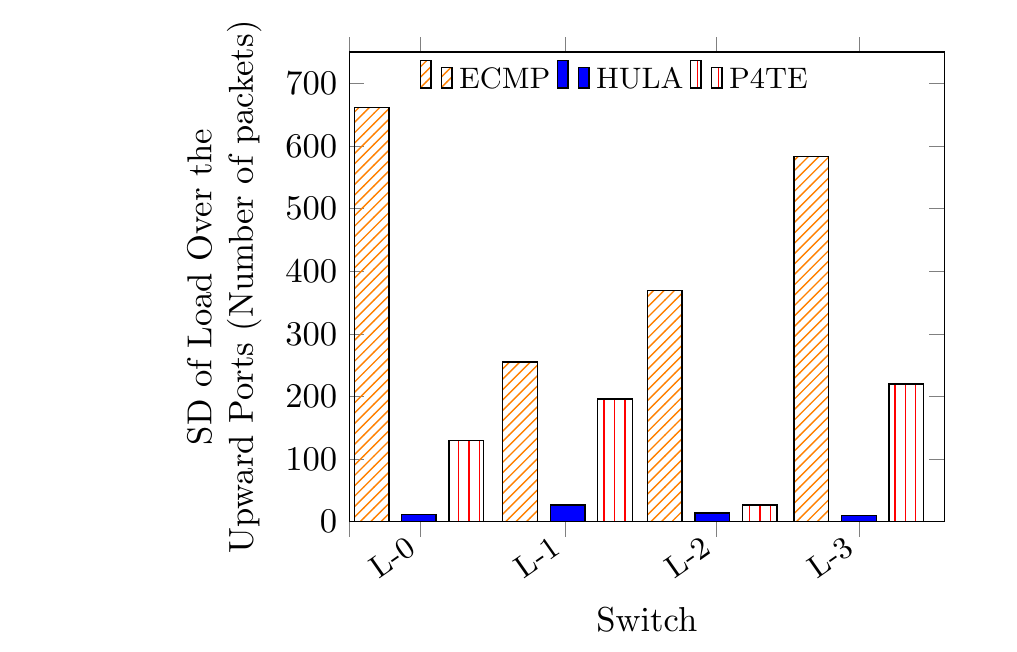}
      

    
    \caption{\centering{ \small{Standard deviation of load distribution over the upward ports of the leaf switches for the web-search workload.}}}
    \label{fig:switch1UpwardPortEgressPacketCounter}
    \end{figure}

    ECMP's path selection algorithm often suffers due to hash collision. On the other hand,
    P4TE’s dynamic path priority reconfiguration mechanism and path selection 
    algorithm helps to distribute traffic over multiple paths more evenly compared to ECMP. 
    We counted the total number of packets forwarded
    through each of the upward links (where path diversity exists) of the leaf switches for the web-search workload at 80\% load. 
    Next, we computed the standard deviation of traffic load over the leaf-to-spine links at each switch. 
    Fig.~\ref{fig:switch1UpwardPortEgressPacketCounter} shows, in the case of
    ECMP, the standard deviation of load over different upward ports at each switch 
    ranges between 196.06 and 661.35 packets. Whereas, in the case of
    P4TE, the values range between 27.86 and 220.72 packets. Therefore,
    P4TE distributes traffic over multiple paths better compared to
    ECMP. However, Fig.~\ref{fig:switch1UpwardPortEgressPacketCounter} shows that HULA can
    distribute traffic more evenly over the links. 
    The reason is that HULA probes the link utilization for every destination at a small time scale (order of round-trip-time) and 
    selects the least utilized path. As a result, a new path is selected for a
    destination at every interval, and eventually, the packets are almost evenly distributed
    over multiple paths. But to achieve even traffic distribution over multiple paths, HULA underperforms compared to 
    P4TE (fig.~\ref{fig:WebSearchFCTforshortFlow} and ~\ref{fig:WebSearchFCTforlargeFlow}).

      \begin{figure}
          \begin{subfigure}[b]{.48\columnwidth}
            \centering
            \includegraphics[trim=0.35in 1.25in 1in .25in, clip,scale=.7]{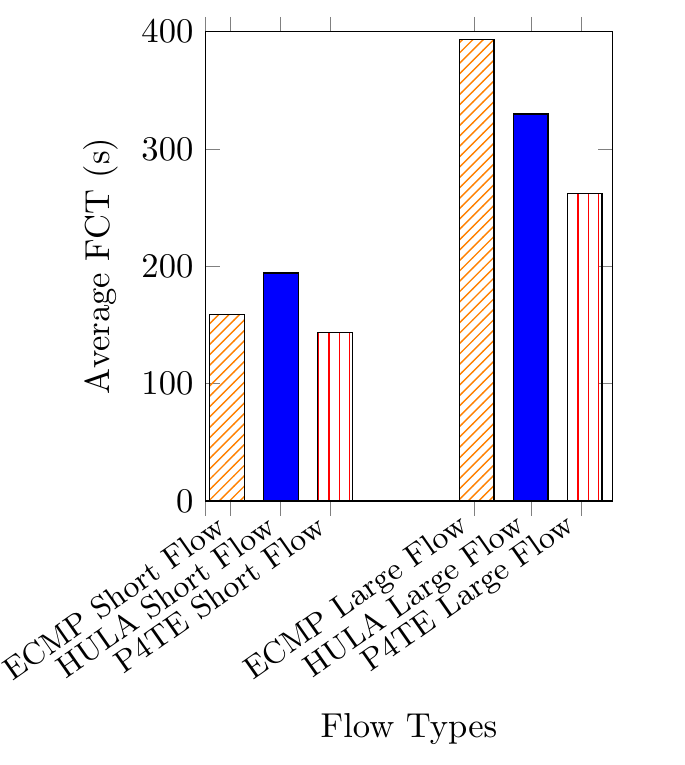}
            \caption{\centering {\small{Average FCT for \hspace{\textwidth}  short \hspace{\textwidth} and large flows \hspace{\textwidth}}}}
            \label{fig:IncastFCT}
          \end{subfigure}
          \begin{subfigure}[b]{.48\columnwidth}
            \centering
            \includegraphics[trim=0.25in 1.25in 1in .25in, clip,scale=.7]{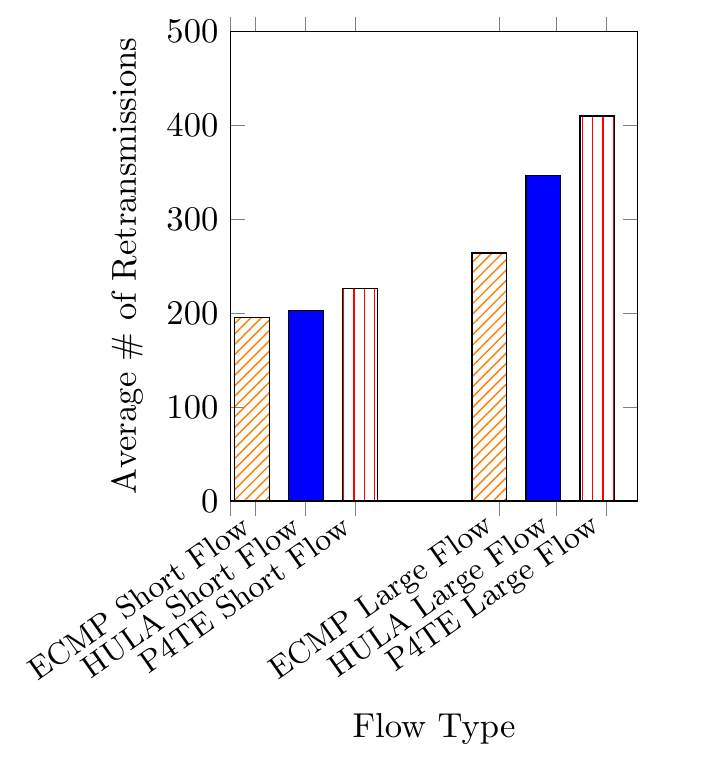}
            \caption{\centering {\small{Average \# of retransmissions for  short and large flows}}}
            \label{fig:IncastRetrans}
          \end{subfigure}
      \label{fig:IncastPerformanceComparison}
      \caption{\centering{\small{Average flow completion time and number of retranmissions in incast scenario.}}}
        \end{figure}

    \subsection{Incast} \label{IncastTrafficScenario}

    \textbf{Traffic design}: In this case,  incast traffic patterns similar to prior 
    works~\cite{alizadeh2014conga,wu2012ictcp}  were generated. 
    However, as our Mininet~\cite{mininet} based testbed is small in scale; we have scaled down the flow sizes in the experiments. 
    Besides this, instead of all the same size flow, we have started a mix of 24 short flows (512KB size) and 8 large flows (1024KB size) 
    from 12 different hosts to a \textit{single host} 
    in a synchronized fashion.   
    A mix of both short and large flows was used to analyze the effectiveness P4TE's traffic class-based forwarding scheme. 
    The source hosts were connected to different leaf switches to not create congestion at different 
    switches. 
    Here, the link between the destination host and the connected leaf switches was the main bottleneck. 
    This scenario was created purposefully 
    to replicate the incast behavior.

    \textbf{Result analysis}: Here, the main bottleneck link is at the last hop of the source-destination path.
    Under all three schemes,  heavy congestion occurs at the bottleneck link, and packets get dropped from the queue. 
    But, in the case of P4TE, the flow sources get the congestion notification earlier. Because 
    P4TE enabled switch imposes rate control at the last leg of the path, and the FACK packets are sent from the last 
    hop switch (the leaf switch directly connected with the destination).  
    On the other hand, under ECMP or HULA, the ECN~\cite{floyd1994tcp} based congestion marking (marked by the same hop) reaches the destination 
    and then reaches back to the source. As a result, compared to P4TE, the congestion notification reaches the flow source 
    lately in RTT time. 
    However, the time gap is really small (because there is only one hop difference). 
    Moreover,  P4TE and ECMP/HULA (with DCTCP) 
    do not penalize a flow more than once within a window of a similar amount of bytes.
    Hence, under P4TE, the flows do not get too much 
    chance to improve.  
    It is reflected by the small improvement in FCT for short flows by P4TE shown in Fig.~\ref{fig:IncastFCT}. 
    However, the large flows run for a longer time, and P4TE gets more chance to control its rate. 
    Hence, the improvement in FCT for the large flows (Fig.~\ref{fig:IncastFCT}) is more compared to the short flows. 
    To control the flow rates, P4TE uses the FACK packets. As a result,
    similar to web search and data mining workload,  P4TE requires more retransmissions (Fig.~\ref{fig:IncastRetrans}) in this scenario also.


    \section{Conclusion}\label{Conclusion}
    This paper presented P4TE, a switch-centric traffic engineering scheme for fat-tree topology-based data center networks. 
    It is deployable using existing PISA switches and able to maintain line-rate performance. P4TE leverages programmable switch-based 
    in-band mechanisms to monitor link performance metrics at fine-grained accuracy. P4TE uses a 
    fully distributed algorithm for link performance-aware routing and rate control of the flows using monitored metrics.
    Our evaluations show that P4TE offers superior performance compared to ECMP without any support from the end-host
    transport layer protocol stack. 
    P4TE is oblivious to the end-host transport layer protocol stack and uses a fake acknowledgment packet as a measure to 
    convey feedback to end hosts. As a result, P4TE faces an increased 
    number of retransmission but improves the flow completion time for the majority of the short and large flows in high load levels.
    We plan to work on designing algorithms to achieve other traffic engineering objectives (energy awareness, QoS, etc.) in the future. 
    We also plan to work on the switch and end-host protocol stack codesigning for traffic engineering in data center networks.


\printcredits

\bibliographystyle{unsrtnat}

\appendix

\section{Appendix}

\subsection{Flowlet Interval}\label{FlowletConfigAppendix}
We have evaluated ECMP, HULA, and P4TE's performance in this work. All three schemes 
select a path for a flow at a granularity of flowlet to reduce the impact of TCP reordering. 
Hence, the configuration of flowlet interval plays a 
significant role here. The flowlet interval  can be configured to a large value $T_{max}\geq$ RTT) or to 
a small value $T_f^{min}=$1/maximum traffic burst rate. In the case of $T_{max}$, all packets of a flow form 
only one flowlet, and none of the three schemes can take advantage of the multi-path capability for a flow. 
On the other hand, configuring the flowlet interval to $T_f^{min}$ creates a new flowlet for every packet of 
a flow. It increases the number of packet reordering, and the performance of the TCP protocol degrades. Therefore, it is
important to configure the flowlet interval to an optimal value from the range $[T_f^{min}, T_f^{max}]$. 
In this work, we have experimentally selected the value of flowlet-interval and used it in the experiments 
described in section~\ref{PerformanceEvaluation}. 

\begin{figure}[h]
    \centering
    \includegraphics[trim=0.0in 0in 2in 0in, clip,scale=.6]{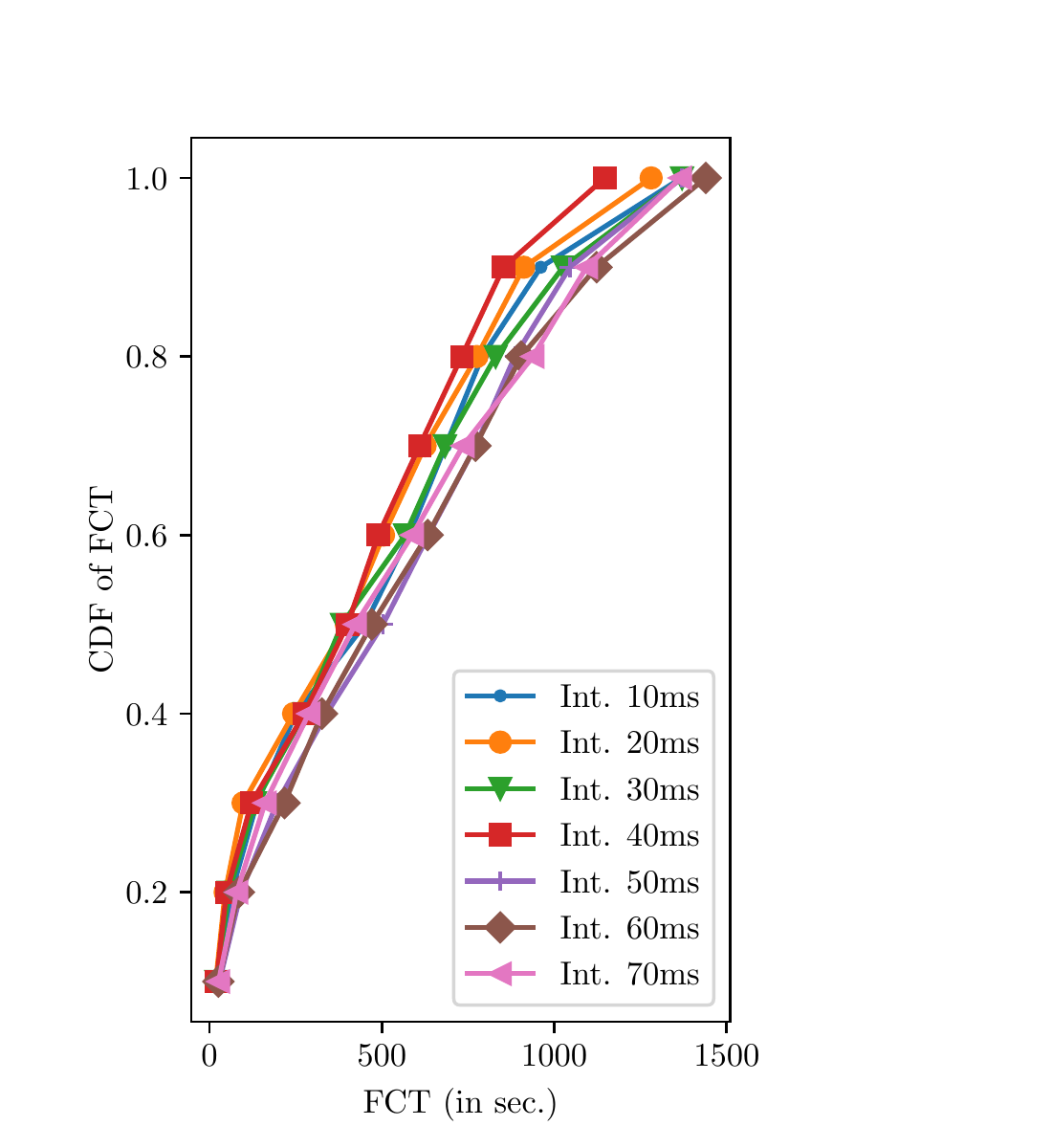}
    
  
  \caption{\centering {\small{CDF of FCT for different values of flowlet interval.}}}
  \label{fig:FCT-FlowletConfiguration}
\end{figure}

To get the value of $T_f^{max}$, we have run the web search workload at 80\% load using the ECMP algorithm and computed the 
average RTT. The average RTT = 70 ms gives the value of $T_f^{max}$. 
Then, we find that TCP sends in bursts of roughly 80 packets in our testbed. It gives the value of $T_f^{min} = 1/80$ ms = 12.5 ms.
Then finally, we have run the web search workload at 80\% load
for different values of flowlet interval within the range $[T_f^{min}=10ms, T_f^{max}=70ms]$.
Fig.~\ref{fig:FCT-FlowletConfiguration} shows the CDF of FCT for different values of flowlet interval. 
The best performance is found at flowlet interval = 40 ms. We used this value as flowlet interval in the experiments for 
evaluating and comparing P4TE's performance. 

\subsection{CDF of FCT}\label{CDFAppendix}

Figure~\ref{fig:DataminingCDF} and~\ref{fig:WebSearchCDF} show how the cumulative distribution function (CDF) of
FCT of P4TE compares with ECMP and HULA for both web search and data mining workload. 
In section~\ref{PerformanceEvaluation}, we have presented the descriptions of the experiments and discussed 
the results using the average FCT.
Here we elaborate on the results of the same experiments through CDF of FCT. 

\subsubsection{Data Mining  Workload}\label{DataMiningCDF}
\begin{figure*}[t]
      \begin{subfigure}[]{.48\columnwidth}
        \centering
        \includegraphics[trim=0in 0in 2in 0in, clip,scale=.5]{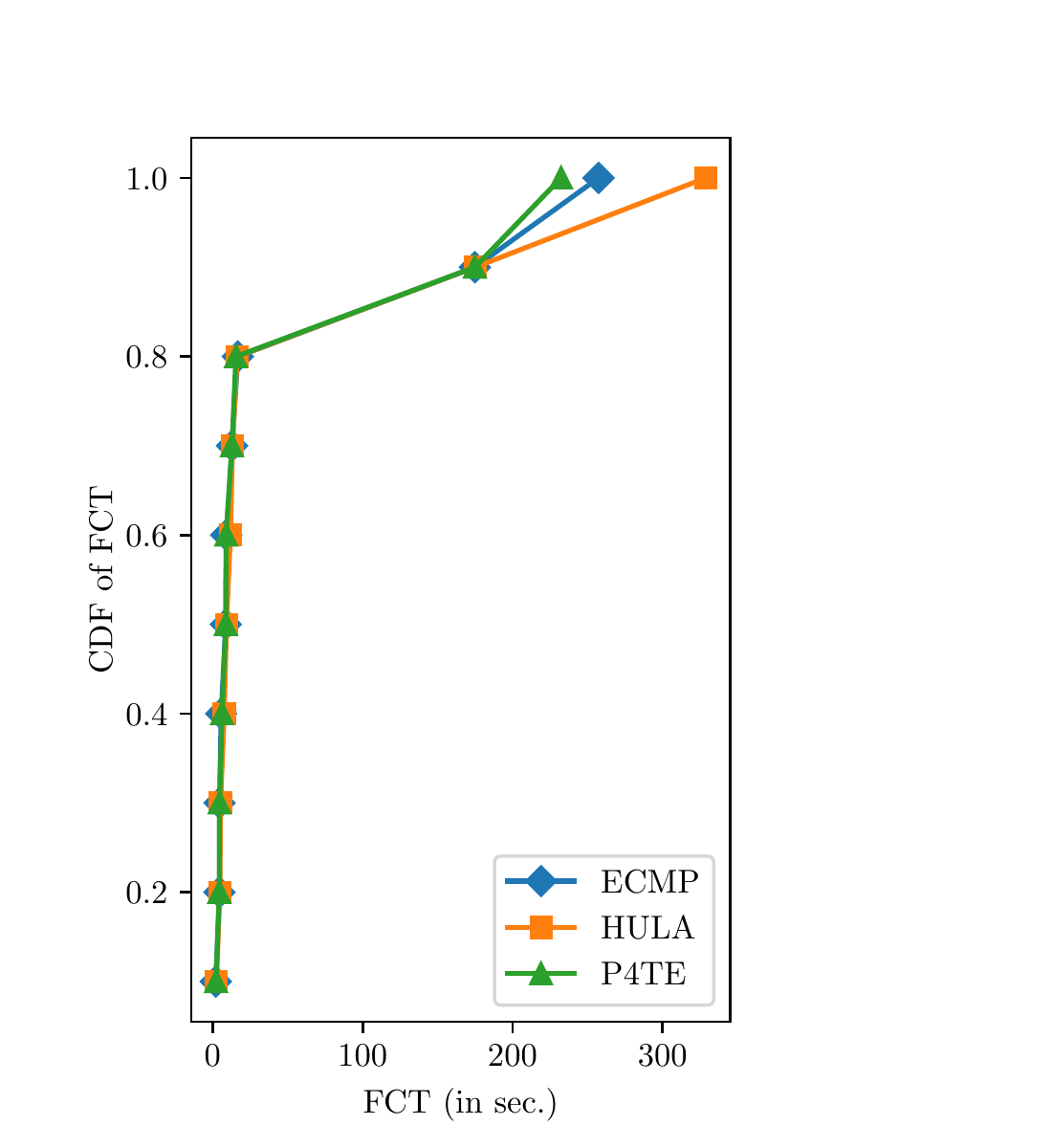}
        \caption{\centering {\small{CDF of FCT for short flows at 20\% load}}}
        \label{fig:1}
      \end{subfigure}
      \begin{subfigure}[]{.48\columnwidth}
        \centering
        \includegraphics[trim=0in 0in 2in 0in, clip,scale=.5]{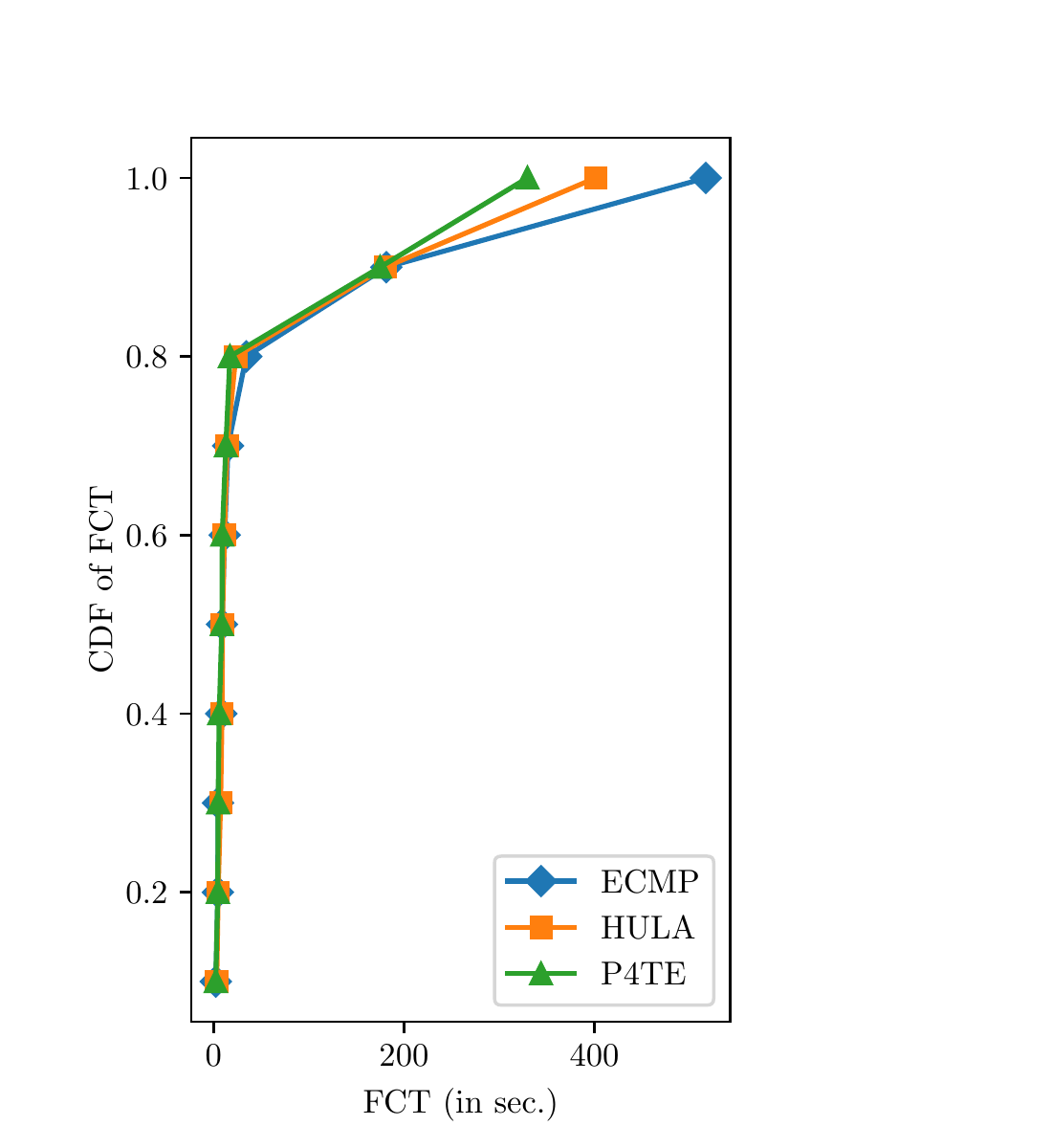}
        \caption{\centering {\small{CDF of FCT for short flows at 40\% load}}}
        \label{fig:2}
      \end{subfigure}
      \begin{subfigure}[]{.48\columnwidth}
        \centering
        \includegraphics[trim=0in 0in 2in 0in, clip,scale=.5]{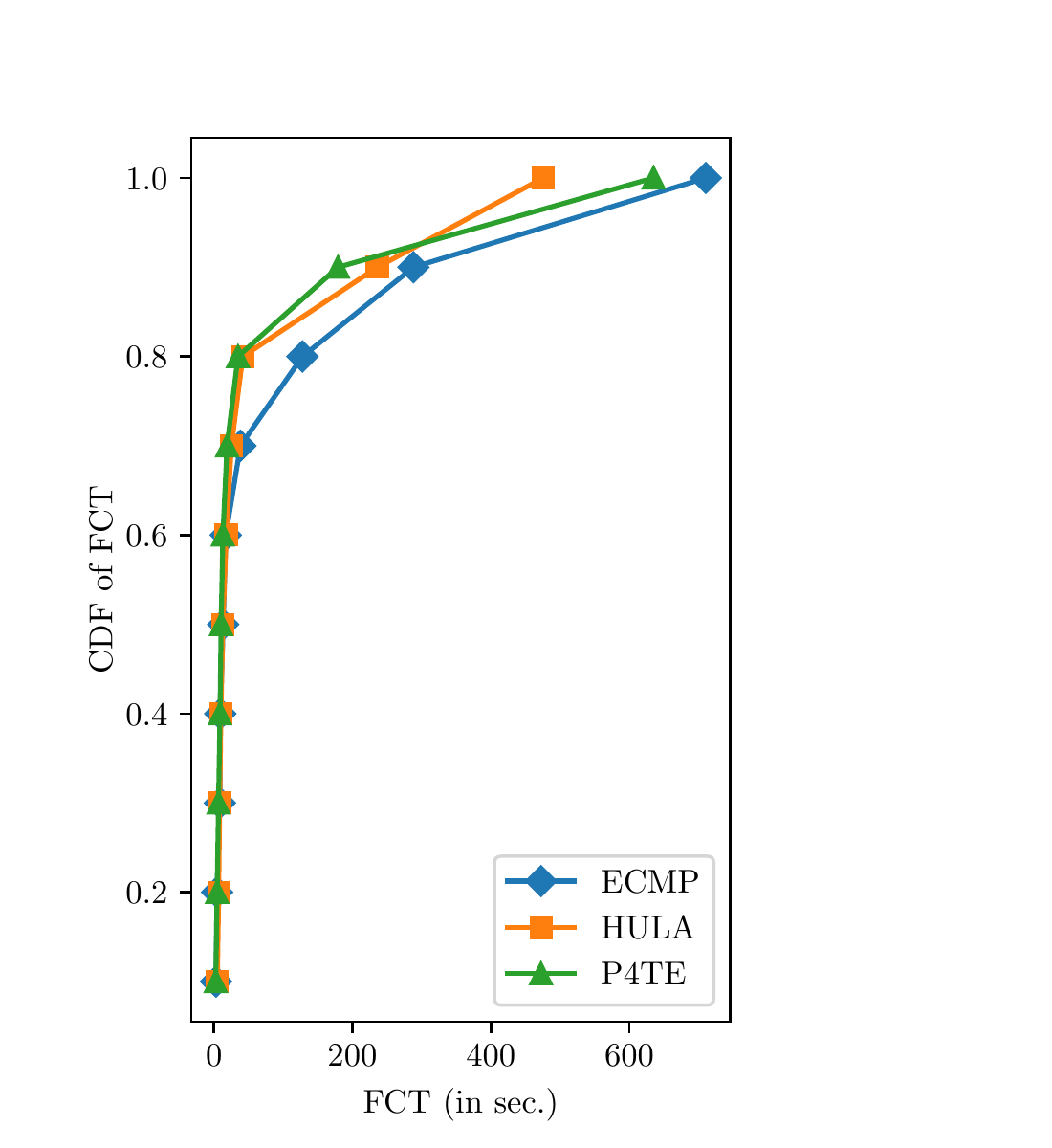}
        \caption{\centering {\small{CDF of FCT for short flows at 60\% load}}}
        \label{fig:3}
      \end{subfigure}
      \begin{subfigure}[]{.48\columnwidth}
        \centering
        \includegraphics[trim=0in 0in 2in 0in, clip,scale=.5]{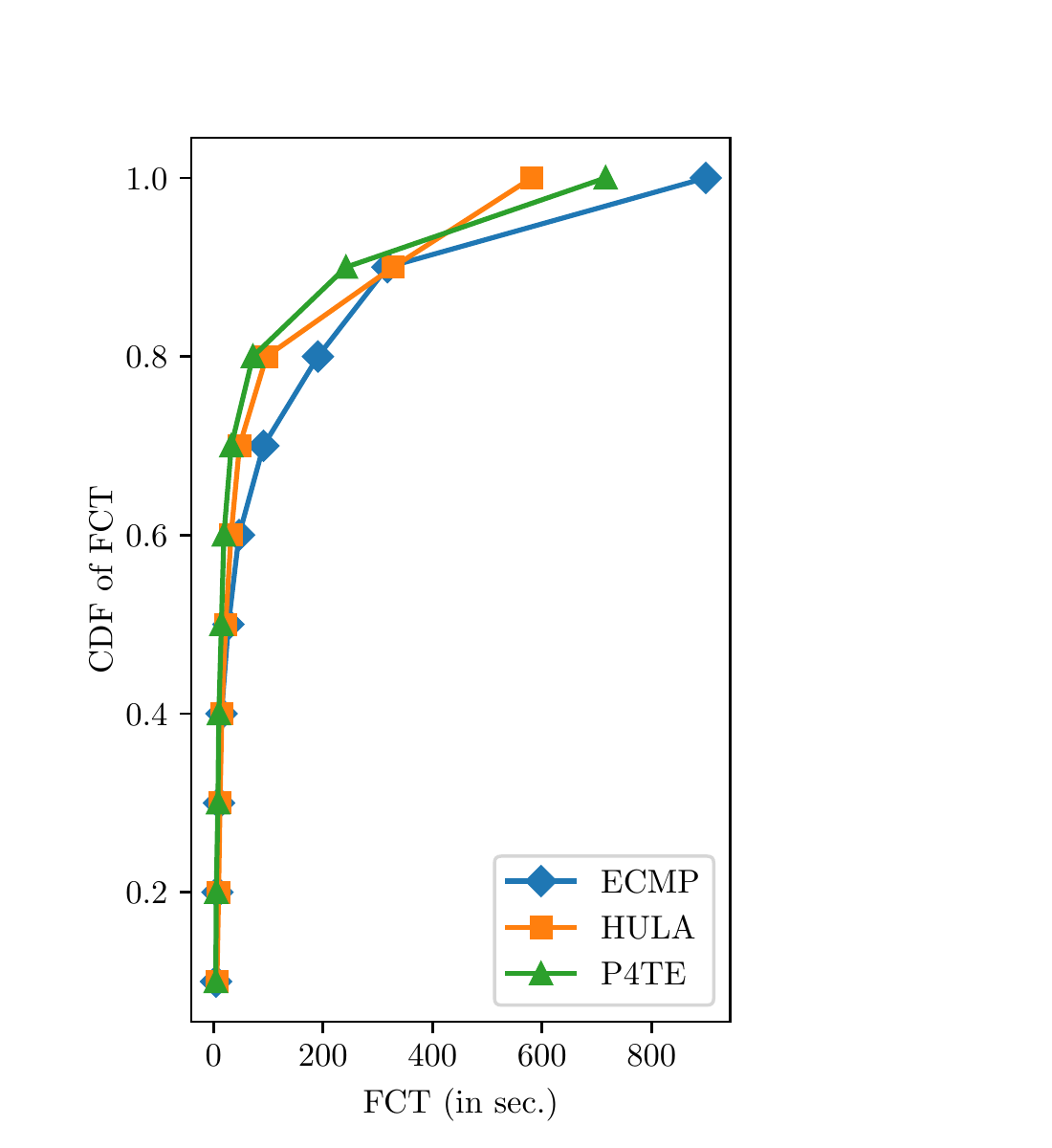}
        \caption{\centering {\small{CDF of FCT for short flows at 80\% load}}}
        \label{fig:4}
      \end{subfigure}
      \begin{subfigure}[]{.48\columnwidth}
        \centering
        \includegraphics[trim=0in 0in 2in 0in, clip,scale=.5]{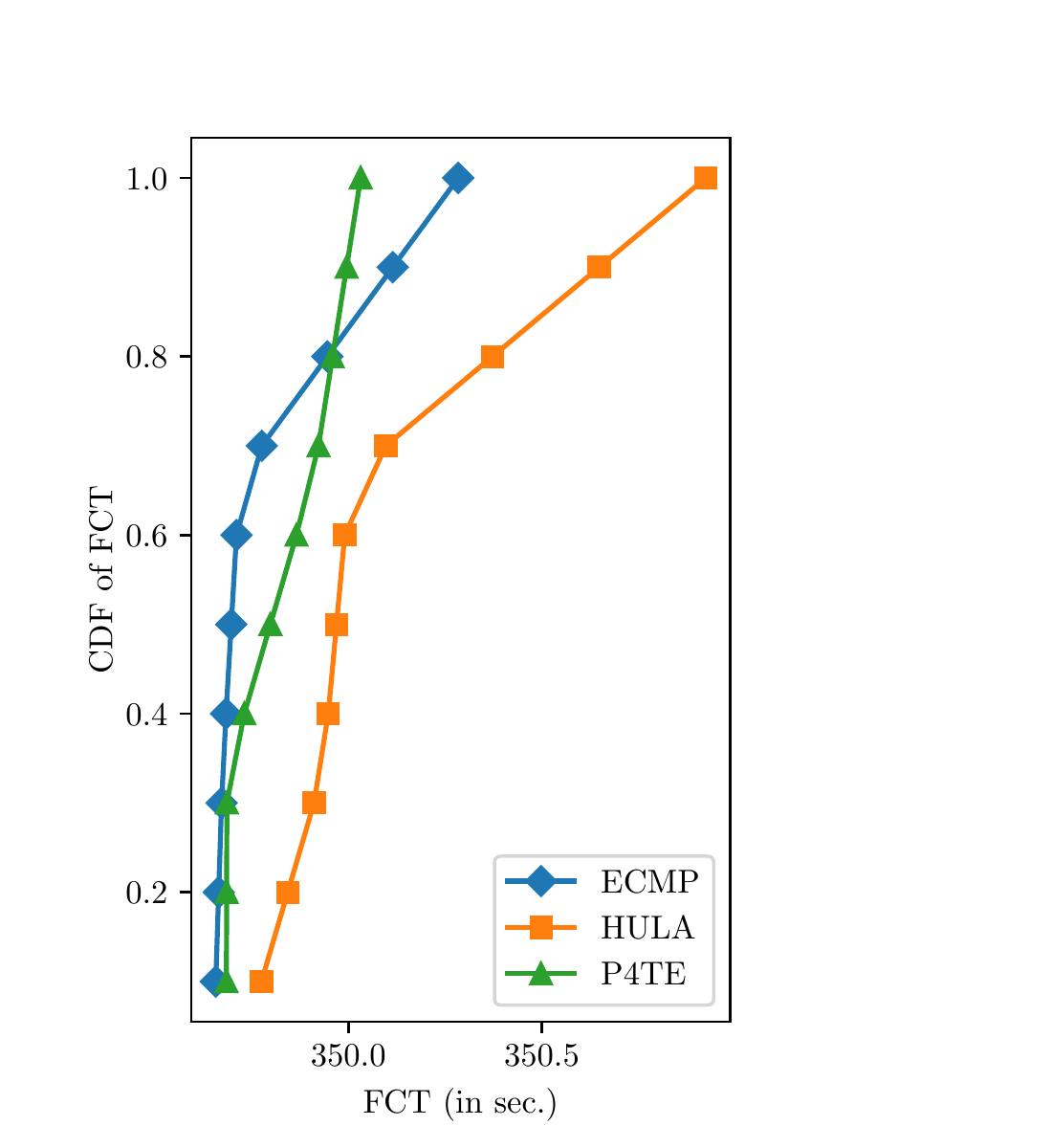}
        \caption{\centering {\small{CDF of FCT for large flows at 20\% load}}}
        \label{fig:5}
      \end{subfigure}
      \begin{subfigure}[]{.48\columnwidth}
        \centering
        \includegraphics[trim=0in 0in 2in 0in, clip,scale=.5]{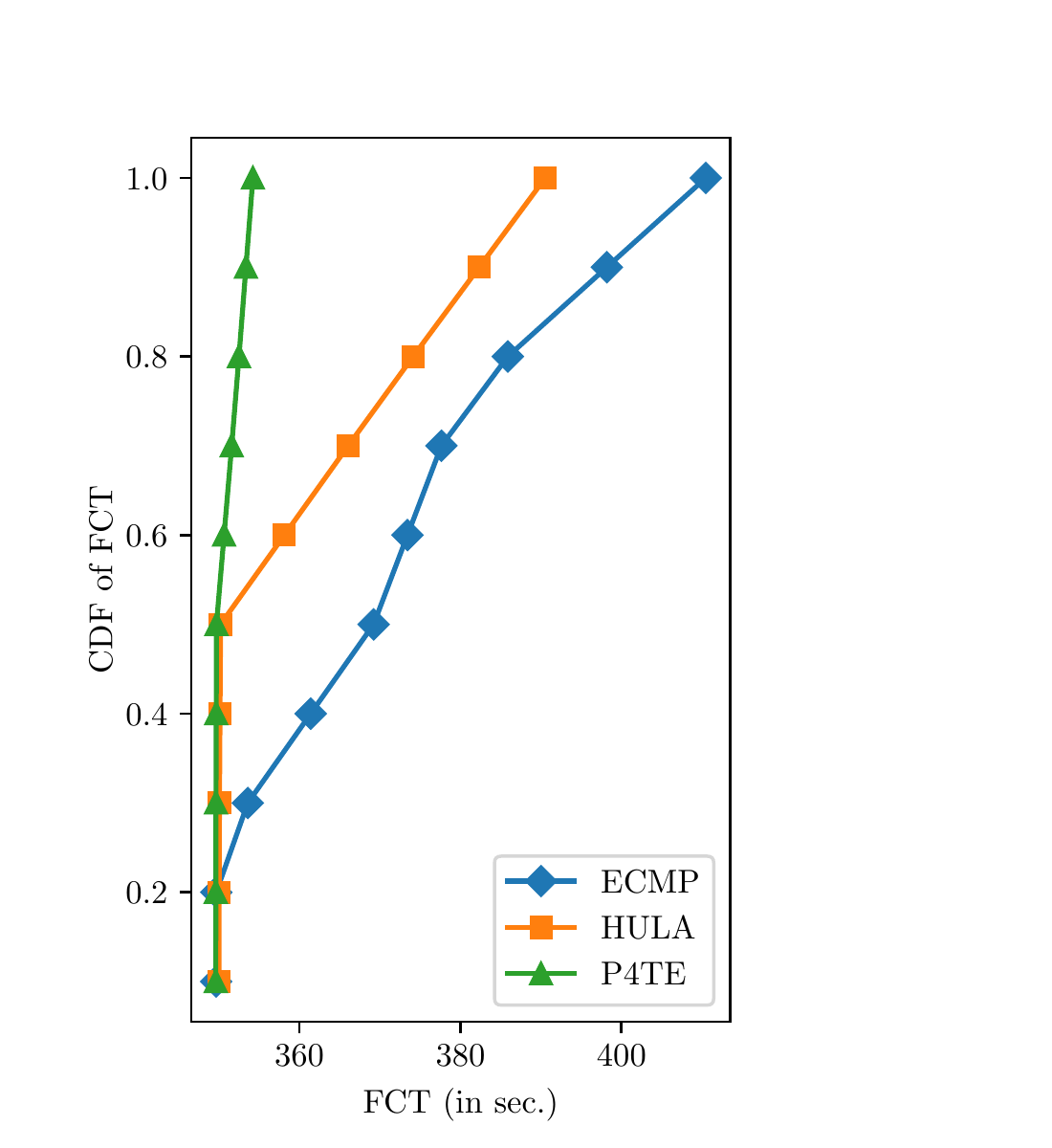}
        \caption{\centering {\small{CDF of FCT for large flows at 40\% load}}}
        \label{fig:6}
      \end{subfigure}
      \begin{subfigure}[]{.48\columnwidth}
        \centering
        \includegraphics[trim=0in 0in 2in 0in, clip,scale=.5]{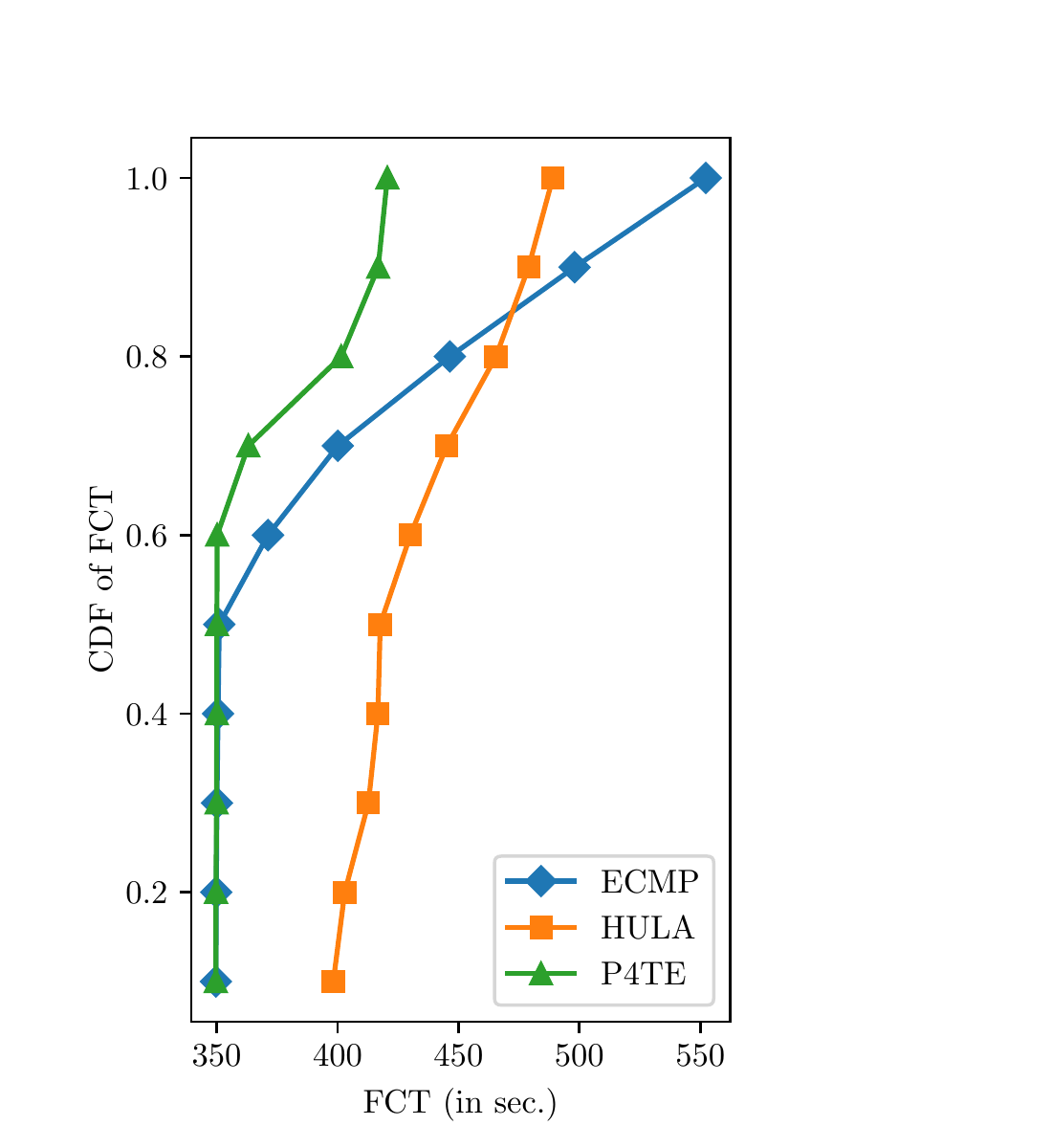}
        \caption{\centering {\small{CDF of FCT for large flows at 60\% load}}}
        \label{fig:7}
      \end{subfigure}
      \begin{subfigure}[]{.48\columnwidth}
        \centering
        \includegraphics[trim=0in 0in 2in 0in, clip,scale=.5]{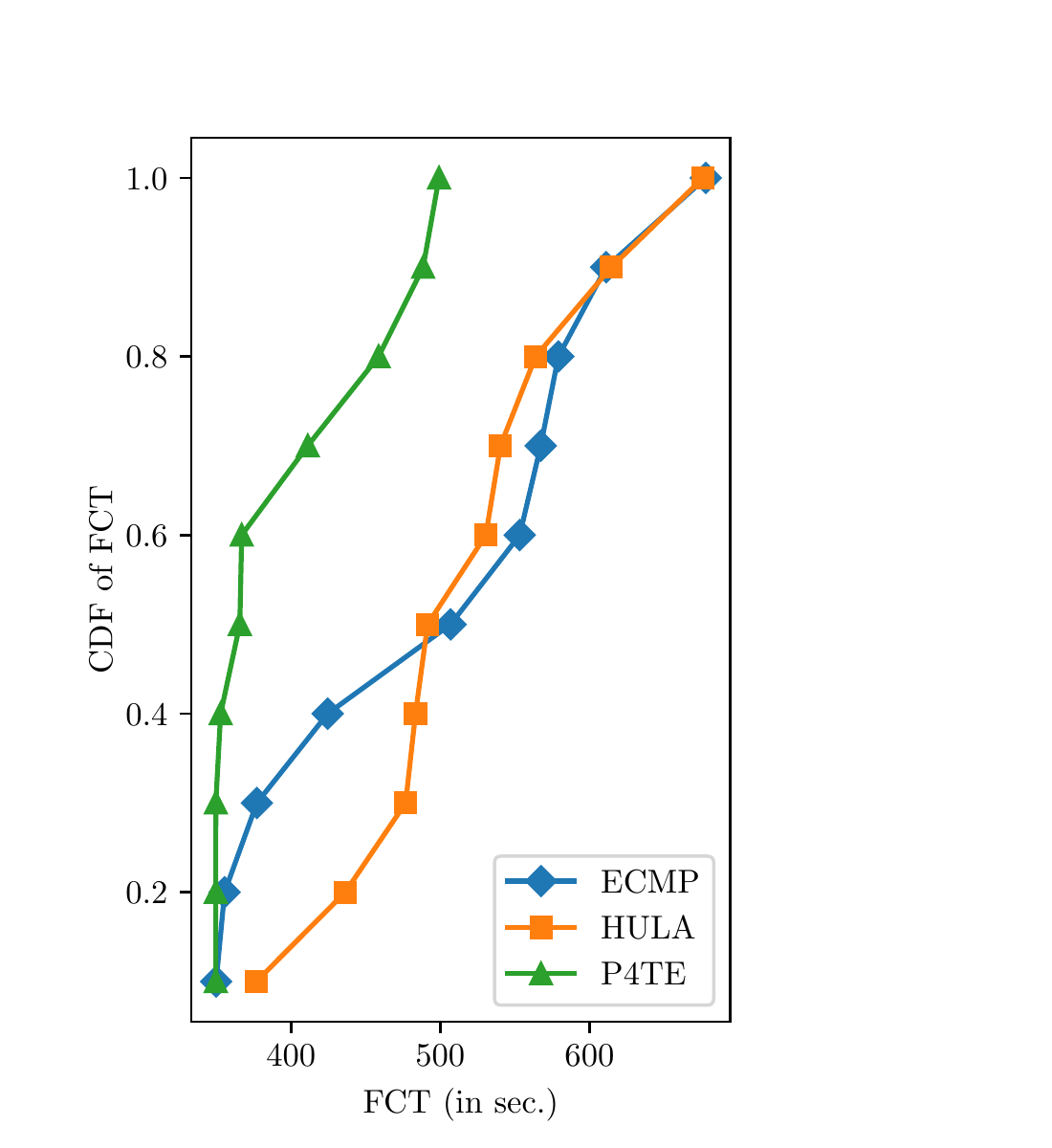}
        \caption{\centering {\small{CDF of FCT for large flows at 80\% load}}}
        \label{fig:8}
      \end{subfigure}
      \begin{subfigure}[]{.48\columnwidth}
        \centering
        \includegraphics[trim=0in 0in 2in 0in, clip,scale=.5]{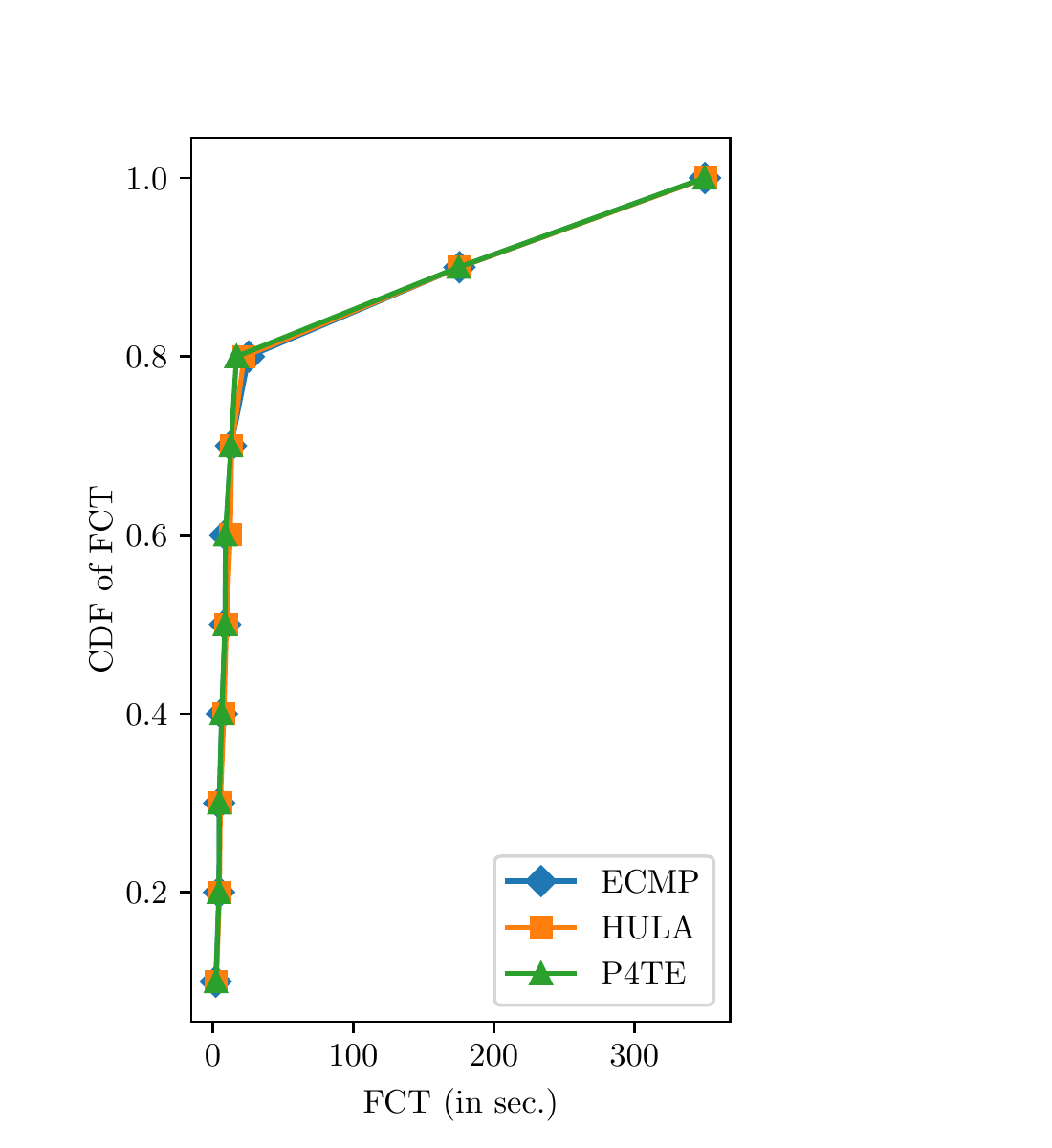}
        \caption{\centering {\small{CDF of FCT for all flows at 20\% load}}}
        \label{fig:9}
      \end{subfigure}
      \begin{subfigure}[]{.48\columnwidth}
        \centering
        \includegraphics[trim=0in 0in 2in 0in, clip,scale=.5]{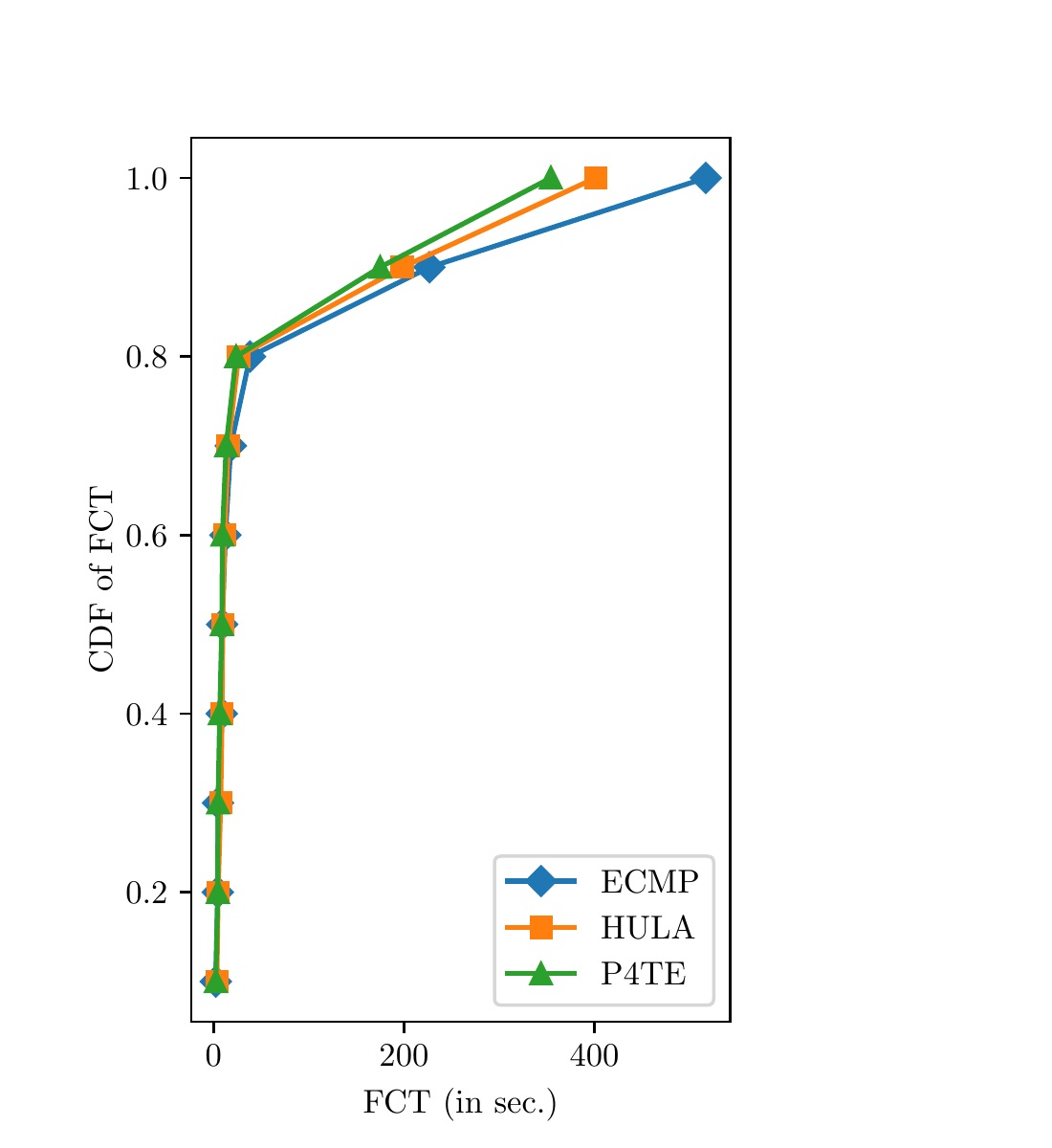}
        \caption{\centering {\small{CDF of FCT for all flows at 40\% load}}}
        \label{fig:10}
      \end{subfigure}
      \begin{subfigure}[]{.48\columnwidth}
        \centering
        \includegraphics[trim=0in 0in 2in 0in, clip,scale=.5]{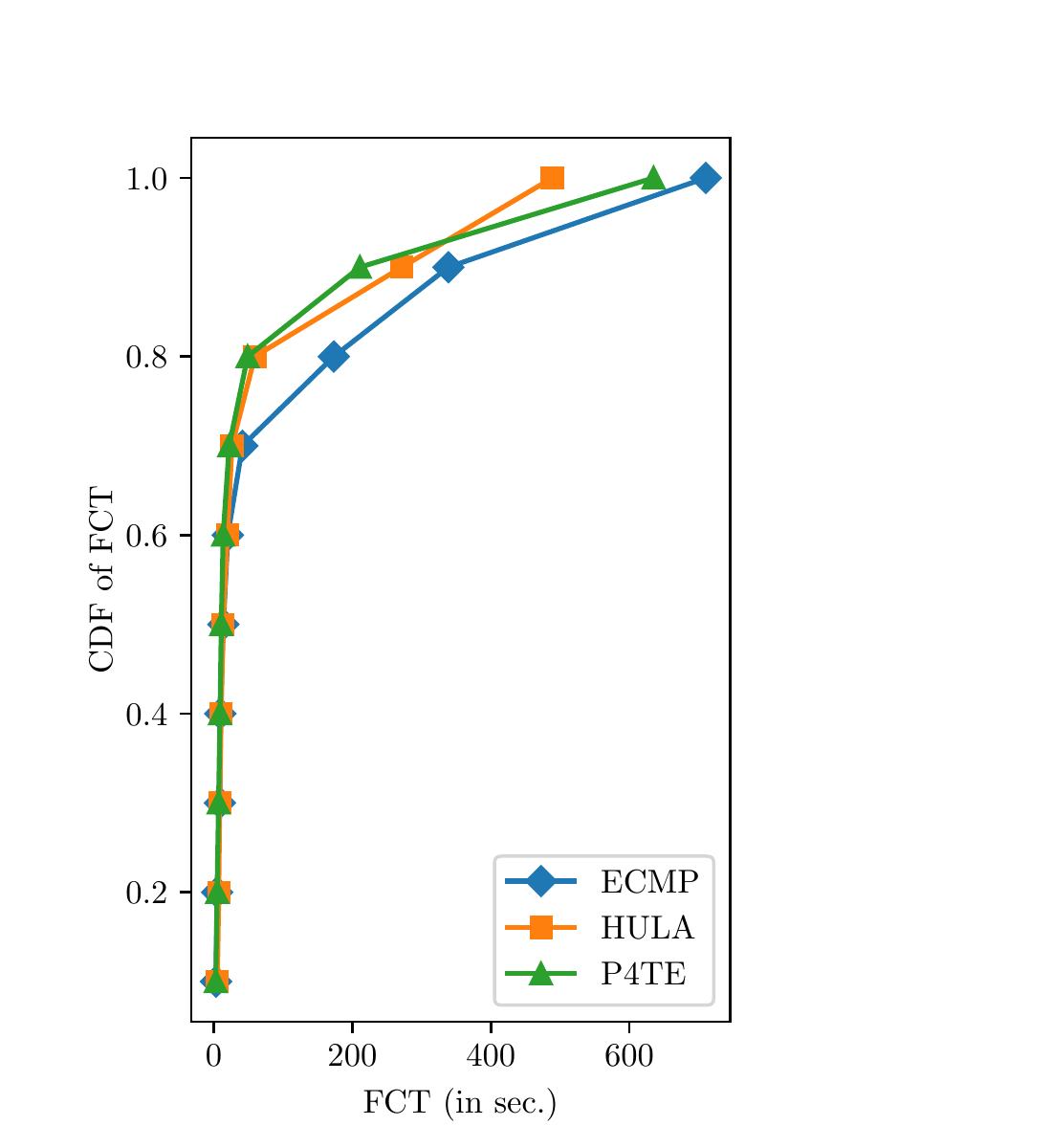}
        \caption{\centering {\small{CDF of FCT for all flows at 60\% load}}}
        \label{fig:11}
      \end{subfigure}
      \begin{subfigure}[]{.48\columnwidth}
        \centering
        \includegraphics[trim=0in 0in 2in 0in, clip,scale=.5]{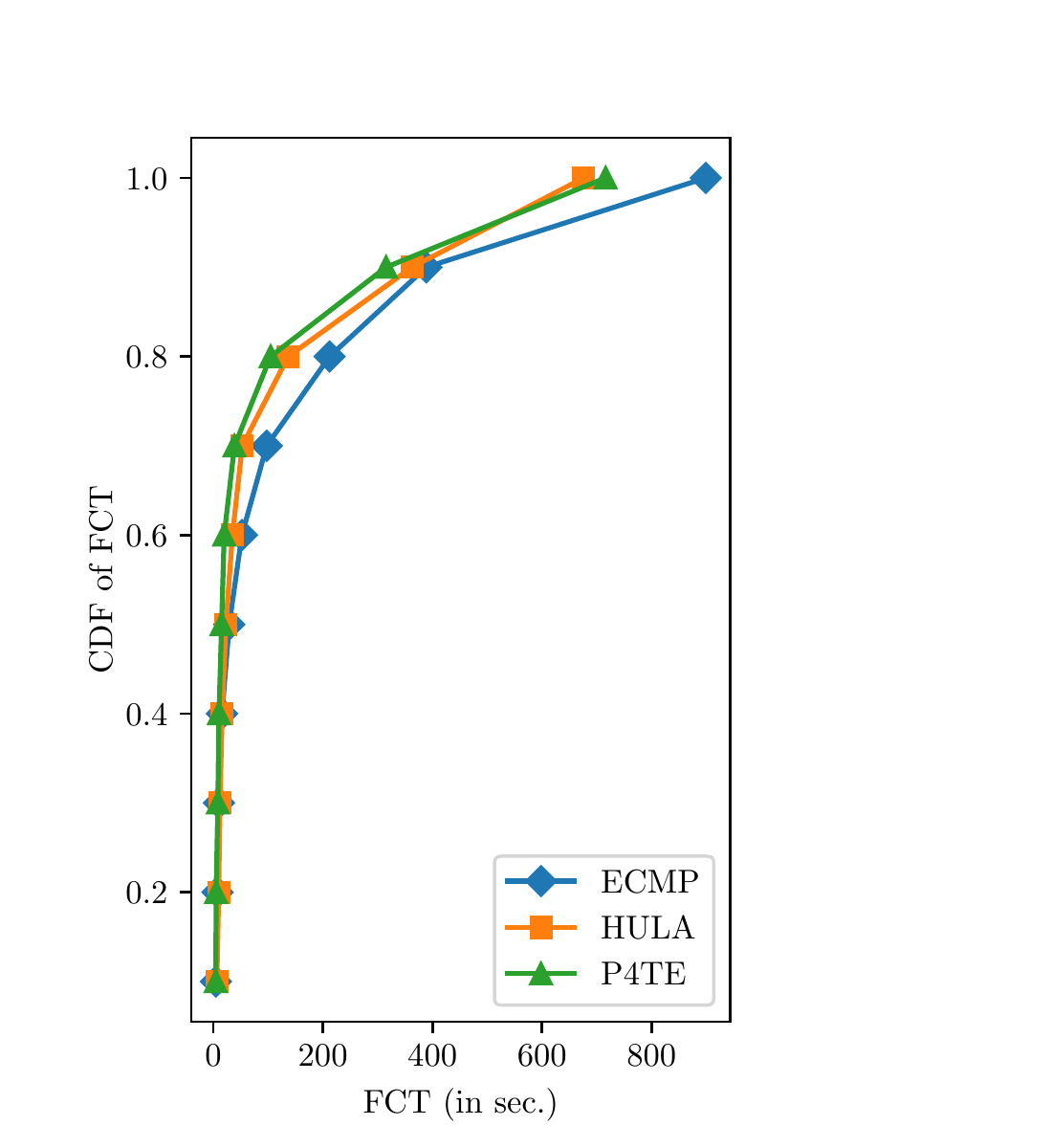}
        \caption{\centering {\small{CDF of FCT for all flows at 80\% load}}}
        \label{fig:12}
      \end{subfigure}
  
  \caption{\centering{\small{CDF of flow completion time (FCT) for data mining workload}}}
  \label{fig:DataminingCDF}
    \end{figure*}

In short flows,
P4TE has shorter FCTs when the load is under 60\% (fig.~\ref{fig:1} and  fig.~\ref{fig:2}). 
In higher loads, HULA has shorter FCTs at the tail; the tail 10\% of flows when the
load is 60\% (fig.~\ref{fig:3}) and the tail 4\% of flows when the load is 80\% (fig.~\ref{fig:4}).
In large flows, P4TE has shorter FCTs across all loads (fig.~\ref{fig:5} through fig.~\ref{fig:8}) 
except for 20\%. When the load is
20\% (fig.~\ref{fig:5}), P4TE has a shorter FCTs in the tail 20\% of the flows.
Across all flows, P4TE has strictly shorter 
FCT compared to ECMP and
HULA  when the load is low (20\% and 40\%).
However, HULA has better tail latencies when the load is high (60\% and 80\%): tail 10\%
when the load is 60\% (fig.~\ref{fig:11}) and tail 4\% when the load is 80\% (fig.~\ref{fig:12}).


\subsubsection{Web Search  Workload}\label{WebSearchCDF}

\begin{figure*}[t]
    \begin{subfigure}[]{.48\columnwidth}
      \centering
      \includegraphics[trim=0in 0in 2in 0in, clip,scale=.5]{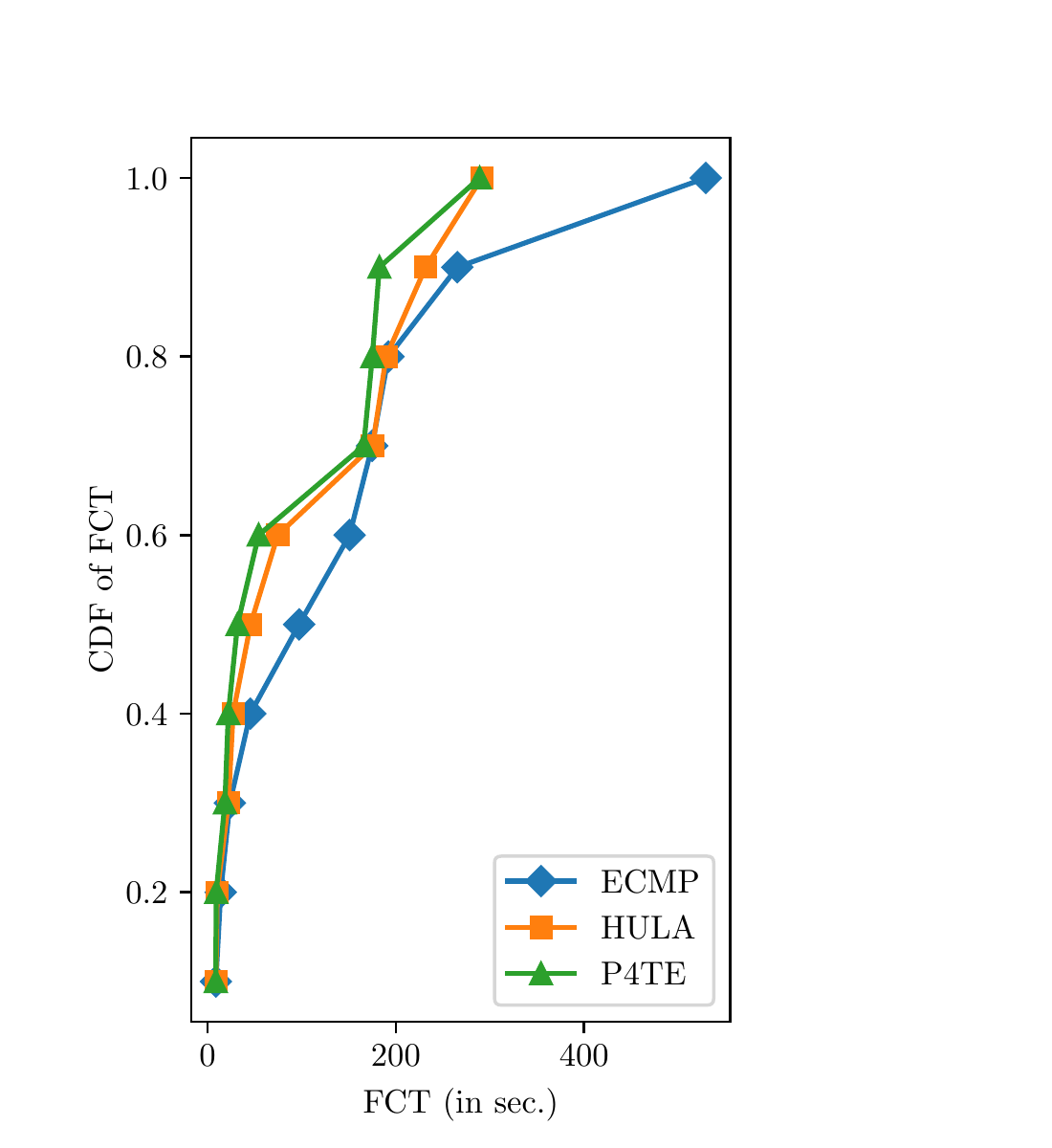}
      \caption{\centering {\small{CDF of FCT for short flows at 20\% load}}}
      \label{fig:13}
    \end{subfigure}
    \begin{subfigure}[]{.48\columnwidth}
      \centering
      \includegraphics[trim=0in 0in 2in 0in, clip,scale=.5]{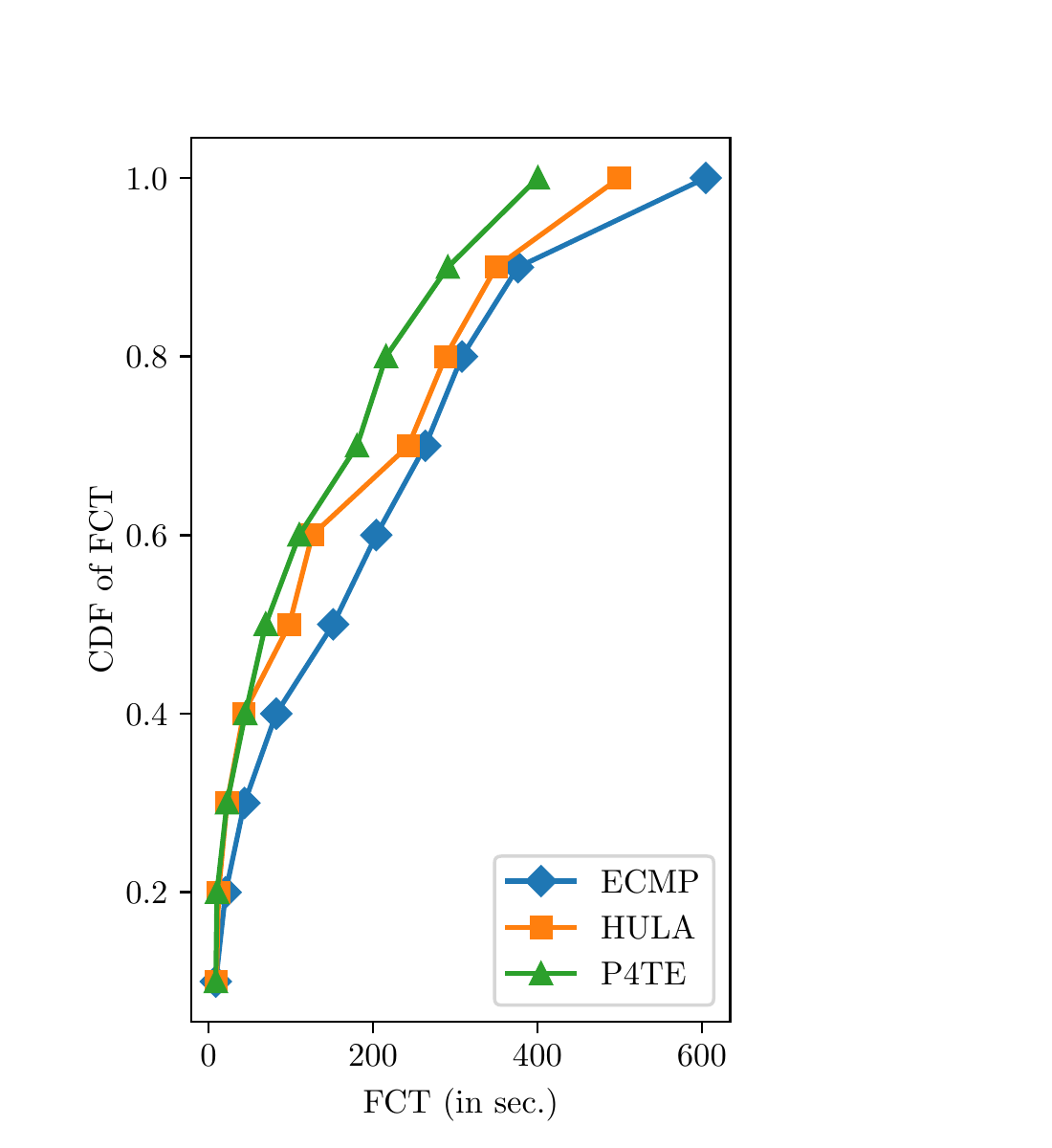}
      \caption{\centering {\small{CDF of FCT for short flows at 40\% load}}}
      \label{fig:14}
    \end{subfigure}
    \begin{subfigure}[]{.48\columnwidth}
      \centering
      \includegraphics[trim=0in 0in 2in 0in, clip,scale=.5]{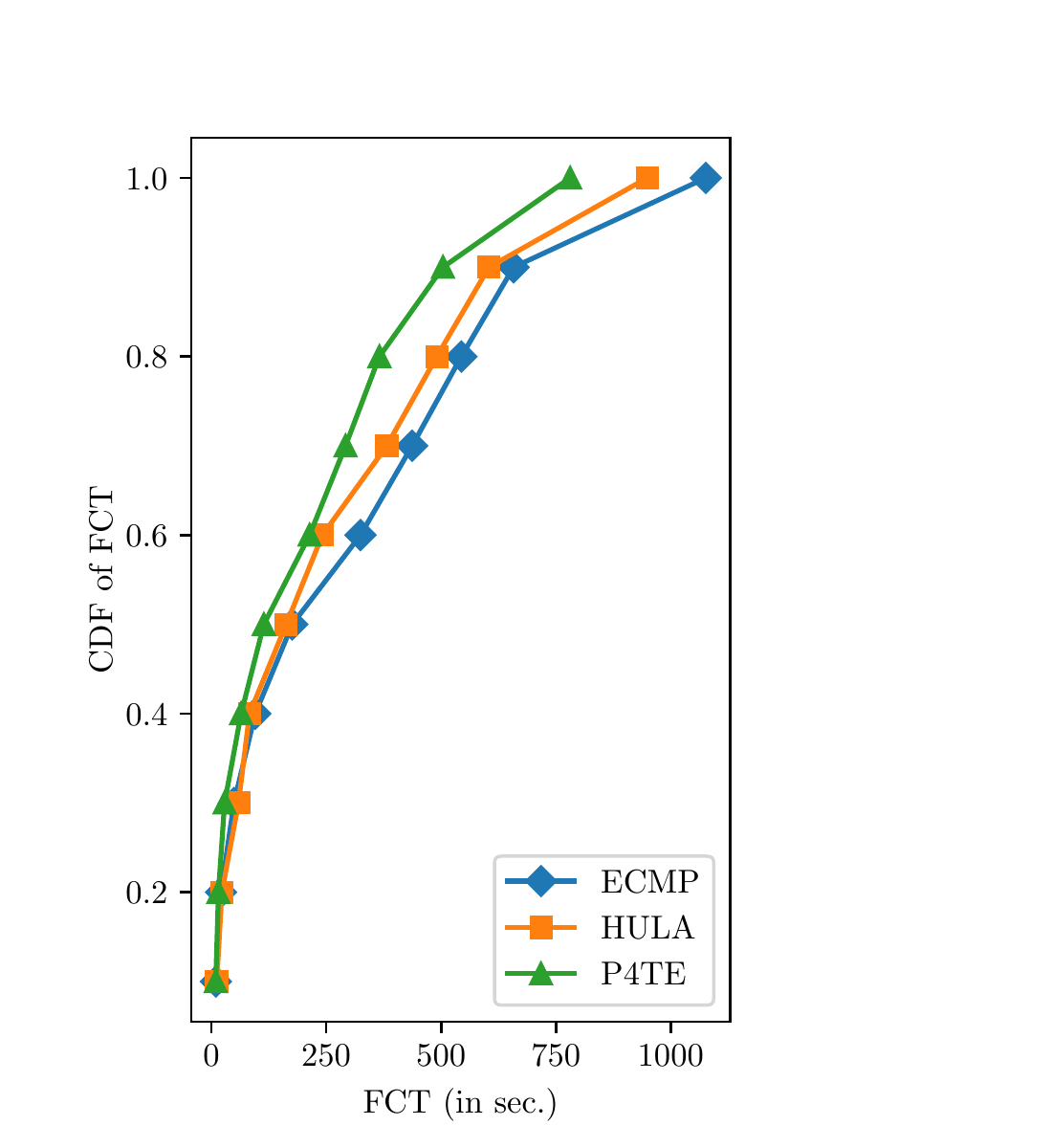}
      \caption{\centering {\small{CDF of FCT for short flows at 60\% load}}}
      \label{fig:15}
    \end{subfigure}
    \begin{subfigure}[]{.48\columnwidth}
      \centering
      \includegraphics[trim=0in 0in 2in 0in, clip,scale=.5]{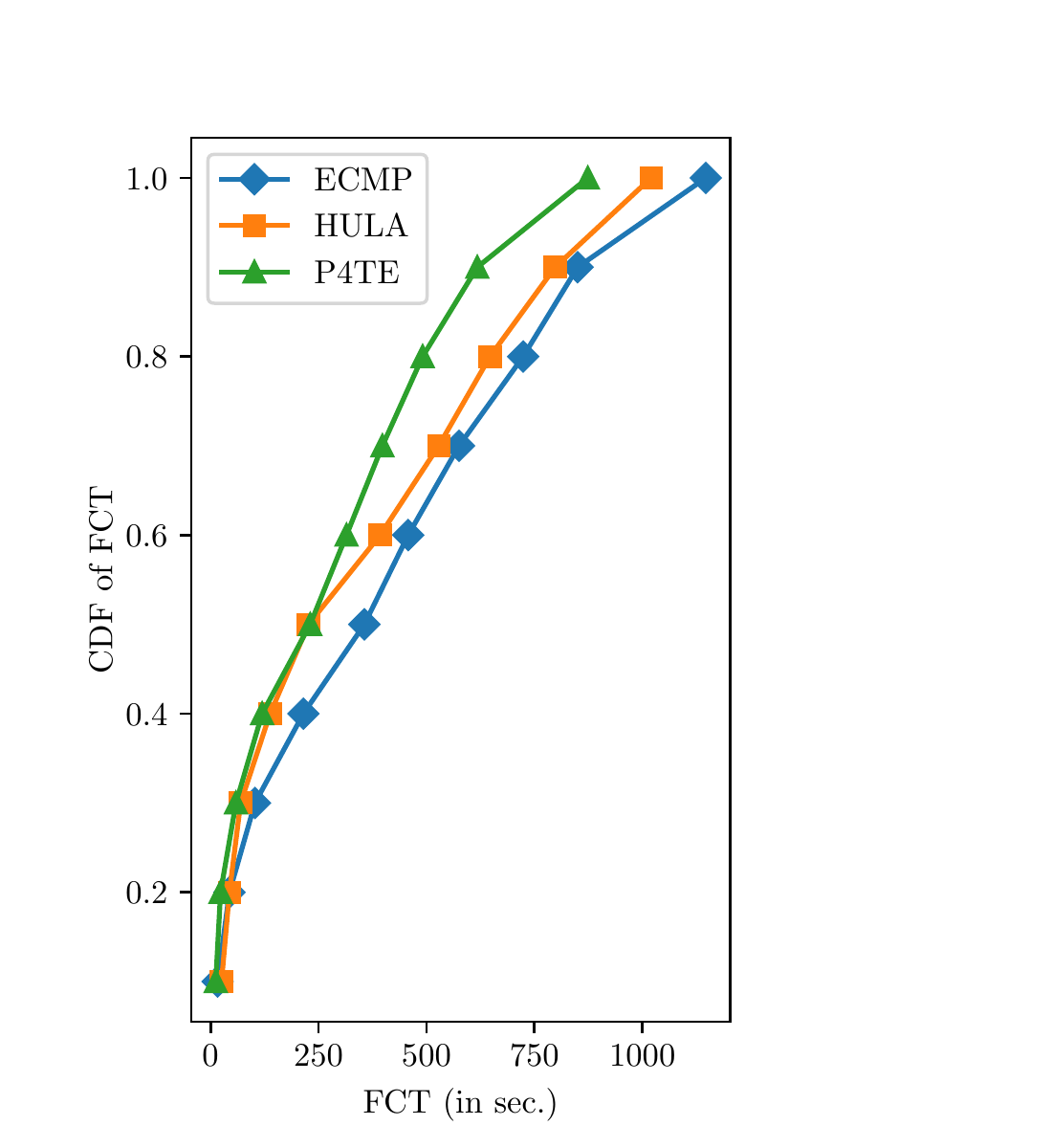}
      \caption{\centering {\small{CDF of FCT for short flows at 80\% load}}}
      \label{fig:16}
    \end{subfigure}
    \begin{subfigure}[]{.48\columnwidth}
      \centering
      \includegraphics[trim=0in 0in 2in 0in, clip,scale=.5]{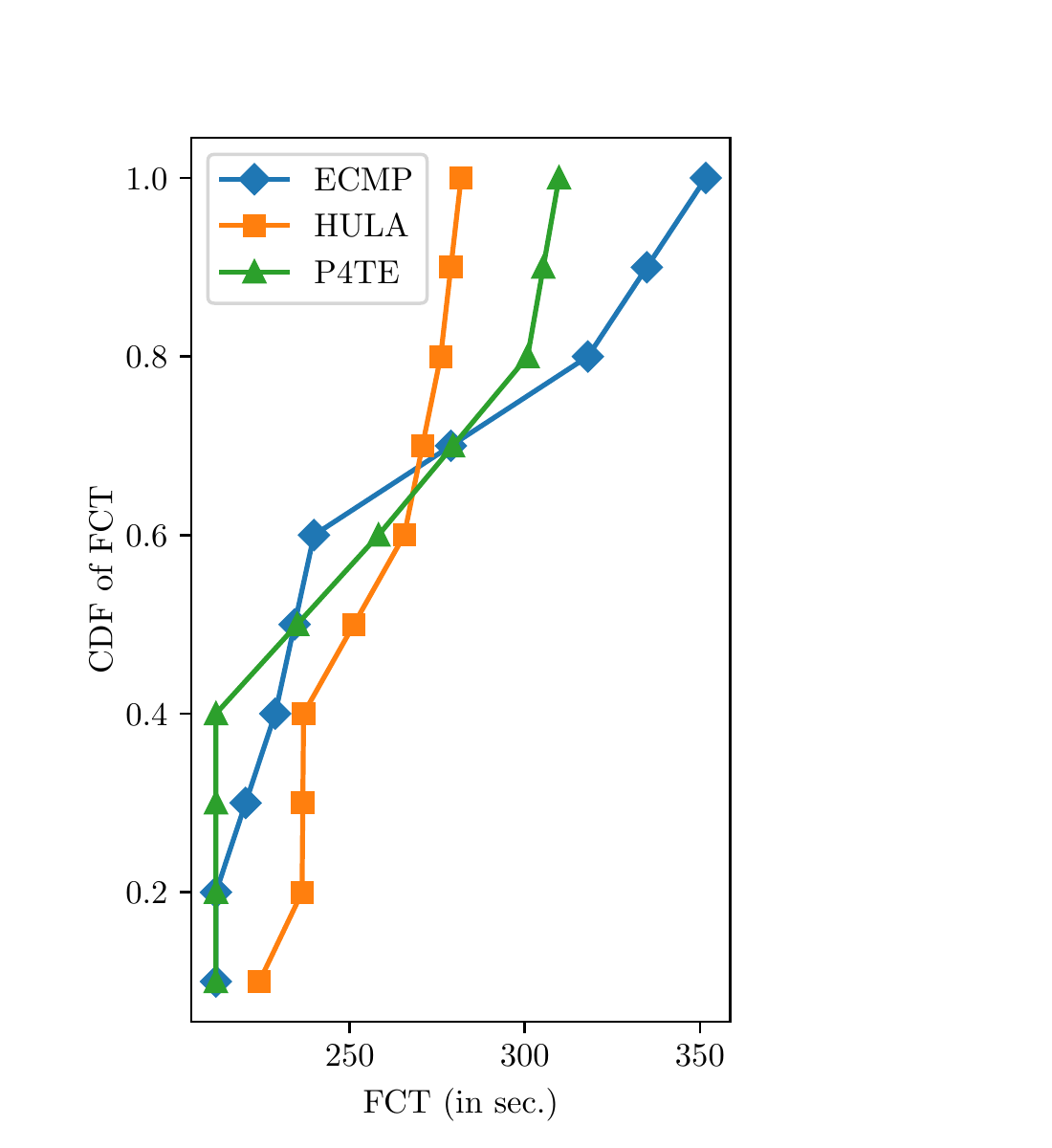}
      \caption{\centering {\small{CDF of FCT for large flows at 20\% load}}}
      \label{fig:17}
    \end{subfigure}
    \begin{subfigure}[]{.48\columnwidth}
      \centering
      \includegraphics[trim=0in 0in 2in 0in, clip,scale=.5]{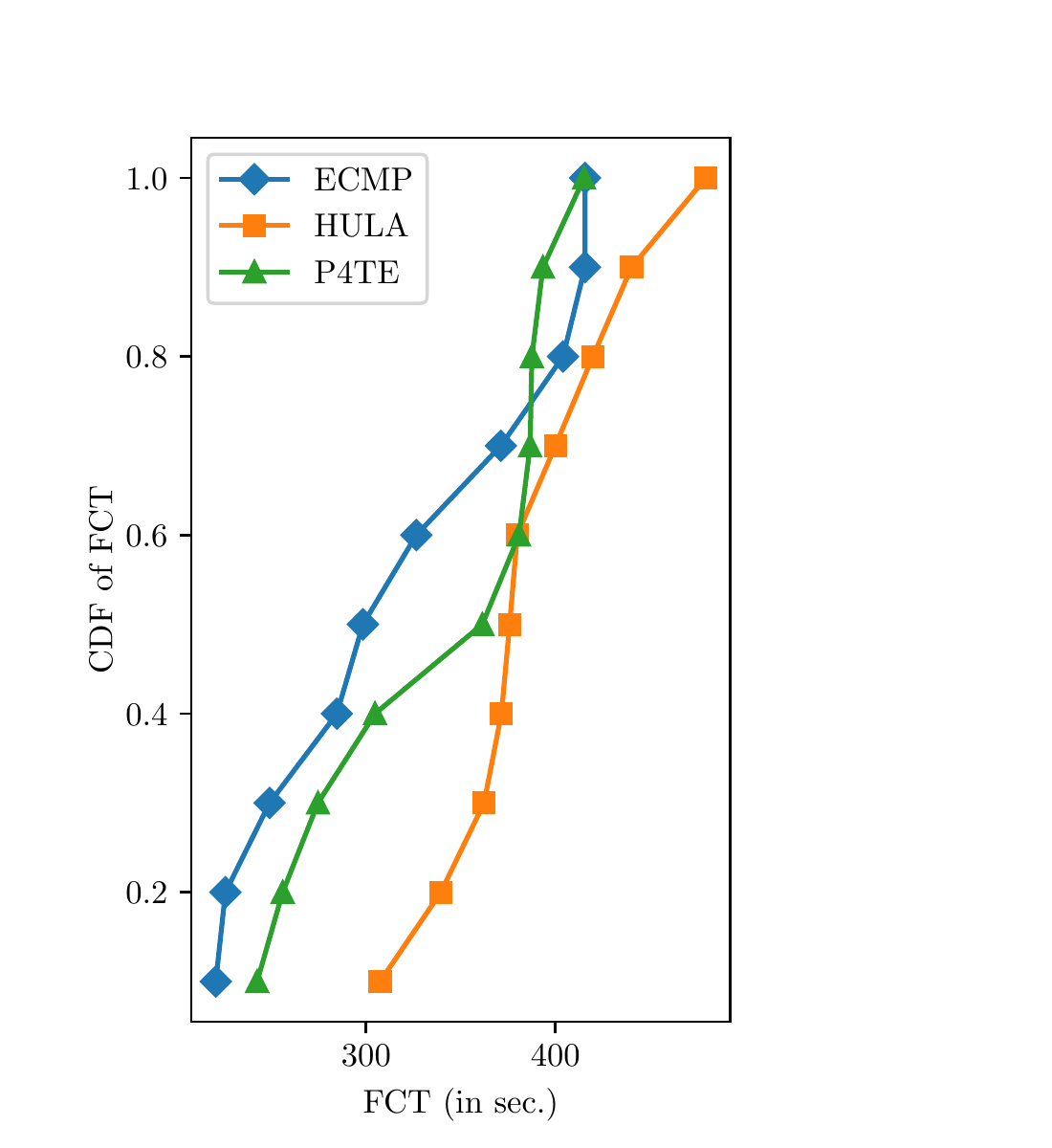}
      \caption{\centering {\small{CDF of FCT for large flows at 40\% load}}}
      \label{fig:18}
    \end{subfigure}
    \begin{subfigure}[]{.48\columnwidth}
      \centering
      \includegraphics[trim=0in 0in 2in 0in, clip,scale=.5]{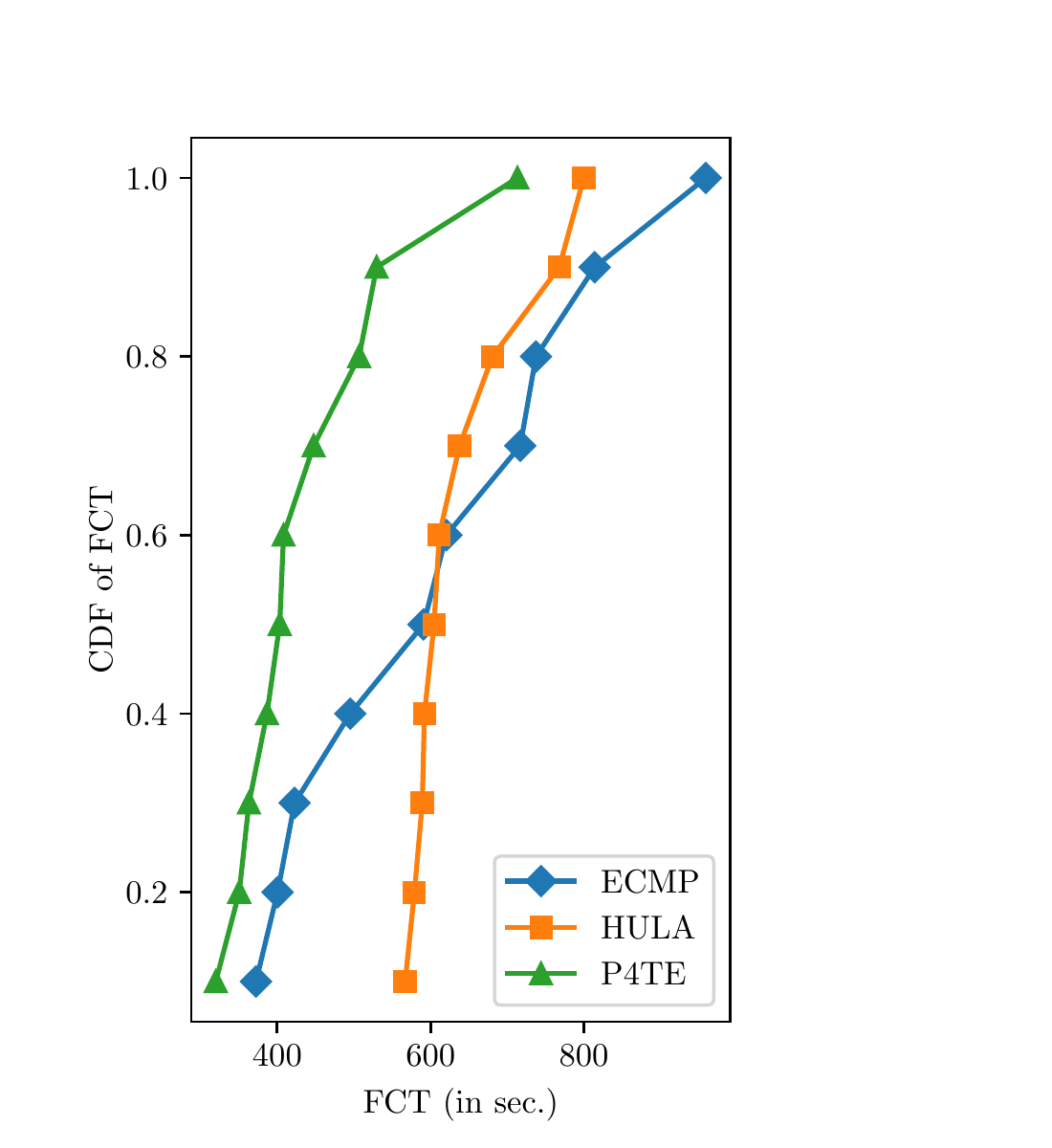}
      \caption{\centering {\small{CDF of FCT for large flows at 60\% load}}}
      \label{fig:19}
    \end{subfigure}
    \begin{subfigure}[]{.48\columnwidth}
      \centering
      \includegraphics[trim=0in 0in 2in 0in, clip,scale=.5]{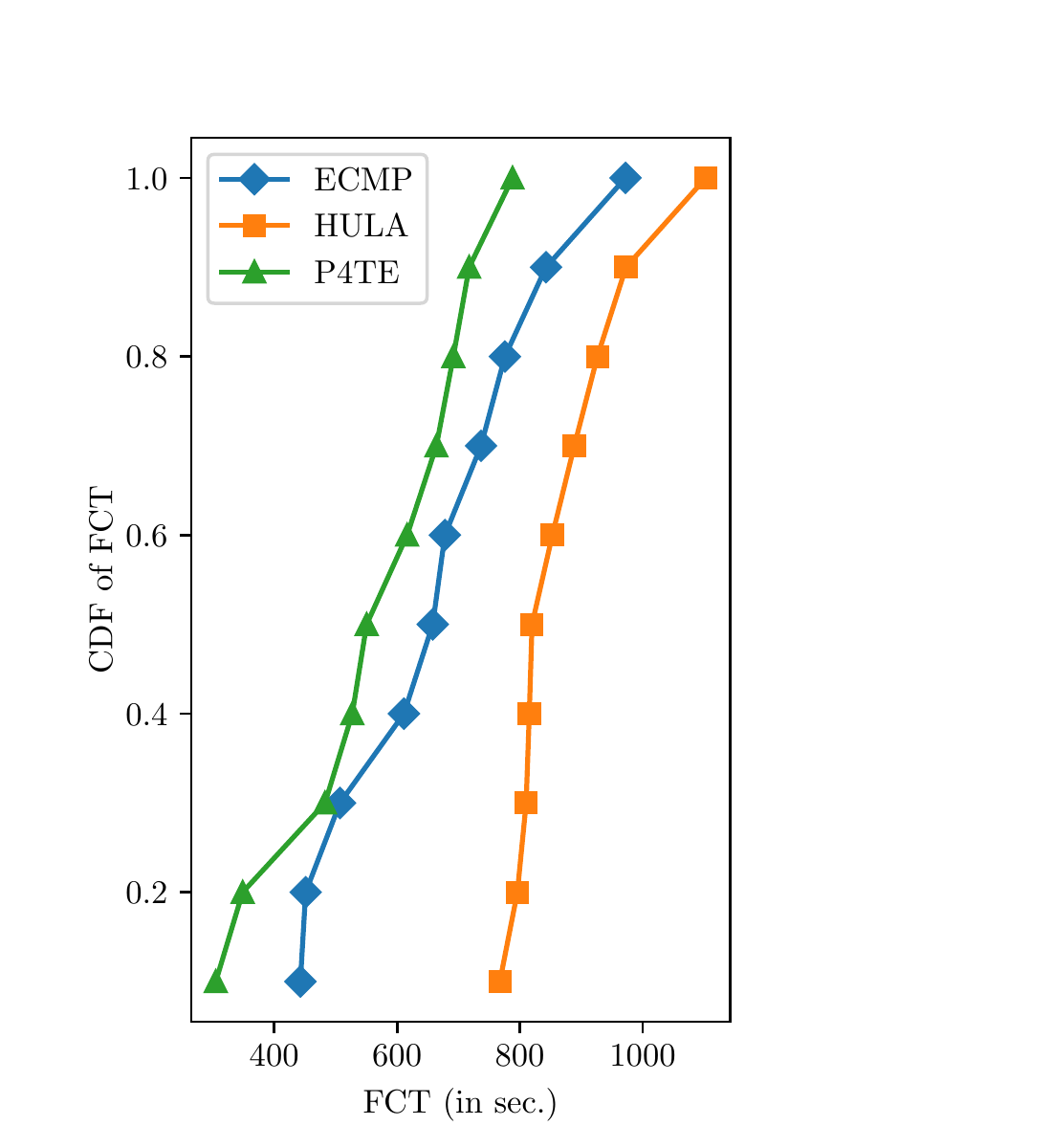}
      \caption{\centering {\small{CDF of FCT for large flows at 80\% load}}}
      \label{fig:20}
    \end{subfigure}
    \begin{subfigure}[]{.48\columnwidth}
      \centering
      \includegraphics[trim=0in 0in 2in 0in, clip,scale=.5]{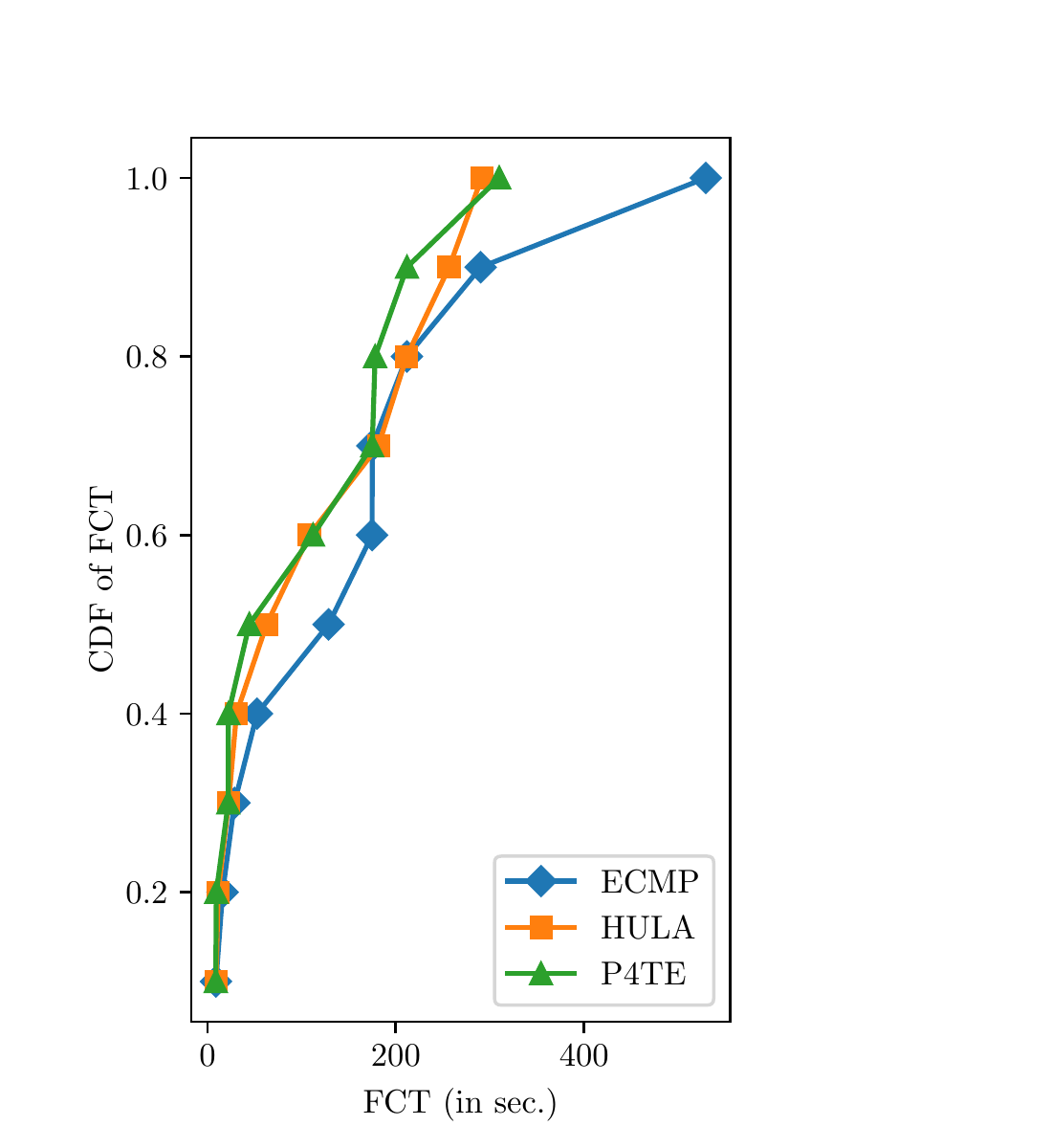}
      \caption{\centering {\small{CDF of FCT for all flows at 20\% load}}}
      \label{fig:21}
    \end{subfigure}
    \begin{subfigure}[]{.48\columnwidth}
      \centering
      \includegraphics[trim=0in 0in 2in 0in, clip,scale=.5]{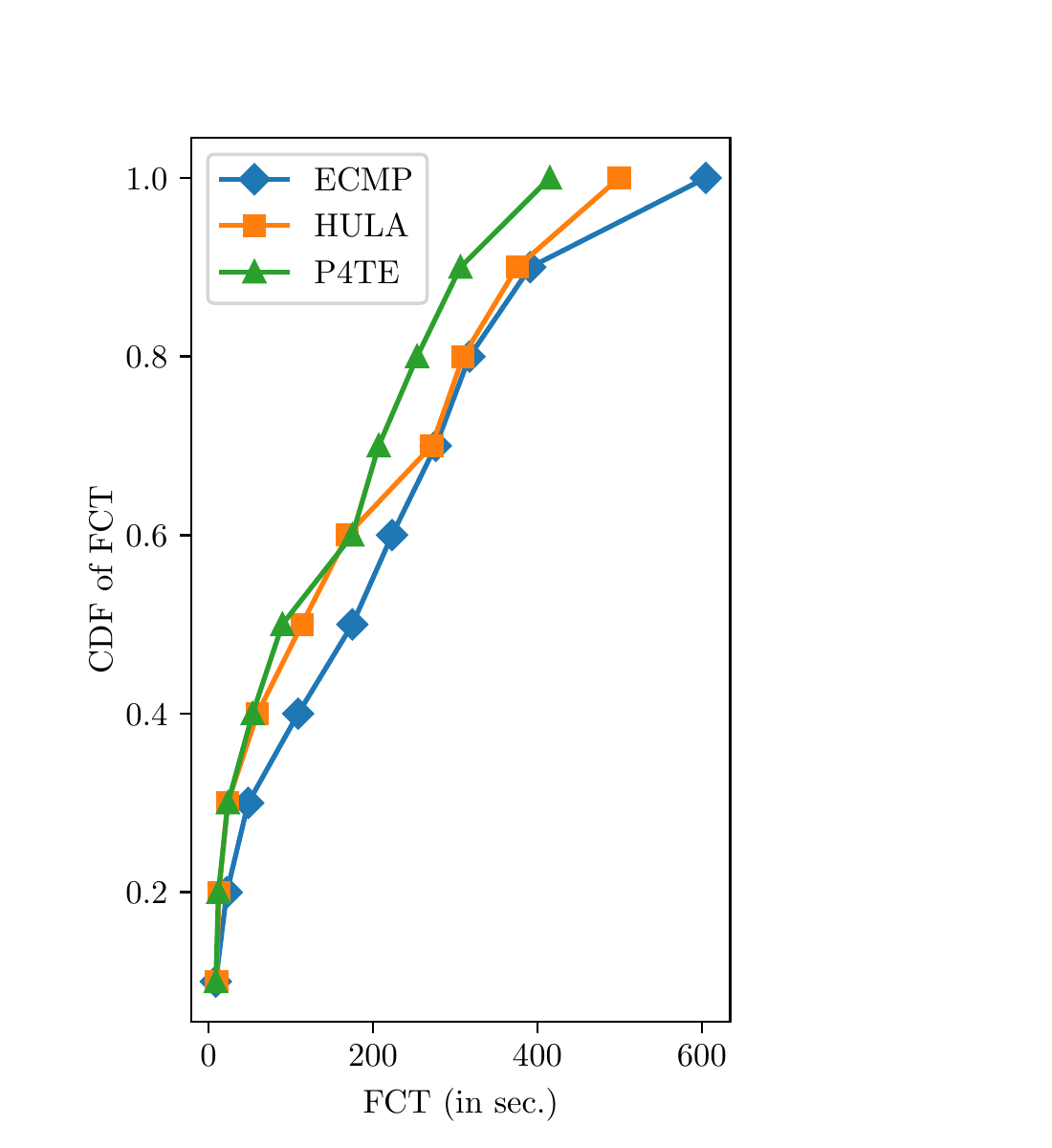}
      \caption{\centering {\small{CDF of FCT for all flows at 40\% load}}}
      \label{fig:22}
    \end{subfigure}
    \begin{subfigure}[]{.48\columnwidth}
      \centering
      \includegraphics[trim=0in 0in 2in 0in, clip,scale=.5]{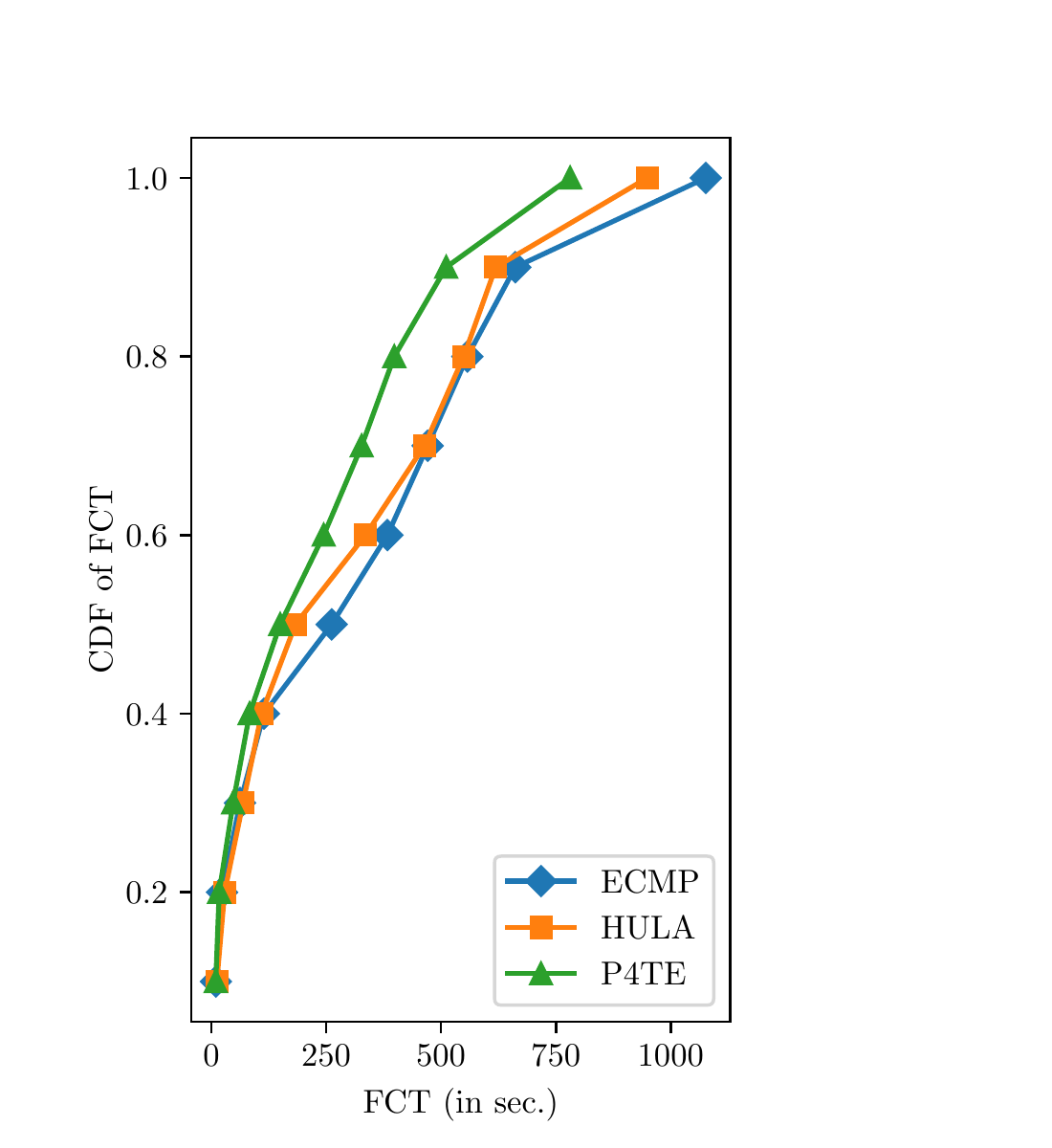}
      \caption{\centering {\small{CDF of FCT for all flows at 60\% load}}}
      \label{fig:23}
    \end{subfigure}
    \begin{subfigure}[]{.48\columnwidth}
      \centering
      \includegraphics[trim=0in 0in 2in 0in, clip,scale=.5]{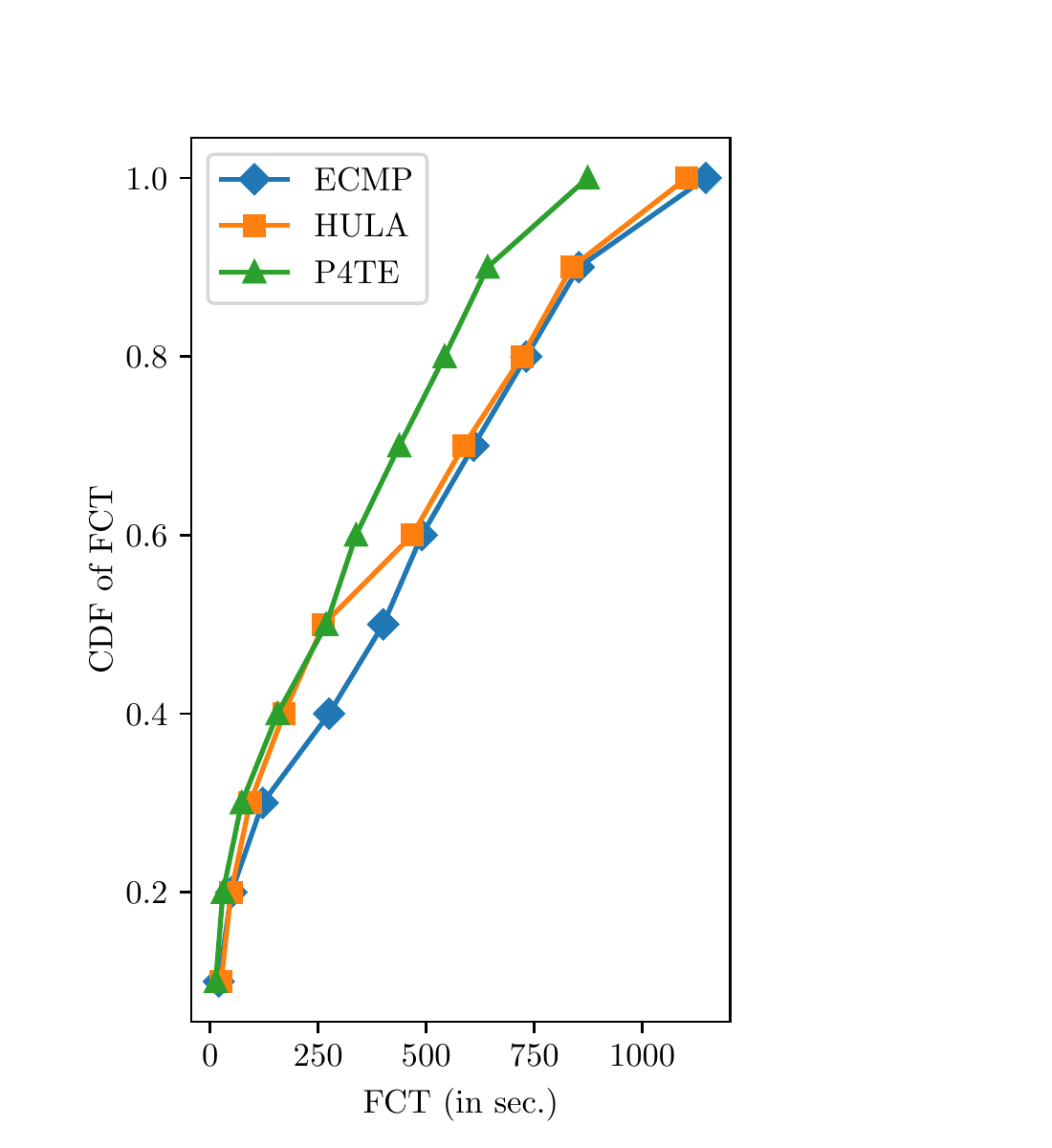}
      \caption{\centering {\small{CDF of FCT for all flows at 80\% load}}}
      \label{fig:24}
    \end{subfigure}

\caption{\centering{\small{CDF of flow completion time (FCT) for web search workload}}}
\label{fig:WebSearchCDF}
  \end{figure*}

In short flows, 
P4TE has strictly shorter FCTs compared to ECMP and HULA across all load levels (fig.~\ref{fig:13}-~\ref{fig:16}). 
In large flows, at 20\% load level,  HULA has a shorter FCT at the tail 30\% of the flows, and ECMP has a shorter 
FCT for a small portion of all the flows (fig.~\ref{fig:17}). 
At 40\% load level, P4TE has a shorter FCT at the tail 20\%, and 
ECMP performs better for the rest of the flows (fig.~\ref{fig:18}). 
However, at  60-80\% load level, P4TE has a strictly shorter 
FCT compared to ECMP and HULA (fig.~\ref{fig:19}-~\ref{fig:20}). 
Across all flows, P4TE has strictly shorter FCTs when the load is medium to high (40-80\%) 
(fig.~\ref{fig:22}-~\ref{fig:24}). 
However, at low load (20\%), HULA has a shorter FCT at the tail 2\%, and P4TE performs either similar to or 
better than HULA for the rest of the flows. 



\pagebreak

\bibliography{P4TE}



  

\end{document}